\documentclass[english]{article}
\usepackage[T1]{fontenc}
\usepackage[latin9]{inputenc}
\usepackage{geometry}
\usepackage{xcolor}
\usepackage{pdfcolmk}
\usepackage{float}
\usepackage{amsmath}
\usepackage{amssymb}
\usepackage{graphicx}
\PassOptionsToPackage{normalem}{ulem}
\usepackage{ulem}

\makeatletter

\providecommand{\tabularnewline}{\\}
\providecolor{lyxadded}{rgb}{0,0,1}
\providecolor{lyxdeleted}{rgb}{1,0,0}
\DeclareRobustCommand{\lyxsout}[1]{\ifx\\#1\else\sout{#1}\fi}

\numberwithin{equation}{section}
\newcommand{\lyxaddress}[1]{
	\par {\raggedright #1
	\vspace{1.4em}
	\noindent\par}
}

\makeatother

\usepackage{babel}
\begin{document}
\title{Leading exponential finite size corrections for non-diagonal form
factors}
\author{Zoltán Bajnok, Márton Lájer, Bálint Szépfalvi and István Vona}
\maketitle

\lyxaddress{\begin{center}
\emph{MTA Lendület Holographic QFT Group, Wigner Research Centre
for Physics}\\
\emph{Konkoly-Thege Miklós u. 29-33, 1121 Budapest, Hungary}\\
\emph{and}\\
\emph{Institute for Theoretical Physics, Roland Eötvös University,
}\\
\emph{Pázmány sétány 1/A, 1117 Budapest, Hungary}
\par\end{center}}
\begin{abstract}
We derive the leading exponential finite volume corrections in two
dimensional integrable models for non-diagonal form factors in diagonally
scattering theories. These formulas are expressed in terms of the
infinite volume form factors and scattering matrices. If the particles
are bound states then the leading exponential finite-size corrections
($\mu$-terms) are related to virtual processes in which the particles
disintegrate into their constituents. For non-bound state particles
the leading exponential finite-size corrections (F-terms) come from
virtual particles traveling around the finite world. In these F-terms
a specifically regulated infinite volume form factor is integrated
for the momenta of the virtual particles. The F-term is also present
for bound states and the $\mu$-term can be obtained by taking an
appropriate residue of the F-term integral. We check our results numerically
in the Lee-Yang and sinh-Gordon models based on newly developed Hamiltonian
truncations.
\end{abstract}

\section{Introduction}

Two dimensional integrable quantum field theories are hoped to be
exactly soluble. Theoretically, solvability allows us to find exact
values for all physical observables including the energy spectrum
and correlation functions at any finite size. However, even in integrable
theories this very progressive task has not been completed yet. Integrability
has only offered us a systematic way to attack these problems so far.

The first step of this systematic solution is to solve the theory
in infinite volume by completing the S-matrix and form factor (FF)
bootstraps \cite{Zamolodchikov:1978xm,Babujian:2003sc,Dorey:1996gd,Smirnov:1992vz}.
In infinite volume the powerful crossing symmetry can be used to derive
restrictive functional relations for the scattering matrix and for
the matrix elements of local operators, i.e. for form factors. Having
solved these functional relations the resulting S-matrix and FFs can
be used to describe all the finite size corrections systematically
as follows.

At finite size, the leading corrections are polynomial in the inverse
of the volume and originate from finite volume momentum quantization
\cite{Luscher:1986pf,Zamolodchikov:1989cf}. Periodicity of the wave
function requires that the scattering phase cancels the translational
phase when a particle is moved around the cylinder and scattered through
all other particles. The leading exponential corrections for bound
states (called $\mu$-terms) are related to the fact that in a finite
volume bound states can virtually decay into their constituents. The
next exponential corrections (F-terms) are caused by the polarization
of the non-trivial finite volume vacuum \cite{Luscher:1985dn}. Pairs
of virtual particles can appear from the vacuum. These travel around
the world and scatter on the physical particles, then annihilate each
other or get absorbed by the operators, such that this amplitude is
described by the infinite volume form factor. There could be any number
of virtual particles, which can also scatter on themselves. Thus,
for an exact description all these virtual processes have to be quantified
and summed up.

For the finite volume energy levels the momentum quantization is given
by the so-called Bethe-Yang equations \cite{Luscher:1986pf,Zamolodchikov:1989cf},
which provide the polynomial corrections. Leading exponential corrections
for standing one-particle states were identified in \cite{Luscher:1985dn}
and were later extended for a single moving particle in \cite{Klassen:1990ub,Janik:2007wt}.
The contribution of a single pair of virtual particles was extended
for multiparticle states in \cite{Bajnok:2008bm}, while the similar
contribution with two pairs of virtual particles was analyzed for
the vacuum in \cite{Ahn:2011xq}, and for multiparticle states in
\cite{Bombardelli:2013yka}. Finally, all virtual processes are summed
up by the Thermodynamic Bethe Ansatz (TBA) equation, which was derived
in the simplest case in \cite{Zamolodchikov:1989cf}. This provides
the exact finite volume ground state energy. Excited states can be
obtained by careful analytic continuations \cite{Dorey:1996re,Bajnok:2010ke}.

For finite volume form factors our understanding is much more restricted.
As far as polynomial corrections are concerned one merely has to take
into account momentum quantization and the corresponding change in
the normalization of states, which was proved for non-diagonal form
factors in \cite{Pozsgay:2007kn}. For diagonal form factors extra
disconnected terms appear \cite{Pozsgay:2007gx}, which can be derived
by carefully evaluating the diagonal limit of a non-diagonal form
factor \cite{Bajnok:2017bfg}. The finite volume one-point functions
can be expressed in terms of the infinite volume connected form factors
and the TBA pseudo energies \cite{Leclair:1999ys,Saleur:1999hq} in
a way summing up the contributions of virtual processes. This result
has been extended by analytic continuation for diagonal matrix elements
in diagonally scattering theories \cite{Pozsgay:2010xd,Pozsgay:2013jua,Pozsgay:2014gza}.
The expansion of these formulae provides the leading exponential corrections
for diagonal form factors. For non-diagonal form factors, however,
even these leading exponential corrections are not known in general.
For the simplest non-diagonal form factor (vacuum-one-particle state)
the leading exponential $\mu$-term corrections were obtained in \cite{Pozsgay:2008bf},
while the $F$-term correction in \cite{Bajnok:2018fzx}. The aim
of the present paper is to extend these analyses for generic non-diagonal
matrix elements in diagonally scattering theories. Although the $F$-term
calculation was based on the form factor expansion of the torus two-point
function \cite{Bajnok:2018fzx}, this method is very difficult to
generalize even considering the interesting developments in \cite{Pozsgay:2018tns,Cubero:2018vyb}.
We thus focus on a formal and direct derivation of the cylinder one-point
function in the crossed channel. We test the conjectured results by
comparing them to the $\mu$-term corrections and to numerical data
obtained from the combination of the Truncated Conformal Space Approach
(TCSA) and mini-superspace approaches newly developed for the sinh-Gordon
theory and from TCSA in the Lee-Yang model.\footnote{We note that very similar ideas appeared in an independent investigation
by Konik, Mussardo et al., see also footnote \ref{fn:footnote_mussardo}
at the beginning of Section 4.}

Our results provide the leading exponential corrections for form factors,
which contribute to the leading exponential correction to correlation
functions, too. These results can be relevant for various branches
of physics including finite temperature and finite volume correlation
functions in statistical and solid state systems as well as in lattice
gauge theories, where the size of the system is inherently finite
and finite size effects are unavoidable. Our results can be useful
in the AdS/CFT correspondence, too, where the calculation of correlation
functions boils down to the calculation of finite volume form factors
of nonlocal operators \cite{Bajnok:2015hla} or, alternatively, it
can be obtained by gluing hexagon \cite{Basso:2015zoa} and octagon
\cite{Bajnok:2017mdf} amplitudes. This gluing procedure is analogous
to the calculation of finite size effects of form factors and requires
a regularization procedure \cite{Basso:2017muf}. Thus, our systematic
method which gives rise to a regulated form factor could be implemented
there as well.

The paper is organized as follows: In Section 2 we review the exact
results for the finite size corrections of the energy spectrum. We
start this by describing the existing excited state TBA equations
for the sinh-Gordon and Lee-Yang models and expanding them iteratively
to second order. We extract the $\mu$- and $F$-term corrections
and demonstrate how the $\mu$- terms can be obtained from the $F$-terms
by calculating appropriate residues. Section 3 deals with the finite
size corrections of non-diagonal form factors. We first review the
asymptotic results for polynomial corrections. Assuming the particles
are bound states the asymptotic results provide the $\mu$-term corrections,
which we derive in a compact form. Afterwards, we provide a formal
derivation of the F-term corrections, relating them to the $\mu$-terms
subsequently. In Section 4 we check numerically our formulas in the
sinh-Gordon and Lee-Yang models and conclude in Section 5. Technical
details are relegated to the Appendices.

\section{Finite volume energy spectrum}

In this section we recall how the TBA equations provide an exact description
of the energy spectrum. We focus on theories with diagonal scatterings.

In the simplest case the theory has a single particle with mass $m$.
Multiparticle scatterings factorize into the product of two-particle
scatterings, with S-matrix $S(\theta)$, which satisfies unitarity
and crossing symmetry 
\begin{equation}
S(-\theta)=S^{-1}(\theta)\quad;\qquad S(i\pi-\theta)=S(\theta)
\end{equation}
Here $\theta$ is the rapidity difference of the particles $\theta=\theta_{1}-\theta_{2}$.
The simplest non-trivial scattering matrix is 
\begin{equation}
S(\theta)=\frac{\sinh\theta-i\sin p\pi}{\sinh\theta+i\sin p\pi}
\end{equation}
For $p>0$ there is no singularity in the physical strip $\Im m(\theta)\in[0,\pi]$
and the scattering matrix corresponds to the sinh-Gordon theory. However,
if $p<0$ bound states have to be introduced to explain the appearing
poles. For $p=-\frac{2}{3}$ the scattering matrix 
\begin{equation}
S(\theta)=\frac{\sinh\theta+i\sin\frac{\pi}{3}}{\sinh\theta-i\sin\frac{\pi}{3}}
\end{equation}
satisfies the relation
\begin{equation}
S(\theta+iu)S(\theta-iu)=S(\theta)\quad;\qquad u=\frac{\pi}{3}\label{eq:fusion}
\end{equation}
which, together with the bound state energy relation
\begin{equation}
m\cosh\theta=m\cosh(\theta+iu)+m\cosh(\theta-iu)\label{eq:boundstate}
\end{equation}
implies that the bound state is the original particle itself. This
theory is a consistent scattering theory \cite{Cardy:1989fw}, called
the scaling Lee-Yang model.

\subsection{Sinh-Gordon finite size spectrum}

The exact finite volume energy spectrum can be obtained by calculating
the continuum limit of an integrable lattice regularization \cite{Teschner:2007ng}.
A finite volume multiparticle state can be described by the pseudo
energy $\epsilon(\theta\vert\{\theta_{j}\})$ and parameters $\{\bar{\theta}_{j}\}_{j=1,\dots,N}$
satisfying the non-linear integral equation 
\begin{equation}
\epsilon(\theta\vert\{\theta\})=mL\cosh\theta+\sum_{j}\log S(\theta-\theta_{j}-\frac{i\pi}{2})-\int_{-\infty}^{\infty}\frac{dv}{2\pi}\phi(\theta-v)\log(1+e^{-\epsilon(v\vert\{\theta\})})\label{eq:shGTBA}
\end{equation}
where $\phi(\theta)=-i\partial_{\theta}\log S(\theta)$ and the particles'
rapidities satisfy the quantization condition 
\begin{equation}
Q_{k}(\{\bar{\theta}\})=2\pi n_{k}\qquad;\qquad Q_{k}(\{\theta\})=-i\epsilon(\theta_{k}+\frac{i\pi}{2}\vert\{\theta\})-\pi\quad;\qquad k=1,\dots,N\label{eq:shGQQ}
\end{equation}
Here and from now on we abbreviate the set of rapidities $\{\theta_{j}\}_{j=1,\dots,N}$
as $\{\theta\}$. Given quantization numbers $n_{k}$, the rapidities
$\{\bar{\theta}\}$ and the pseudo energy $\epsilon(\theta\vert\{\bar{\theta}\})$
can be determined, which provide the finite volume energy of the multiparticle
state via
\begin{equation}
E_{\{n\}}(L)=m\sum_{j}\cosh\bar{\theta}_{j}-m\int_{-\infty}^{\infty}\frac{dv}{2\pi}\,\cosh v\,\log(1+e^{-\epsilon(v\vert\{\bar{\theta}\})})\label{eq:shGE}
\end{equation}
We note that both in(\ref{eq:shGTBA}) and (\ref{eq:shGE}) the terms
with the sum can be absorbed into the integral term by choosing a
contour which goes around the singularities of the integrands at $v=\bar{\theta}_{j}+i\frac{\pi}{2}$.
These zero of logarithm singularities are actually encoded in the
quantization conditions (\ref{eq:shGQQ}).

\subsubsection{Polynomial energy corrections}

The TBA equations admit a systematic large volume expansion. At leading
order, indicated by a superscript $(0)$, we drop the exponentially
small integral terms and arrive at 
\begin{equation}
\epsilon^{(0)}(\theta\vert\{\theta\})=mL\cosh\theta+\sum_{j}\log S(\theta-\theta_{j}-\frac{i\pi}{2})\label{eq:shGeps0}
\end{equation}
Asymptotic rapidities satisfy the Bethe-Yang equations 
\begin{equation}
Q_{k}^{(0)}(\{\bar{\theta}^{(0)}\})=2\pi n_{k}\quad;\qquad Q_{k}^{(0)}(\{\theta\})=mL\sinh\theta_{k}-i\sum_{j:j\neq k}\log S(\theta_{k,j})\label{eq:shGQ0}
\end{equation}
where $\theta_{k,j}=\theta_{k}-\theta_{j}$. This equation has a very
transparent meaning. Periodicity of the multiparticle wavefunction
requires that, when moving particle $k$ around the circle, the acquired
phase -- consisting of the translational and the scattering phases
-- has to be a multiple of $2\pi$.

The energy at leading order is simply the sum of the one-particle
energies
\begin{equation}
E_{\{n\}}^{(0)}(L)=m\sum_{j}\cosh\bar{\theta}_{j}^{(0)}\label{eq:shGE0}
\end{equation}
 incorporating all finite volume corrections, which are \emph{polynomial}
in the inverse of the volume.

\subsubsection{Leading exponential volume corrections}

The leading \emph{exponential} volume correction can be obtained by
iterating the exact equations once. At this order, denoted by superscript
$(1)$, we have 
\begin{equation}
\epsilon^{(1)}(\theta\vert\{\theta\})=mL\cosh\theta+\sum_{i}\log S(\theta-\theta_{i}-\frac{i\pi}{2})-\int_{-\infty}^{\infty}\frac{dv}{2\pi}\phi(\theta-v)e^{-\epsilon^{(0)}(v\vert\{\theta\})}\label{eq:shGeps1}
\end{equation}
and the quantization conditions get modified as 
\begin{equation}
Q_{k}^{(1)}(\{\bar{\theta}^{(1)}\})=2\pi n_{k}\quad;\qquad Q_{k}^{(1)}(\{\theta\})=Q_{k}^{(0)}(\{\theta\})+\partial_{k}\Phi(\{\theta\})\label{eq:shGQQ1}
\end{equation}
where $\partial_{i}\equiv\partial_{\theta_{i}}\equiv\frac{\partial}{\partial\theta_{i}}$
and 
\begin{equation}
\Phi(\{\theta\})=\int_{-\infty}^{\infty}\frac{dv}{2\pi}\prod_{j}S(v+i\frac{\pi}{2}-\theta_{j})e^{-mL\cosh v}\label{eq:shGPhi}
\end{equation}
The exponentially corrected energy is 
\begin{equation}
E_{\{n\}}^{(1)}(L)=m\sum_{i}\cosh\bar{\theta}_{i}^{(1)}-m\int_{-\infty}^{\infty}\frac{dv}{2\pi}\,\cosh v\,\prod_{j}S(v+i\frac{\pi}{2}-\bar{\theta}_{j}^{(0)})e^{-mL\cosh v}\label{eq:shGE1}
\end{equation}
which can be expressed also in terms of $\bar{\theta}_{i}^{(0)}$
as in \cite{Bajnok:2008bm}. The integral terms in all formulae above
are called the $F$-term corrections.

\subsection{Scaling Lee-Yang finite size spectrum}

The ground state TBA equation was derived in \cite{Zamolodchikov:1989cf},
and careful analytical continuation in the volume lead to the TBA
equations of excited states \cite{Dorey:1996re}. The same TBA equations
can be derived from a continuum limit of a lattice model as well \cite{Bajnok:2014fca}.
The TBA equations are formally the same as in the sinh-Gordon theory
except that each particle with rapidity $\bar{\theta}_{j}$ is represented
as a bound state of two 'elementary' particles of rapidities $\bar{\theta}_{j\pm}=\bar{\theta}_{j}\pm i\bar{u}_{j}$\footnote{Here both $\bar{\theta}_{j}$ and $\bar{u}_{j}$ are real parameters.}.
Thus, the pseudo energy equations are
\begin{equation}
\epsilon(\theta\vert\{\theta_{\pm}\})=mL\cosh\theta+\sum_{j,s=\pm}\log S(\theta-\theta_{js}-\frac{i\pi}{2})-\int_{-\infty}^{\infty}\frac{dv}{2\pi}\phi(\theta-v)\log(1+e^{-\epsilon(v\vert\{\theta_{\pm}\})})
\end{equation}
where $\{\theta_{\pm}\}$ is the shorthand for $\{\theta_{j\pm}\}$
and the quantization conditions are 
\begin{equation}
Q_{k\pm}(\{\bar{\theta}_{\pm}\})=2\pi n_{k\pm}\quad;\qquad Q_{k\pm}(\{\theta_{\pm}\})=-i\epsilon(\theta_{k\pm}+\frac{i\pi}{2}\vert\{\theta_{\pm}\})-\pi\quad;\qquad k=1,\dots,N\label{eq:LYQQ}
\end{equation}
It is advantageous to introduce the symmetric and antisymmetric combinations
of these equations

\begin{equation}
Q_{k}(\{\theta_{\pm}\})=Q_{k+}(\{\theta_{\pm}\})+Q_{k-}(\{\theta_{\pm}\})\quad;\qquad\bar{Q}_{k}(\{\theta_{\pm}\})=Q_{k+}(\{\theta_{\pm}\})-Q_{k-}(\{\theta_{\pm}\})
\end{equation}
The energy formula is also analogous to the sinh-Gordon theory: 
\begin{equation}
E_{\{n_{\pm}\}}(L)=m\sum_{js}\cosh\bar{\theta}_{js}-m\int_{-\infty}^{\infty}\frac{dv}{2\pi}\,\cosh v\,\log(1+e^{-\epsilon(v\vert\{\bar{\theta}_{\pm}\})})
\end{equation}

\subsubsection{Polynomial and $\mu$-term energy corrections}

Let us expand the equations as before by dropping the integral terms.
We indicate this order by a superscript $(\mu)$ on $\bar{\theta}_{j\pm}^{(\mu)}=\bar{\theta}_{j}^{(\mu)}\pm i\bar{u}_{j}^{(\mu)}$
as it contains both polynomially and exponentially small volume corrections.
Similarly to the sinh-Gordon case we assign the superscript $(0)$
for polynomial corrections only. The pseudo energy at this order is:
\begin{equation}
\epsilon^{(0)}(\theta\vert\{\theta_{\pm}\})=mL\cosh\theta+\sum_{js}\log S(\theta-\theta_{js}-\frac{i\pi}{2})
\end{equation}
while the BY equations read as $Q_{k\pm}^{(0)}(\{\bar{\theta}_{\pm}^{(\mu)}\})=2\pi n_{k\pm}$
with 
\begin{equation}
Q_{k\pm}^{(0)}(\{\theta_{\pm}\})=mL\sinh(\theta_{k\pm})-i\log S(\theta_{k\pm,k\mp})-i\sum_{j:j\neq k,s}\log S(\theta_{k\pm,js})\label{eq:LYQQ0}
\end{equation}
Focusing on the imaginary part of the equations we see that in the
$L\to\infty$ limit the term $imL\cosh\bar{\theta}_{j}^{(\mu)}\sin\bar{u}_{j}^{(\mu)}$
goes to $i\infty$. This can be compensated only by the bound state
pole of the scattering matrix
\begin{equation}
S(\theta)=i\frac{\Gamma^{2}}{\theta-2iu}+S_{0}+O(\theta-2iu),\quad\Gamma^{2}=-2\sqrt{3}
\end{equation}
which forces $\bar{u}_{j}$ to approach $u$ in the large volume limit.
Let us parametrize $\bar{u}_{j}^{(\mu)}$ as\footnote{We could indicate the relevant order by using $\delta\bar{u}_{j}^{(\mu)}$
instead of $\delta\bar{u}_{j}$, but since we do not go to higher
orders in $\delta\bar{u}_{j}$ we drop its superscript.} 
\begin{equation}
\bar{u}_{j}^{(\mu)}=u+\delta\bar{u}_{j}
\end{equation}
The relation $S(2i\bar{u}_{j}^{(\mu)})=\frac{\Gamma^{2}}{2\delta\bar{u}_{j}}+\dots$
together with (\ref{eq:LYQQ0}) imply that $\delta\bar{u}_{j}$ is
actually exponentially small in the volume. We can then expand the
equations for large volume in $\delta\bar{u}_{j}$.

At leading order we set $\delta\bar{u}_{j}$ to be zero, i.e. we keep
only the polynomial corrections and take $\bar{\theta}_{j\pm}^{(0)}=\bar{\theta}_{j}^{(0)}\pm iu$.
Using the fusion property of the scattering matrix one can see that
$Q_{k}^{(0)}(\{\bar{\theta}_{\pm}^{(0)}\})=Q_{k}^{(0)}(\{\bar{\theta}^{(0)}\})$.
The resulting formulas are exactly the same as the sinh-Gordon equations
(\ref{eq:shGeps0}-\ref{eq:shGE0}) with quantization numbers $n_{j}=n_{j+}+n_{j-}$.

At the leading non-vanishing order in $\delta\bar{u}_{k}$ the equation
for $\bar{Q}_{k}^{(0)}(\{\bar{\theta}\})$ determines $\delta\bar{u}_{k}$
as 
\begin{equation}
\delta\bar{u}_{k}=(-1)^{n_{k}}\frac{\Gamma^{2}}{2}e^{-mL\sin u\cosh\bar{\theta}_{k}^{(0)}}\prod_{j:j\neq k}\sqrt{\frac{S(\bar{\theta}_{k,j}^{(0)}+iu)}{S(\bar{\theta}_{k,j}^{(0)}-iu)}}
\end{equation}
Clearly this expression is at least as small as $e^{-\mu L}$ with
$\mu=m\sin u$ and that is why we only kept the polynomial corrections
in the $\theta$s, using $\bar{\theta}_{j}^{(0)}$ here. Alternatively,
we could determine $\delta\bar{u}_{k}$ from the expansion of the
two equations for $Q_{k\pm}^{(0)}(\{\theta_{\pm}\})$. By introducing
\begin{equation}
\delta u_{k\pm}(\{\theta\})=\frac{\Gamma^{2}}{2}e^{\pm im_{a}L\sinh(\theta_{k}\pm iu)}\prod_{j:j\neq k}S(\theta_{k,j}\pm iu)^{\pm1}
\end{equation}
the solutions of the Bethe-Yang equations will be
\begin{equation}
\delta\bar{u}_{k}=\delta u_{k+}(\{\bar{\theta}^{(0)}\})=\delta u_{k-}(\{\bar{\theta}^{(0)}\})
\end{equation}
Using these quantities the Bethe-Yang equations for $\bar{\theta}_{j}$
at first order in $\delta\bar{u}_{j}$, i.e. at order $(\mu)$, takes
the form 
\begin{equation}
Q_{j}^{(\mu)}(\{\bar{\theta}^{(\mu)}\})=2\pi n_{j}\quad;\qquad Q_{j}^{(\mu)}(\{\theta\})=Q_{j}^{(0)}(\{\theta\})+\partial_{j}\sum_{k}(\delta u_{k+}(\{\theta\})+\delta u_{k-}(\{\theta\}))\label{eq:QQmu}
\end{equation}
where we used that $\partial_{j}\delta u_{k\pm}(\{\theta\})=\pm i\delta u_{j\pm}(\{\theta\})\partial_{j}Q_{k\pm}^{(0)}(\{\theta\pm iu\})$
(and the bound state relations (\ref{eq:fusion}-\ref{eq:boundstate})).

Thus, dropping the integral terms in the TBA equations not only gives
the polynomial corrections, but also provides the leading exponential
$\mu$-term corrections. This can be seen in the energy formula as
well, which at leading order reads as
\begin{equation}
E_{\{n\}}^{(\mu)}(L)=m\sum_{j,s}\cosh\bar{\theta}_{js}^{(\mu)}=m\sum_{j}\cosh\bar{\theta}_{j}^{(\mu)}-2m\sin u\sum_{j}\cosh\bar{\theta}_{j}^{(0)}\delta\bar{u}_{j}\label{eq:mu1}
\end{equation}
We note that here $\bar{\theta}_{j}^{(\mu)}$ also contains exponentially
small corrections coming form the quantization condition (\ref{eq:QQmu}),
which involves $\mu$-terms.

\subsubsection{F-term energy correction}

To iterate the integral equations once we use the leading order term
in the integrand. These formulas are completely equivalent to (\ref{eq:shGeps1}-\ref{eq:shGE1})
except that each rapidity comes in pairs, $\theta_{j\pm}$. These
equations contain both the $O(e^{-mL})$ and $O(e^{-(\mu+m)L})$ corrections.
In the following we are only interested in the $O(e^{-\mu L})$ and
$O(e^{-mL})$ corrections thus we put $\delta u_{j}=0$ in the integrands.
At this order, denoted by superscript $(1)$, we have 
\begin{eqnarray}
\epsilon^{(1)}(\theta\vert\{\theta\}) & = & mL\cosh\theta+\sum_{j}\log S(\theta-\theta_{j}-\frac{i\pi}{2})+\sum_{j}\phi(\theta-iu-\frac{i\pi}{2}-\theta_{j})\delta u_{j-}(\{\theta\})\label{eq:LYeps1}\\
 &  & -\sum_{j}\phi(\theta+iu-\frac{i\pi}{2}-\theta_{j})\delta u_{j+}(\{\theta\})-\int_{-\infty}^{\infty}\frac{dv}{2\pi}\phi(\theta-v)\prod_{j}S(v+i\frac{\pi}{2}-\theta_{j})e^{-mL\cosh(v)}\nonumber 
\end{eqnarray}
Since the quantization condition $\bar{Q}_{k}$ modifies $\delta u_{k}$
only at order $O(e^{-(m+\mu)L})$ we focus on $Q_{k}$. In addition
to (\ref{eq:QQmu}) we also get an integral term 
\begin{equation}
Q_{k}^{(1)}(\{\bar{\theta}\})=2\pi n_{k\pm}\quad;\qquad Q_{k}^{(1)}(\{\theta\})=Q_{k}^{(0)}(\{\theta\})+\partial_{k}\sum_{j}(\delta u_{j+}(\{\theta\})+\delta u_{j-}(\{\theta\}))+\partial_{k}\Phi(\{\theta\})\label{eq:LYQQ1}
\end{equation}
where $\Phi(\{\theta\})$ is the same as (\ref{eq:shGPhi}). The exponentially
corrected energy also gets the integral term 
\begin{equation}
E_{\{n\}}^{(1)}(L)=m\sum_{j}\cosh\bar{\theta}_{j}^{(1)}-2m\sin u\sum_{j}\cosh\bar{\theta}_{j}^{(0)}\delta u_{j}-m\int_{-\infty}^{\infty}\frac{d\theta}{2\pi}\,\cosh\theta\,\prod_{j}S(\theta+i\frac{\pi}{2}-\bar{\theta}_{j}^{(0)})e^{-mL\cosh\theta}\label{eq:LYE1}
\end{equation}

\begin{figure}
\begin{centering}
\includegraphics[width=10cm]{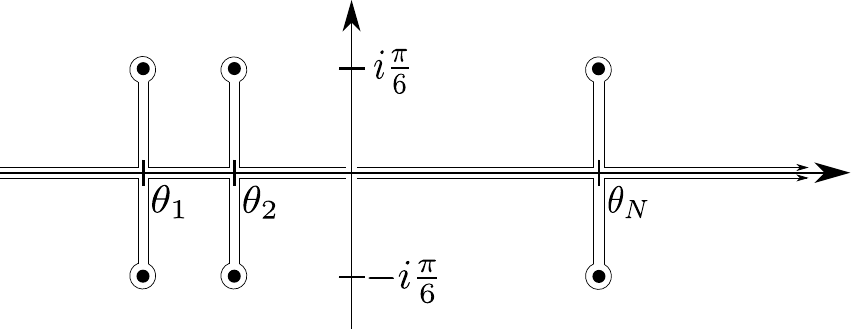}
\par\end{centering}
\caption{Integration contour, which contains both the $F$- and the $\mu$-terms.The
integrand is half the $F$-term integrand both on the upper and on
the lower contour.}

\label{fig:contour}
\end{figure}

In all the formulas (\ref{eq:LYeps1},\ref{eq:LYQQ1},\ref{eq:LYE1})
terms containing $\delta u_{j}$ are the $\mu$-terms, while the integral
terms are the $F$-terms. Note that the $F$-terms are universal in
the sense that they are the same for both theories once the corresponding
S-matrix is used. It is also very important for our further study
to point out that in the Lee-Yang theory the two corrections are not
independent: the $\mu$ terms can be obtained as appropriate residues
of the $F$-terms. Indeed, the scattering matrix not only has a pole
at $\theta=2iu=i\frac{2\pi}{3}$ but also at $\theta=i\pi-2iu=i\frac{\pi}{3}$
with opposite residue. This implies that $e^{-\epsilon^{(0)}(\theta\vert\{\theta\})}$
has poles at $\theta=\theta_{j}\pm i\frac{\pi}{6}$ with residues
\begin{equation}
\text{Res}_{\theta=\theta_{j}\pm i\frac{\pi}{6}}e^{-\epsilon^{(0)}(\theta\vert\{\theta\})}=\pm2i\delta u_{j\mp}(\{\theta\})
\end{equation}
 We can think of taking the real contour and deforming half of it
onto the upper half-plane and the other half to the lower half-plane.
Then we can subtract the two residues, which appear with opposite
orientations. As a result we can recover the $\mu$-terms from the
$F$-terms in all the formulas (\ref{eq:LYeps1}-\ref{eq:LYE1}).
Alternatively, we can choose the contours of integration as shown
in Figure \ref{fig:contour} and keep only the $F$-term integral,
which is universal and is the same for both theories.

\section{Finite volume form factors}

In this section we summarize the results for finite volume form factors.
We start by reviewing the definition of these quantities together
with the available results for the polynomial finite size corrections.
We then derive the leading exponential $\mu$- and $F$-term corrections
for general nondiagonal finite volume form factors. Technical details
are presented in Appendices A and B. In Appendix C we also show how
the $\mu$-term correction can be obtained from the $F$-term corrections.

Finite volume form factors are the matrix elements of local operators
${\cal O}(x,t)$ between finite volume energy eigenstates, which can
be labeled either by the quantization numbers $\{n_{i}\}$ or by the
corresponding rapidities $\{\bar{\theta}_{i}\}$:
\begin{equation}
\vert\bar{\theta}_{1},\dots,\bar{\theta}_{N}\rangle_{L}=\vert\{\bar{\theta}\}\rangle_{L}\equiv\vert n_{1},\dots,n_{N}\rangle_{L}=\vert\{n\}\rangle_{L}
\end{equation}
These rapidities satisfy the exact quantization conditions (\ref{eq:shGQQ})
or the related equations for the Lee-Yang theory (\ref{eq:LYQQ}).

Our aim is to express the \emph{finite} volume form factors in terms
of the scattering matrix and the \emph{infinite} volume elementary
form factors defined by\footnote{Infinite volume form factors are normalized as $\langle\theta\vert\theta^{'}\rangle=2\pi\delta(\theta-\theta^{'})$.}
\begin{equation}
\langle0\vert{\cal O}(0,0)\vert\theta_{1},\dots,\theta_{N}\rangle=F_{N}^{{\cal O}}(\theta_{1},\dots,\theta_{N})
\end{equation}
These infinite volume form factors satisfy the monodromy axioms: 
\begin{equation}
F_{N}(\theta_{1},\dots,\theta_{N})=F_{N}(\theta_{2},\dots,\theta_{N},\theta_{1}-2i\pi)=S(\theta_{i,i+1})F_{N}(\theta_{1},\dots,\theta_{i+1},\theta_{i},\dots,\theta_{N})\label{eq:FFax}
\end{equation}
which together with their known analytic properties allows one to
find the relevant physical solutions.

Form factors have pole singularities, with either kinematical or dynamical
origin. The kinematical pole is related to disconnected diagrams and
appear whenever an outgoing particle coincides with an incoming one.
At the level of the elementary form factor this implies that
\begin{equation}
F_{N+2}(\theta+i\pi+\frac{\epsilon}{2},\theta-\frac{\epsilon}{2},\{\theta\})=\frac{i}{\epsilon}(1-\prod_{j}S(\theta-\theta_{j}))F_{N}(\{\theta\})+F_{N+2}^{r}(\theta+i\pi,\theta,\{\theta\})+O(\epsilon)
\end{equation}
where we introduced a specific symmetric evaluation, since the $O(1)$
piece defined this way, that we call the \emph{regulated} form factor,
will be relevant in the further discussions. The notation $\{\theta\}$
abbreviates the ordered set $\{\theta_{1},\dots,\theta_{N}\}$.

Dynamical pole singularities are only present for theories in which
the scattering matrix has a bound state pole. They relate the form
factors of elementary particles to those of bound states. For the
Lee-Yang model they read as 
\begin{equation}
F_{N+2}(\theta+iu+\frac{\epsilon}{2},\theta-iu-\frac{\epsilon}{2},\{\theta\})=\frac{i\Gamma}{\epsilon}F_{N+1}(\theta,\{\theta\})+F_{N+1}^{b}(\theta,\{\theta\})+O(\epsilon)
\end{equation}
where the symmetrically evaluated $O(1)$ piece will be used later
on. In particular, we will need the expansion 
\begin{align}
F_{2N}(\theta_{1}+iu+\frac{\epsilon_{1}}{2},\theta_{1}-iu-\frac{\epsilon_{1}}{2},\dots,\theta_{N}+iu+\frac{\epsilon_{N}}{2},\theta_{N}-iu-\frac{\epsilon_{N}}{2})=\,\,\,\,\,\,\,\,\,\,\,\,\,\,\,\,\,\,\,\,\,\,\,\label{eq:bkff}\\
\,\,\,\,\,\,\,\,\,\,\,\,\,\,\,\,\,\prod_{j}\left(\frac{i\Gamma}{\epsilon_{j}}\right)\left\{ F_{N}(\{\theta\})-i\Gamma^{-1}\sum_{k}\epsilon_{k}F_{N,k}^{b}(\{\theta\})+O(\epsilon^{2})\right\} \nonumber 
\end{align}

In diagonally scattering theories with a single species the form factors
take the form 
\begin{equation}
F_{N}^{{\cal O}}(\theta_{1},\dots,\theta_{N})=H_{N}^{{\cal O}}\prod_{i<j}\frac{f(\theta_{i,j})}{x_{i}+x_{j}}P_{N}^{{\cal O}}(x_{1},\dots,x_{N})\quad;\qquad x_{i}=e^{\theta_{i}}\label{eq:FFparam}
\end{equation}
where $f(\theta)$ is the minimal two particle form factor, which
satisfies $f(\theta)=S(\theta)f(-\theta)=f(2i\pi-\theta)$ and has
the right dynamical pole. In the sinh-Gordon theory it does not have
any singularity in the physical strip:
\begin{equation}
f(\theta)=\exp\left\{ -4\int_{0}^{\infty}\frac{dx}{x}\frac{\sinh(\frac{xp}{2})\sinh(\frac{x}{2}(1-p))\sinh(\frac{x}{2})}{\sinh^{2}(x)}\cos\left(\frac{x}{\pi}(i\pi-\theta)\right)\right\} 
\end{equation}
 while in the Lee-Yang theory it can be obtained by an analytic continuation
of this:
\begin{equation}
f(\theta)=\frac{\cosh\theta-1}{\cosh\theta+\frac{1}{2}}\exp\left\{ 4\int_{0}^{\infty}\frac{dx}{x}\frac{\sinh(\frac{x}{2})\sinh(\frac{x}{3})\sinh(\frac{x}{6})}{\sinh^{2}(x)}\cos\left(\frac{x}{\pi}(i\pi-\theta)\right)\right\} 
\end{equation}

\subsection{Polynomial finite volume corrections}

Let us analyze the following finite volume form factor
\begin{equation}
\langle\bar{\vartheta}_{1},\dots,\bar{\vartheta}_{M}\vert\mathcal{O}\vert\bar{\theta}_{1},\dots,\bar{\theta}_{N}\rangle_{L}\equiv\langle\{\bar{\vartheta}\}\vert\mathcal{O}\vert\{\bar{\theta}\}\rangle_{L}
\end{equation}
As far as polynomial corrections are concerned the rapidities $\{\bar{\theta}_{i}\}$
satisfy the Bethe-Yang equations (\ref{eq:shGQ0}) and there are analogous
equations for $\vartheta$s with quantization numbers $m_{k}$. As
it was shown in \cite{Pozsgay:2007kn} the polynomial finite volume
corrections of form factors solely come from the normalization change
of the finite volume states. These states are normalized to Kronecker
delta functions, while infinite volume states are normalized to Dirac
delta functions. Additionally, the phase of finite volume states is
usually chosen such that it is symmetric for the exchanges of rapidities,
as opposed to the phase of infinite volume states, which pick up the
scattering matrix whenever two neighboring particles are exchanged
(\ref{eq:FFax}). Thus, the finite volume form factor at order $(0)$
is 
\begin{equation}
\langle\{m\}\vert\mathcal{O}\vert\{n\}\rangle_{L}=\frac{F_{N+M}(\{\bar{\vartheta}^{(0)}+i\pi\},\{\bar{\theta}^{(0)}\})}{\sqrt{\prod_{i<j}S(\bar{\vartheta}_{j,i}^{(0)})\rho_{M}^{(0)}(\{\bar{\vartheta}^{(0)}\})\prod_{i<j}S(\bar{\theta}_{i,j}^{(0)})\rho_{N}^{(0)}(\{\bar{\theta}^{(0)}\})}}\label{eq:FF0}
\end{equation}
where the density of states $\rho_{N}^{(0)}(\{\bar{\theta}^{(0)}\})$
is defined as the determinant of the matrix $\partial_{i}Q_{j}^{(0)}(\{\theta\})$:
\begin{equation}
\rho_{N}^{(0)}(\{\theta\})=\det\left|\partial_{j}Q_{i}^{(0)}(\{\theta\})\right|
\end{equation}
which is the Jacobian for changing the variables from $\{n\}$ to
$\{\bar{\theta}^{(0)}\}$ via $Q_{k}^{(0)}(\bar{\theta}_{j}^{(0)})=2\pi n_{k}$.
This form for the polynomial finite size corrections is correct only
if the matrix element is non-diagonal, i.e. if the quantum numbers
$\{n\}$ and $\{m\}$ are not exactly the same. For diagonal form
factors a more complicated formula is valid including disconnected
terms \cite{Pozsgay:2007gx}, which can be obtained from the non-diagonal
form factor by taking an appropriate limit \cite{Bajnok:2017bfg}.
Expression (\ref{eq:FF0}) contains all polynomial corrections in
the inverse of the volume and is valid for any theory with a single
particle type, in particular, for both the sinh-Gordon and the scaling
Lee-Yang models.

\subsection{Leading exponential corrections: the $\mu$-term}

In this subsection we calculate the $\mu$-terms for finite volume
form factors in the scaling Lee-Yang model. This is analogous to the
order $(\mu)$ calculation of the energy in subsubsection 2.2.1. As
we have shown there the results at this order can be obtained by taking
the order $(0)$ correction with the additional requirement that each
particle is a bound state represented by its constituents. Expanding
the bound states' equations to the leading exponential order provided
the $\mu$-terms for the energy. Based on this observation Pozsgay
suggested in \cite{Pozsgay:2008bf} that the $\mu$-terms for form
factors can be calculated by taking the form (\ref{eq:FF0}) with
the additional assumption that each particle with rapidity $\bar{\theta}_{j}$
is represented as a bound state of particles with rapidities $\bar{\theta}_{j\pm}$.
He also carried out this calculation for the simplest one-particle
form factor $\langle0\vert\mathcal{O}\vert\theta\rangle_{L}$ and
checked the results numerically with the TCSA method. In the following
we calculate the $\mu$-term correction for a generic $N$-particle
state based on this idea.

We start with the order $\left(0\right)$ form factor in which only
incoming particles are present, composed from the constituents $\theta_{j\pm}$:
\begin{equation}
\langle0\vert\mathcal{O}\vert\{\theta_{\pm}\}\rangle_{L}=\frac{F_{2N}(\theta_{1+},\theta_{1-},\ldots\theta_{N+},\theta_{N-})}{\sqrt{\prod\limits _{k}S(\theta_{k+,k-})\rho_{2N}(\{\theta_{\pm}\})\prod\limits _{i<j,r,s}S(\theta_{ir,js})}}\label{eq:Fconstituent}
\end{equation}
We evaluate this expression at $\theta_{j\pm}=\bar{\theta}_{j\pm}^{(\mu)}=\bar{\theta}_{j}^{(\mu)}\pm i(u+\delta\bar{u}_{j})$
and expand to leading order in $\delta\bar{u}_{j}$. Since at the
leading order both the numerator and the denominator is proportional
to $\prod_{k}\delta\bar{u}_{k}^{-1}$ we multiply them with the factor
$\left(\prod_{k}\frac{2\delta\bar{u}_{k}}{\Gamma}\right)$. This ensures
that the numerator has the expansion starting with the form factor
$F_{N}(\{\bar{\theta}^{(\mu)}\})$:

\begin{equation}
\prod_{k}\left(\frac{2\delta\bar{u}_{k}}{\Gamma}\right)F_{2N}(\{\bar{\theta}_{\pm}^{(\mu)}\})=F_{N}(\{\bar{\theta}^{(\mu)}\})+\sum_{k}\left(\frac{2\delta\bar{u}_{k}}{\Gamma}\right)F_{N,k}^{b}(\{\bar{\theta}^{(0)}\})+O(\delta\bar{u}^{2})
\end{equation}
where we used the quantity introduced in (\ref{eq:bkff}). The calculation
of the denominator can be done in two steeps. The detailed derivation
is relegated to Appendix \ref{sec:muterm} and we provide an outline
here. In the first step we derive that 
\begin{equation}
\prod_{k}\left(\frac{2\delta\bar{u}_{k}}{\Gamma}\right)^{2}S(2i(u+\delta\bar{u}_{k}))\rho_{2N}(\{\bar{\theta}_{\pm}^{(\mu)}\})=\rho_{N}^{\left(\mu\right)}(\{\bar{\theta}^{(\mu)}\})\left(1+\sum_{k}\partial_{k}Q_{k}^{(0)}(\{\bar{\theta}^{(0)}\})\delta\bar{u}_{k}\right)
\end{equation}
where we introduced the density of states $\rho_{N}^{(\mu)}(\{\theta\})$
corresponding to the quantization $Q_{k}^{(\mu)}(\{\bar{\theta}^{(\mu)}\})=2\pi n_{j}$
as 
\begin{equation}
\rho_{N}^{(\mu)}(\{\theta\})=\text{det}\left[\partial_{i}Q_{j}^{(\mu)}(\{\theta\})\right]=\text{det}\left[\partial_{i}Q_{j}^{(0)}(\{\theta\})+\partial_{i}\partial_{j}\sum_{k}(\delta u_{k+}(\{\theta\})+\delta u_{k-}(\{\theta\}))\right]\label{eq:rhomu}
\end{equation}
 In the second step one uses that 
\begin{align}
\prod_{i<j}S(\theta_{i+,j+})S(\theta_{i+,j-})S(\theta_{i-,j+})S(\theta_{i-,j-})=\prod_{i<j}S(\bar{\theta}_{i,j}^{(\mu)})\left(1+\bar{\partial}_{i}Q_{j}^{(0)}(\{\theta_{\pm}\})\delta\bar{u}_{i}-\partial_{i}\bar{Q}_{j}^{\left(0\right)}(\{\theta_{\pm}\})\delta\bar{u}_{j}\right)
\end{align}
 where $\bar{\partial}_{j}=\partial_{j+}-\partial_{j-}$. By collecting
all factors the finite volume form factor including the $\mu$-term
order can be found and it takes the form 
\begin{equation}
\langle0\vert\mathcal{O}\vert\{n\}\rangle_{L}=\frac{F_{N}(\{\bar{\theta}^{(\mu)}\})+\delta^{(\mu)}F_{N}(\{\bar{\theta}^{(0)}\})}{\sqrt{\prod\limits _{k<j}S(\bar{\theta}_{k,j}^{(\mu)})\rho_{N}^{(\mu)}(\{\bar{\theta}^{(\mu)}\})}}
\end{equation}
where 
\begin{align}
\delta^{(\mu)}F_{N}(\{\bar{\theta}^{(\mu)}\}) & =\sum_{k}\left\{ \frac{2}{\Gamma}F_{N,k}^{b}(\{\bar{\theta}^{(0)}\})-\frac{1}{2}\partial_{k}Q_{k}^{(0)}(\{\bar{\theta}^{(0)}\})F_{N}(\{\bar{\theta}^{(0)}\})\right\} \delta\bar{u}_{k}\label{eq:FFmu}\\
 & \,\,\,\,\,\,\,\,+\frac{1}{2}\sum_{j<k}\left[\phi_{-}(\bar{\theta}_{j,k}^{(0)})\left(\delta\bar{u}_{j}+\delta\bar{u}_{k}\right)\right]F_{N}(\{\bar{\theta}^{(0)}\})\nonumber 
\end{align}
with the notation$\phi_{-}(\theta)=\phi(\theta+2iu)-\phi(\theta-2iu)$
used.

In case of both incoming and outgoing particles the form factor takes
the form 
\begin{eqnarray}
\langle\{m\}\vert\mathcal{O}\vert\{n\}\rangle_{L}=\hspace{12cm}\label{eq:inoutmu}\\
\frac{F_{N+M}(\{\bar{\vartheta}^{(\mu)}+i\pi\},\{\bar{\theta}^{(\mu)}\})+\delta_{u}^{(\mu)}F_{N+M}(\{\bar{\vartheta}^{(0)}+i\pi\},\{\bar{\theta}^{(0)}\})+\delta_{v}^{(\mu)*}F_{N+M}(\{\bar{\vartheta}^{(0)}+i\pi\},\{\bar{\theta}^{(0)}\})}{\sqrt{\prod\limits _{k<j}S(\bar{\vartheta}_{j,k}^{(\mu)})\rho_{M}^{(\mu)}(\{\bar{\vartheta}^{(\mu)}\})\prod\limits _{k<j}S(\bar{\theta}_{k,j}^{(\mu)})\rho_{N}^{(\mu)}(\{\bar{\theta}^{(\mu)}\})}}\nonumber 
\end{eqnarray}
where $\bar{\vartheta}_{j\pm}=\bar{\vartheta}_{j}\pm i\bar{v}_{j}$.
The quantity $\delta_{u}^{(\mu)}F_{N+M}(\{\bar{\vartheta}^{(0)}+i\pi\},\{\bar{\theta}^{(0)}\})$
can be obtained from $\delta^{(\mu)}F_{N}(\{\bar{\theta}^{(0)}\})$
by replacing $F_{N}(\{\bar{\theta}^{(0)}\})$ with $F_{N+M}(\{\bar{\vartheta}^{(0)}+i\pi\},\{\bar{\theta}^{(0)}\})$
and $\delta_{v}^{(\mu)}F_{N+M}(\{\bar{\vartheta}^{(0)}+i\pi\},\{\bar{\theta}^{(0)}\})$
from $\delta_{u}^{(\mu)}F_{N+M}(\{\bar{\vartheta}^{(0)}+i\pi\},\{\bar{\theta}^{(0)}\})$
by replacing $u$ with $v$, respectively. The quantity $\delta^{(\mu)*}F_{N}(\{\bar{\theta}^{(\mu)}\})$
can be obtained from $\delta^{(\mu)}F_{N}(\{\bar{\theta}^{(\mu)}\})$
by changing the sign of the second line in (\ref{eq:FFmu}), which
for real form factors means complex conjugation.

\subsection{F-term correction for form factors\label{subsec:Fterm}}

\begin{figure}
\begin{centering}
\includegraphics[width=3cm]{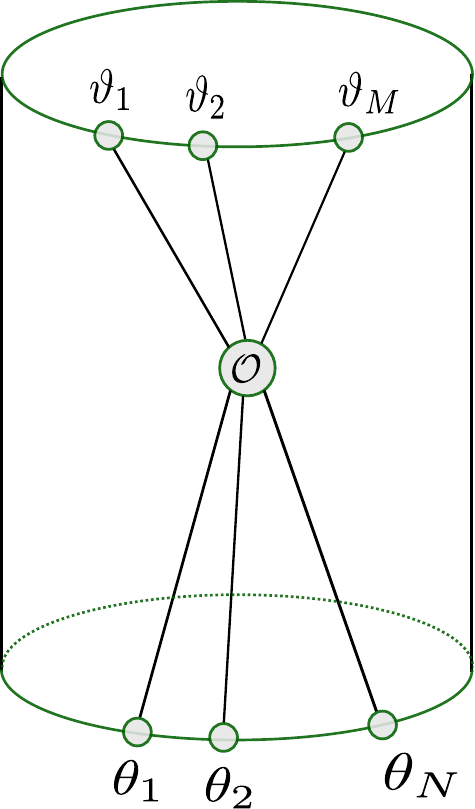}~~~~~~~~~~~~~~~~~\includegraphics[width=5cm]{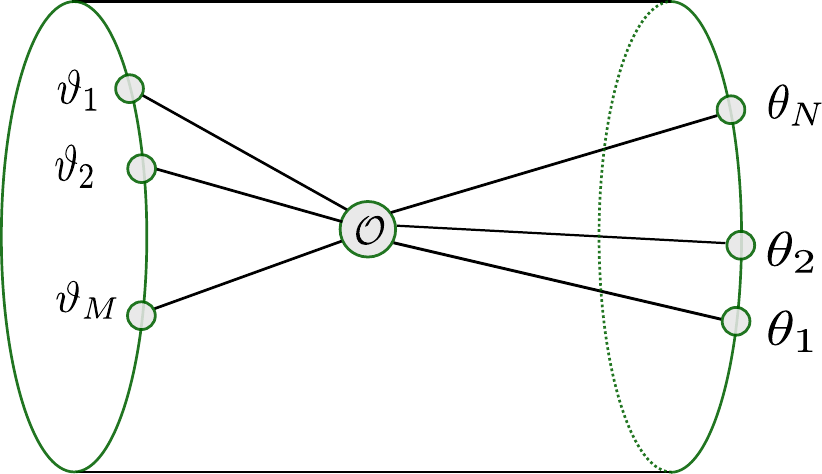}
\par\end{centering}
\caption{On the left: graphical representation of a finite volume form factor
with incoming state $\vert\theta_{1},\dots,\theta_{N}\rangle$ and
outgoing state $\vert\vartheta_{1},\dots,\vartheta_{M}\rangle$. Particles'
trajectories are schematically drawn as solid lines and the operator
${\cal O}$ is represented as a circle. On the right: the same form
factor is represented after double Wick rotation in the thermal channel,
i.e. when the Euclidean time is finite and space is infinite.}

\label{finitevolumeff}
\end{figure}

In Appendix \ref{sec:Fterm} we give a formal derivation for the $F$-term
of the form factors' finite size correction. Here we just summarize
our method. In the sinh-Gordon theory this is the leading exponential
correction and in the Lee-Yang theory it is intimately related to
the previously calculated $\mu$-term correction. We parametrize the
form factor as 
\begin{equation}
\langle\{m\}\vert\mathcal{O}\vert\{n\}\rangle_{L}=\frac{F_{N+M}(\{\bar{\vartheta}^{(1)}+i\pi\},\{\bar{\theta}^{(1)}\})+\delta^{(F)}F_{N+M}(\{\bar{\vartheta}^{(0)}+i\pi\},\{\bar{\theta}^{(0)}\})}{\sqrt{\prod_{i<j}S(\bar{\vartheta}_{j,i}^{(1)})\rho_{M}^{(1)}(\{\bar{\vartheta}^{(1)}\})\prod_{i<j}S(\bar{\theta}_{i,j}^{(1)})\rho_{N}^{(1)}(\{\bar{\theta}^{(1)}\})}}\label{eq:FF_F0}
\end{equation}
The denominator is simply related to the normalization of states originating
from the Bethe-Yang equation (\ref{eq:shGQQ1}). The numerator is
represented graphically on the left part of Figure \ref{finitevolumeff}.
Exchanging the role of Euclidean space and time leads to the picture
on the right of Figure \ref{finitevolumeff}, where we have to calculate
a normalized trace:

\begin{equation}
\frac{\mbox{Tr}(e^{-LH}\mathcal{O}_{N,M})}{\sqrt{\mbox{Tr}_{N}(e^{-LH})\mbox{Tr}_{M}(e^{-LH})}}
\end{equation}
In this channel $\mathcal{O}_{N,M}$ is not a local operator as moving
a particle with rapidity $v$ around it picks up the phase $\prod_{j}S(v-\theta_{j}-\frac{i\pi}{2})\prod_{k}S(\vartheta_{k}+i\frac{\pi}{2}-v)$.
This is the reason why we cannot apply a finite volume regularization
as the system cannot be made periodic: the past/future or the left/right
asymptotics are different. Particularly, in case of diagonal form
factors, there is no monodromy and we could impose a periodic boundary
condition in a finite volume. Clearly the normalization in this case
would be the excited state partition function: $Z_{N}=\mbox{Tr}_{N}(e^{-LH})$.
A particle with rapidity $\theta$ act in this channel as a defect
operator with transmission factor $T(v)=S(\frac{i\pi}{2}+\theta-v)$.

In evaluating the trace we insert two complete systems of states

\begin{equation}
\mbox{Tr}(e^{-LH}\mathcal{O}_{N,M})=\sum_{\nu,\mu}\langle\nu\vert O_{N,M}\vert\mu\rangle\langle\mu\vert\nu\rangle e^{-E_{\nu}L}
\end{equation}
and keep only the vacuum and one-particle states for $\mu$ and $\nu$
with rapidities $u$ and $v$. Infinite volume states are normalized
as $\langle u\vert v\rangle=2\pi\delta(u-v)$ and the matrix element
of the defect operator ${\cal O}_{N,M}$ can be expressed in terms
of the infinite volume form factor as 
\begin{eqnarray}
\langle v\vert O_{N,M}\vert u\rangle & = & F_{N+M+2}(v+i\pi-i\epsilon,\{\vartheta+\frac{i\pi}{2}\},u,\{\theta-\frac{i\pi}{2}\})+\\
 &  & 2\pi\delta(v-u)\prod_{j}S(\frac{i\pi}{2}+\vartheta_{j}-u)F_{N+M}(\{\vartheta+\frac{i\pi}{2}\},\{\theta-\frac{i\pi}{2}\})\nonumber 
\end{eqnarray}
Therefore, we are faced with the square of the $\delta$-function.
Using our experience from evaluating the finite temperature 2-point
function \cite{Bajnok:2018fzx} we regulate the $\delta$-function
as 
\begin{equation}
2\pi\delta(u-v)=\frac{i}{u-v+i\epsilon}-\frac{i}{u-v-i\epsilon}
\end{equation}
For the 2-point function this regularization was equivalent to finite
volume regularizations \cite{Bajnok:2018fzx}. Then, we shift the
$v$ contour from the real line to above $i\epsilon$. Taking the
$\epsilon\to0$ limit, in the shifted integral no contribution will
survive thus we merely pick up the residue term at $v=u+i\epsilon$.
For the excited state partition function this results in 
\begin{equation}
\mbox{Tr}_{N}(e^{-LH})=1+\int\frac{du}{2\pi}\frac{1}{\epsilon}\prod_{j}S(\frac{i\pi}{2}+\theta_{j}-u)e^{-mL\cosh u}+O(\epsilon)
\end{equation}
Note that there is no $O(1)$ term. This is consistent with the usual
evaluation of the partition function: if we calculated the contribution
via finite volume regularization we would get $mR\cosh u$ instead
of $\frac{1}{\epsilon}$. By repeating the same steps for the numerator
we can check that the singular $\frac{1}{\epsilon}$ terms cancel
with the square rooted product of the excited state partition functions
and we find that the $O(1)$ piece is: 
\begin{equation}
\delta^{(F)}F_{N+M}(\{\vartheta+i\pi\},\{\theta\})=\int\frac{du}{2\pi}F_{N+M+2}^{r}(u+i\pi,\{\vartheta+\frac{i\pi}{2}\},u,\{\theta-\frac{i\pi}{2}\})e^{-mL\cosh u}\label{eq:FF_Fterm}
\end{equation}
where the previously introduced regulated form factor appears. In
Appendix \ref{sec:muFrel} we show that taking an appropriate residue
of this $F$-term, the $\mu$-term (\ref{eq:FFmu}) can be obtained.

\section{Numerical comparison}

In this section we check our results numerically by using the TCSA
method in the sinh-Gordon\footnote{Note that very similar ideas to the present numerical method were
investigated independently by Robert Konik, Giuseppe Mussardo and
collaborators, as mentioned in a talk of Mussardo presented at the
IHES workshop \textquotedblleft Hamiltonian methods in strongly coupled
Quantum Field Theory\textquotedblright , 8-12 January 2018, https://www.youtube.com/watch?v=pyDXNlXu-2w.
More recently, a collaboration started between one of the present
authors (M.L.) and the aforementioned team, leading to further progress
regarding the proper numerical treatment of sinh-Gordon theory at
stronger couplings. In the present article we focus on the evaluation
of finite volume form factors at relatively small couplings, while
more general numerical results will be published elsewhere. We thank
Gábor Takács for pointing our attention towards the IHES talk.\label{fn:footnote_mussardo}} and scaling Lee-Yang models. The TCSA method was first implemented
for the Lee-Yang model by Yurov and Zamolodchikov \cite{Yurov:1989yu}.
Recently there has been a renewed interest in implementing it for
the relevant perturbations of the noncompact free boson, see e.g.
\cite{Coser:2014}, \cite{Hogervorst:2015} and \cite{PhysRevD.91.085011}.

Both in the sinh-Gordon and scaling Lee-Yang models we compare the
analytical and TCSA results for the finite size energy spectrum then
extract the finite volume matrix elements. After having checked the
polynomial corrections we check the $F$-term corrections in the sinh-Gordon
model and the $\mu$- and $F$- term corrections in the scaling Lee-Yang
model.

\subsection{Sinh-Gordon theory}

The sinh-Gordon theory can be defined as
\begin{equation}
S=\int dt\int_{0}^{L}dx\,\left\{ \frac{g}{2}(\partial_{\mu}\varphi)^{2}-2\mu\cosh(b\varphi)\right\} 
\end{equation}
When quantizing the theory we need to choose a scheme, which separates
the free part and the perturbation. The free part can be either the
free massless or massive boson and then the perturbing operator should
be normal ordered wrt. the chosen free theory.

\subsubsection{Conformal scheme}

In the conformal scheme the free part is simply the kinetic term and
the field has the following mode expansion 
\begin{equation}
\varphi(x,t)=\varphi_{0}+\frac{\pi_{0}}{gL}t+\frac{i}{\sqrt{4\pi g}}\sum_{n\neq0}\frac{1}{n}(a_{n}e^{ik_{n}(x-t)}+\bar{a}_{n}e^{-ik_{n}(x+t)});\quad k_{n}=\frac{2\pi n}{L}\label{eq:field}
\end{equation}
where the nonzero commutators for the oscillators are $[a_{n},a_{m}]=n\delta_{n+m}$
and $[\bar{a}_{n},\bar{a}_{m}]=n\delta_{n+m}$. The zero mode is the
free motion on the line with $[\varphi_{0},\pi_{0}]=i$. The Hilbert
space is the superposition of the quantum mechanical zero mode with
a continuous spectrum and the Fock space of left- and right-moving
particles: 
\begin{equation}
\mathcal{H}=\{a_{-n_{1}}\dots a_{-n_{k}}\bar{a}_{-m_{1}}\dots\bar{a}_{-m_{j}}\vert0\rangle\otimes\Psi(\varphi_{0})\}
\end{equation}
The free Hamiltonian is then 
\begin{equation}
H_{0}=\frac{2\pi}{L}(L_{0}+\bar{L}_{0}-\frac{1}{12})+\frac{\pi_{0}^{2}}{2gL}\quad;\qquad L_{0}=\sum_{n>0}a_{-n}a_{n}
\end{equation}
This theory has conformal invariance. When the perturbing operators
$:e^{\pm b\varphi}:$ are normal ordered wrt. this theory, they are
primary fields of dimension $h=\bar{h}=-b^{2}\left(8\pi g\right)^{-1}$.
The mass-gap relation \cite{Zamolodchikov:1995aa} 
\begin{equation}
-\frac{\pi\mu\Gamma\left(1+\frac{b^{2}}{8\pi g}\right)}{\Gamma\left(-\frac{b^{2}}{8\pi g}\right)}=\left[\frac{m}{4\sqrt{\pi}}\Gamma\left(\frac{1-p}{2}\right)\Gamma\left(1+\frac{p}{2}\right)\right]^{\frac{2}{1-p}}=(m\kappa)^{\frac{2}{1-p}}
\end{equation}
 connects the perturbation parameter of the Lagrangian $\mu$ to the
mass, $m$, of the sinh-Gordon scattering particle, while the scattering
parameter $p$ is related to $b$ as $p=\frac{b^{2}}{8\pi g+b^{2}}$.

In order to use the TCSA method, a discrete spectrum needs to be truncated
at a given energy cut such that the full Hamiltonian can be diagonalized
on the truncated space. To ensure this we separate the zero mode into
a mini-Hilbert space with Hamiltonian
\[
H_{\textrm{mini}}=\frac{1}{4\pi g}\pi_{0}^{2}+\mu\biggl(\frac{L}{2\pi}\biggr)^{2+\frac{b^{2}}{4\pi g}}2\pi(e^{b\varphi_{0}}+e^{-b\varphi_{0}})
\]
where the volume-dependent coefficient comes from the conformal mapping
of the Hamiltonian between the cylinder and the plane

\begin{equation}
H=\frac{2\pi}{L}\left(L_{0}+\bar{L}_{0}-\frac{1}{12}+H_{\mathrm{mini}}+\mu\biggl(\frac{L}{2\pi}\biggr)^{2+\frac{b^{2}}{4\pi g}}2\pi\delta_{P}\left\{ e^{b\varphi_{0}}(:e^{b\hat{\varphi}}:-1)+e^{-b\varphi_{0}}(:e^{-b\hat{\varphi}}:-1)\right\} \right)\label{eq:Hm0}
\end{equation}
Here $\delta_{P}$ projects to matrix elements which do not change
the momentum $P=\frac{2\pi}{L}(L_{0}-\bar{L}_{0})$ and $\hat{\varphi}=\varphi(0,0)-\varphi_{0}$.
Technically, we solve numerically the mini-Hilbert space problem first.
This can be done either in the basis of plane waves in a box or using
the eigenvectors of the harmonic oscillator. For small volumes even
the Liouville reflection factor can be used to get an approximation
of the spectrum \cite{Zamolodchikov:1995aa,Zamolodchikov:2000kt}.
We found that using 100 unperturbed vectors we got a reliable spectrum
up to 5 digits in the range we were interested in. We kept 6-8 vectors
from this zero mode space and calculated the matrix elements of $e^{\pm b\varphi_{0}}$.
By taking the tensor product with the Fock spaces and truncating in
the energy with only the zero mode perturbation added we obtained
a finite Hilbert space. We then diagonalized the full Hamiltonian
and calculated the eigenvalues and eigenvectors. These provided the
finite size spectrum and finite volume form factors.

\subsubsection{Massive boson scheme}

Alternatively, one can start by perturbing the free boson of mass
$M$. The free massive boson on the cylinder can be considered as
a perturbation of the massless one, i.e.
\begin{equation}
H_{0}^{\left(M\right)}=H_{0}+g\frac{M^{2}}{2}\intop_{0}^{L}:\varphi\left(x,0\right)^{2}:dx\label{eq:H0m}
\end{equation}
This operator can be diagonalized by solving the zero mode harmonic
oscillator and applying a Bogoliubov transformation to the finite
momentum oscillators. Technical details are relegated to Appendix
\ref{sec:relate_m_mless}. As a result the field operator (\ref{eq:field})
is expressed in terms of the new massive creation operators $d_{n}$
as
\begin{equation}
\varphi\left(x,0\right)=\varphi_{0}+\sum_{n\neq0}\frac{1}{\sqrt{2L\omega_{n}g}}\left(d_{n}e^{ik_{n}x}+d_{n}^{\dagger}e^{-ik_{n}x}\right)\quad;\qquad\omega_{n}=\sqrt{M^{2}+k_{n}^{2}}\label{eq:phi_d}
\end{equation}
while the Hamiltonian (\ref{eq:H0m}) becomes
\begin{equation}
H_{0}^{\left(M\right)}=\sum_{m\in\mathbb{Z}}\omega_{m}d_{m}^{\dagger}d_{m}+\tilde{E}_{0}^{\prime}\label{eq:free_massive_H}
\end{equation}
These operators satisfy $[d_{m}^{\dagger},d_{n}]=\delta_{n,m}$ and
the vacuum energy contribution $\tilde{E}_{0}^{\prime}$ appears due
to the difference between the normal ordering with respect to the
mode operators $a_{n}$ or $d_{n}$.

When considering the sinh-Gordon model as a perturbation of the massive
boson (e.g. to make a Feynman graph expansion), one may first define
the Hamiltonian in infinite volume 
\begin{align*}
H^{\left(L\rightarrow\infty\right)} & =H_{0}^{\left(M,L\rightarrow\infty\right)}+\frac{gM^{2}}{b^{2}}\intop_{-\infty}^{\infty}:\cosh\left(b\varphi\left(x\right)\right):_{M,\infty}dx-\frac{gM^{2}}{2}\intop_{-\infty}^{\infty}:\varphi^{2}\left(x\right):_{M,\infty}dx
\end{align*}
\begin{equation}
H_{0}^{\left(M,L\rightarrow\infty\right)}=\intop_{-\infty}^{\infty}\sqrt{M^{2}+k^{2}}d_{k}^{\dagger}d_{k}dk\label{eq:massive_scheme}
\end{equation}
Here $::_{M,\infty}$ means normal ordering with respect to the modes
$d_{k}$ in infinite volume. We first connect the bare parameter $M$
to the bare coupling $\mu$ in the conformal plus zero mode scheme.
As a first step, $H^{\left(L\rightarrow\infty\right)}$ is connected
to the Hamiltonian on the cylinder. This is achieved by requiring
that the perturbation has the same behavior in the UV, i.e. the Hamiltonian
density expressed in terms of bare fields takes the same form for
all volumes: 
\begin{equation}
H=H_{0}^{\left(M\right)}+\frac{gM^{2}}{b^{2}}\intop_{0}^{L}:\cosh\left(b\varphi\left(x\right)\right):_{M,\infty}dx-\frac{gM^{2}}{2}\intop_{0}^{L}:\varphi^{2}\left(x\right):_{M,\infty}dx+\tilde{E}_{0}\label{eq:Hmass}
\end{equation}

By introducing a UV momentum cutoff in Appendix \ref{sec:relate_m_mless}
we show that 
\begin{equation}
:e^{b\varphi\left(0,0\right)}:_{M,\infty}=e^{\frac{b^{2}}{2g}\rho\left(ML\right)}:e^{b\varphi\left(0,0\right)}:_{M,L}\quad;\qquad\rho\left(x\right)=\intop_{-\infty}^{\infty}\frac{du}{2\pi}\frac{1}{e^{x\cosh u}-1}\label{eq:rho}
\end{equation}
Note that, similarly to Landau-Ginsburg theories, and as opposed to
the sine-Gordon theory, the coefficient diverges in the limit $L\rightarrow0$.
Now, we bring the zero mode exponentials out of normal ordering and
obtain the relation
\begin{equation}
\mu=\frac{gM^{2+\frac{b^{2}}{4\pi g}}}{2b^{2}}\left(\frac{e^{\gamma_{E}}}{2}\right)^{\frac{b^{2}}{4\pi g}}\label{eq:muM}
\end{equation}
together with the vacuum energy contribution
\begin{equation}
\tilde{E}_{0}=-\frac{M}{4}+\intop_{0}^{1}du\left(\frac{u}{\tanh\left(\frac{MLu}{2}\right)}-\frac{1}{2\left(1+u\right)^{2}\tanh\left(\frac{ML}{2\sqrt{1-u^{2}}}\right)}\right)\label{eq:E0M}
\end{equation}
Following the same line of thought, the normalization of vertex operators
can be related in the massive and massless schemes as well, as:
\begin{align}
\left\langle e^{a\varphi}\right\rangle  & \equiv\left(\frac{L}{2\pi}\right)^{\frac{a^{2}}{4\pi g}}\left\langle \mathbf{0}\left|:e^{a\varphi\left(z=1,\bar{z}=1\right)}:_{0}\right|\mathbf{0}\right\rangle =e^{-\frac{a^{2}\gamma_{E}}{4g\pi}}\left(\frac{2}{M}\right)^{\frac{a^{2}}{4g\pi}}\left\langle \mathbf{0}\left|:e^{a\varphi\left(0\right)}:_{M,\infty}\right|\mathbf{0}\right\rangle 
\end{align}
where $\left|\mathbf{0}\right\rangle $ denotes the interacting ground
state.

We applied Hamiltonian truncation in this scheme, too. The zero mode
problem is again treated separately, similarly to \cite{Rychkov:2015vap}
and \cite{Bajnok:2015bgw}. Comparing the results obtained from the
massive scheme to the results in the massless scheme provided a tool
to estimate the numerical error of our approach. Results are presented
in the next subsection.

\subsubsection{Numerical checks}

For the numerical checks, we fixed $g=1$, such that $b_{\text{self-dual}}=\sqrt{8\pi}$.
The UV coupling was fixed to $\mu=0.2$. The mini Hilbert space was
chosen to be diagonalized on the particle in a rigid box basis with
$800$ states per parity sector. It was sufficient to keep only the
$6$ states of lowest energy out of these \cite{Rychkov:2015vap,Bajnok:2015bgw}.
In the Fock subspace, a conformal cutoff at chiral levels up to $9$
(in the finite momentum sector, $9$ and $10$) was used. The dominant
cutoff dependence of the results came from this chiral cutoff. This
means that the actual computations involved matrices of up to about
$12000$ dimensions. Since the overall momentum is conserved as well
as there is a parity $\mathbb{Z}_{2}$ symmetry present, it suffices
to search the lowest lying eigenpairs of the appropriate sub-Hamiltonians
restricted to the different symmetry sectors. The above cutoff should
be understood separately in each subsector.

\begin{figure}[H]
\begin{centering}
\begin{tabular}{cc}
\includegraphics[width=7cm]{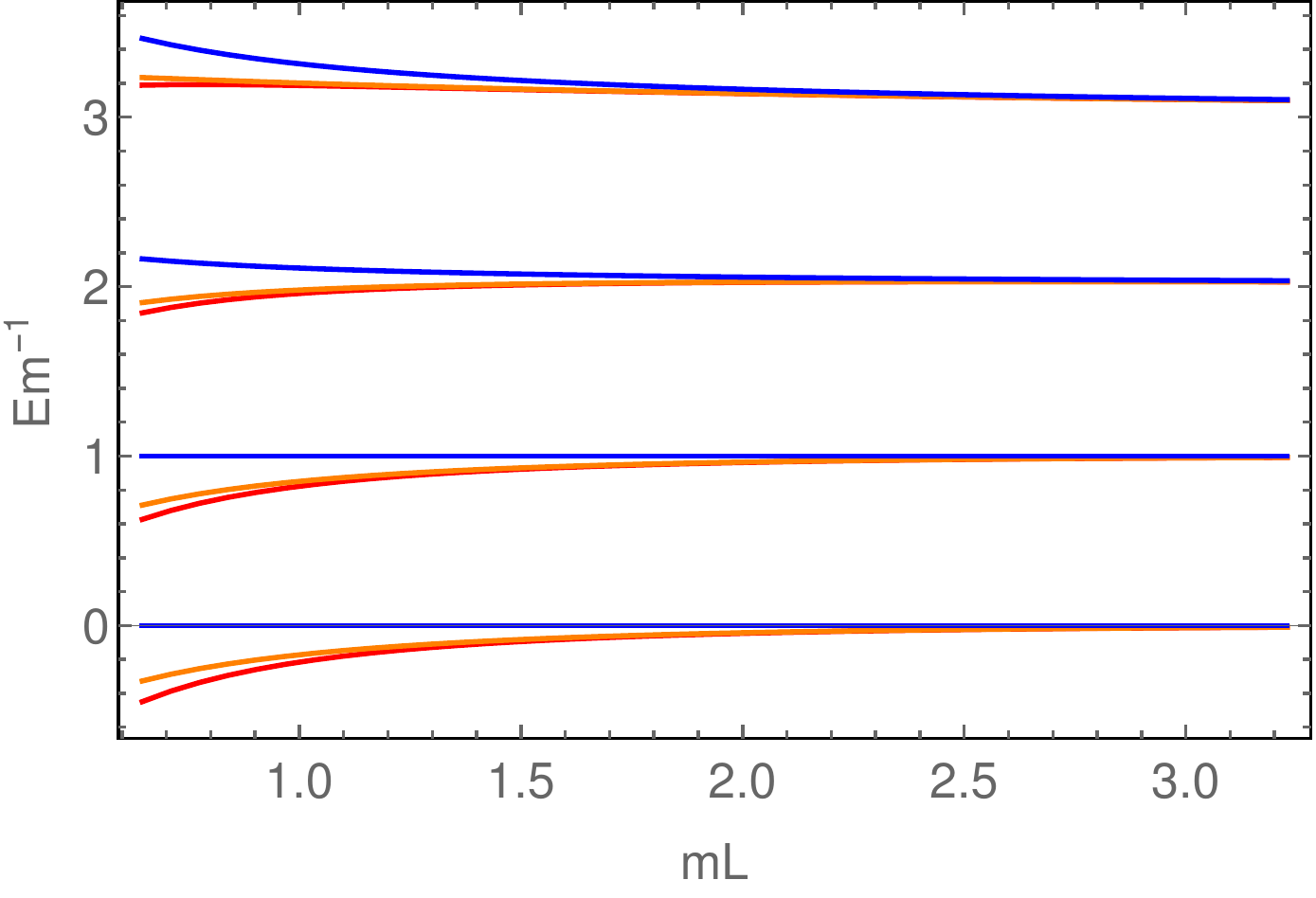} & \includegraphics[width=7cm]{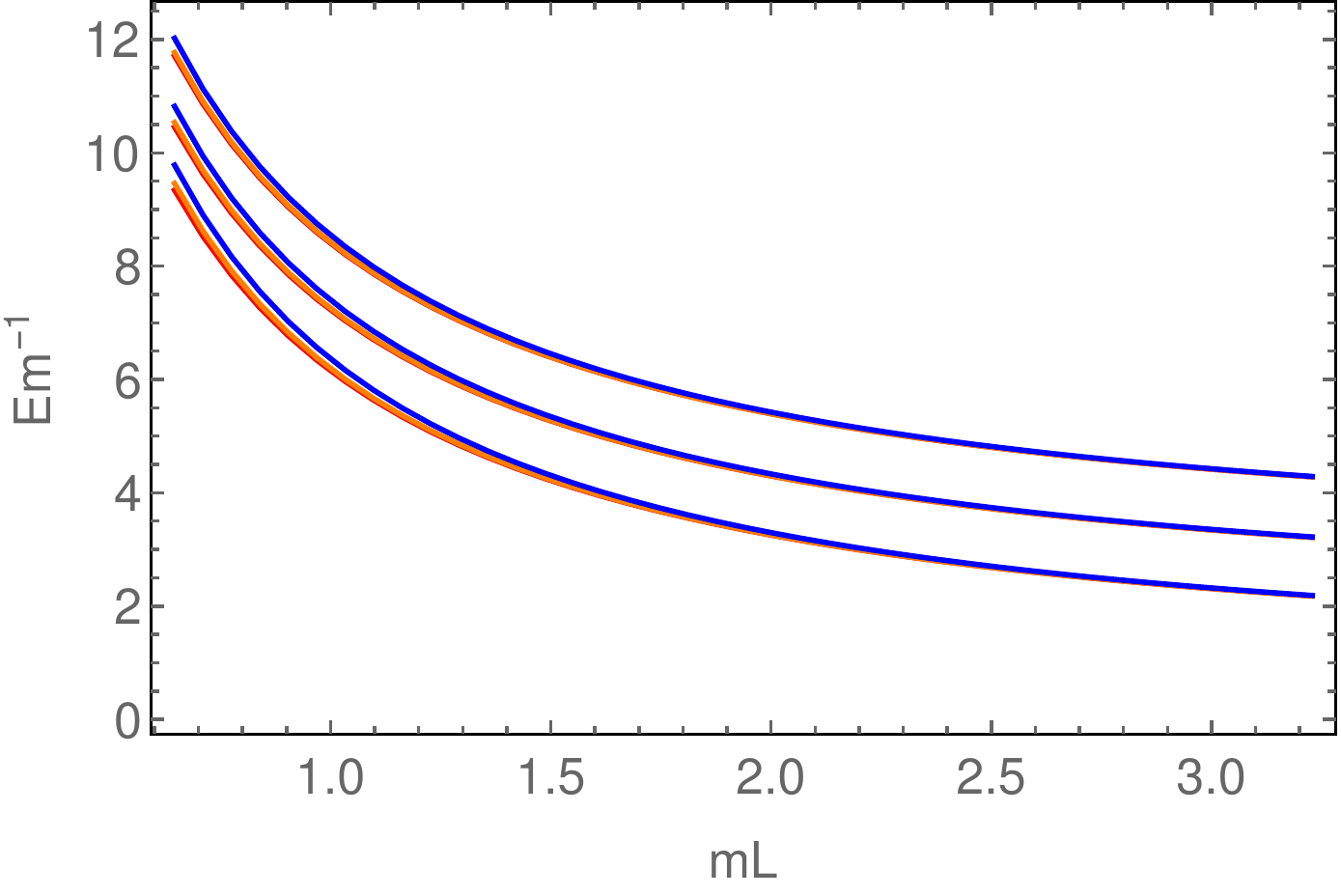}\tabularnewline
\end{tabular}
\par\end{centering}
\caption{Theoretical low-lying energy spectrum of sinh-Gordon model at $b=1$.
The vacuum energy density is subtracted. Left: states in the sector
of overall momentum $0$ (from the bottom up: $\left|\mathrm{vac}\right\rangle $,
$\left|0\right\rangle ,$ $\left|00\right\rangle $ and $\left|000\right\rangle $).
Right: states of overall momentum $1\cdot2\pi L^{-1}$ (from the bottom
up: $\left|1\right\rangle $, $\left|01\right\rangle $ and $\left|001\right\rangle $).
Note that we label the states by the corresponding bosonic Bethe Ansatz
quantization numbers. Bethe-Yang lines are drawn with blue curves.
The leading Lüscher correction is depicted by orange curves. The exact
TBA result is shown with red curves. \label{fig:fig2}}
\end{figure}
The strategy of the computations was to diagonalize the Hamiltonian
(\ref{eq:Hm0}) or (\ref{eq:massive_scheme}) for a number of volumes,
and then plot the volume dependence of the results. A typical spectrum
can be seen on Figure \ref{fig:fig2}.

\subsubsection*{Finite volume energies}

In order to compare the TCSA energies to those obtained by solving
the integral equation (\ref{eq:shGTBA})-(\ref{eq:shGE}), it is important
to keep in mind that the latter is renormalized fixing both the infinite
volume energy and energy density to $0$. This scheme can be connected
to the numerics obtained by TCSA by subtracting the (exactly known)
vacuum energy density of the sinh-Gordon model
\begin{equation}
\mathcal{E}_{0}=\frac{m^{2}}{8\sin\pi p}\quad,\quad\quad p=\frac{b^{2}}{b^{2}+8\pi g}\label{eq:edens}
\end{equation}

\begin{figure}[h]
\begin{centering}
\begin{tabular}{cc}
\includegraphics[width=7cm]{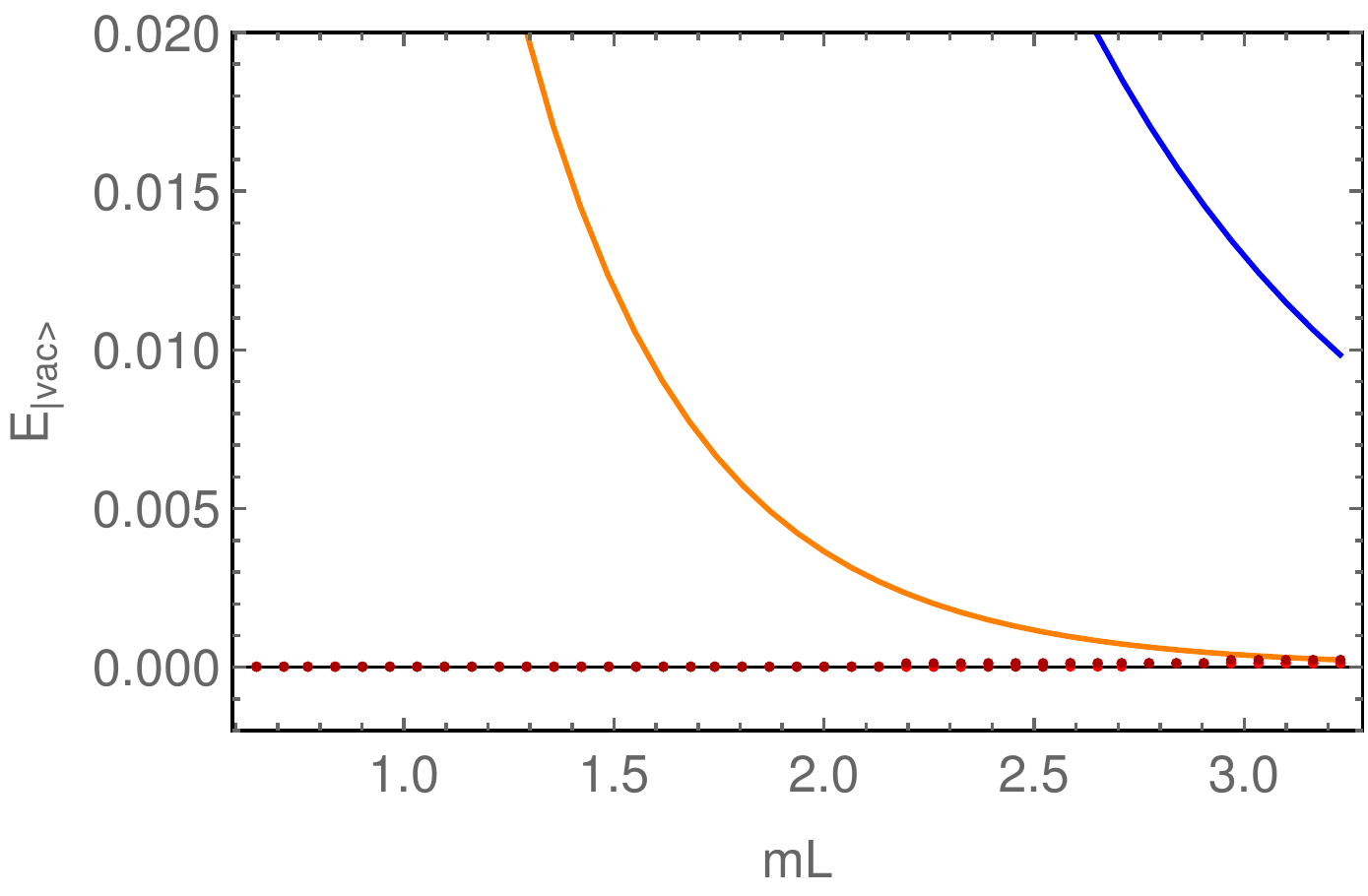} & \includegraphics[width=7cm]{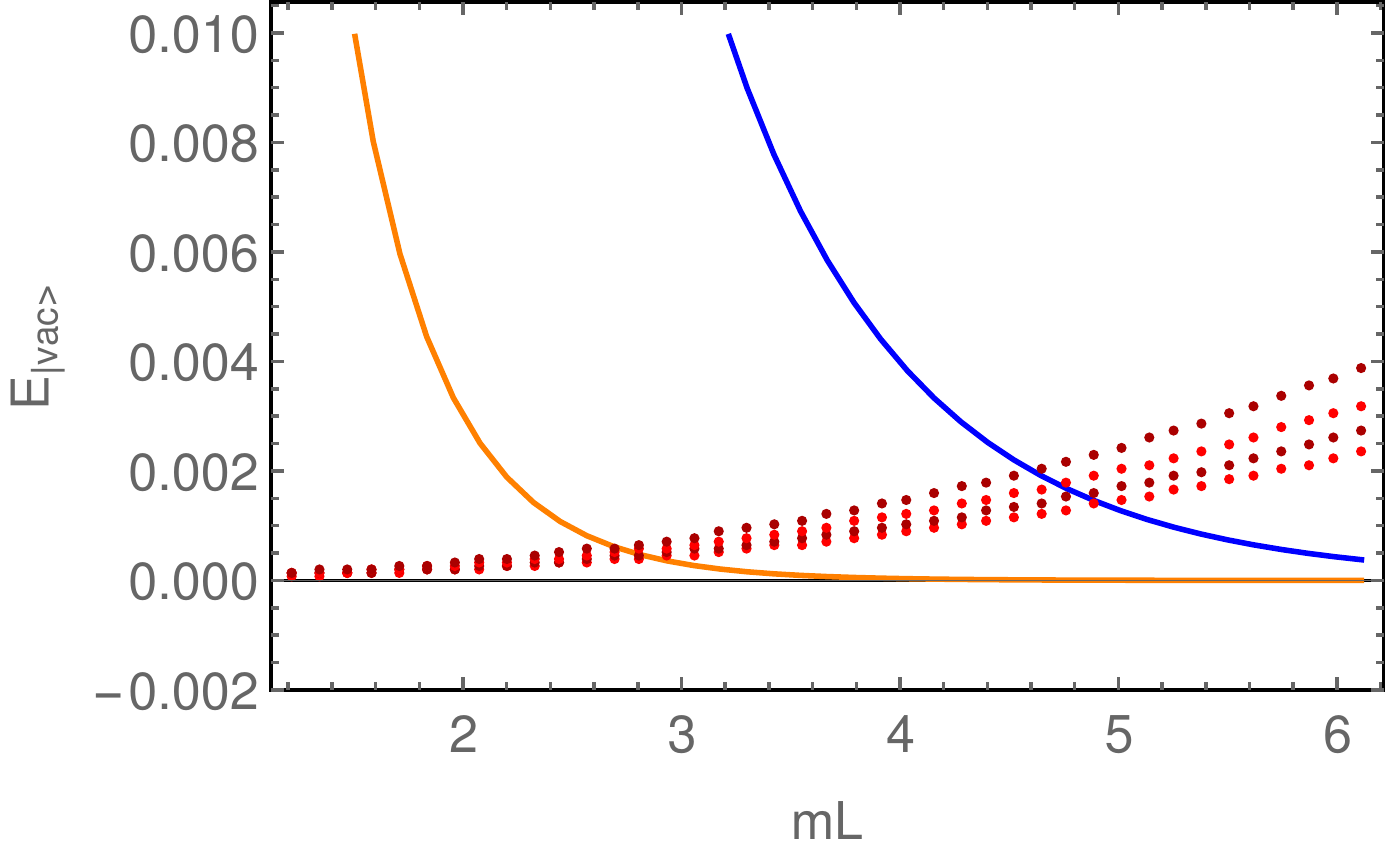}\tabularnewline
\end{tabular}
\par\end{centering}
\caption{Volume dependence of the ground state energy: comparison of TCSA data
with BY, Lüscher and TBA predictions for $b=1$ (left) and $b=2$
(right). On these plots, the TBA predictions are subtracted from each
data set. TCSA measurement points are shown with red and brown dots,
corresponding to different chiral cutoffs (from bottom up: $9,8,7$
and $6$). The difference between the BY lines and TBA (blue) and
between BY+leading Lüscher and TBA (orange) are also indicated. For
$b=1$, TCSA results agree with TBA data to a remarkable precision
(the difference is negligible, as well as the cutoff dependence thereof).
For $b=2$ truncation errors become more significant.\label{fig:fig3}}
\end{figure}

We note that the TCSA numerics produce reliable results at $b=1$
for both the energy levels and the finite volume form factors. For
stronger couplings, e.g. $b=2$, the truncation errors become more
significant, see fig. \ref{fig:fig3}. Experience suggests the general
rule of thumb, using the massive oscillator basis (keeping the same
excitation content of the basis) is equivalent to increasing the chiral
cutoff of the massless TCSA basis by one. Therefore, for the present
work, we mostly consider the case $b=1$.

\begin{figure}[H]
\begin{centering}
\includegraphics[width=15cm]{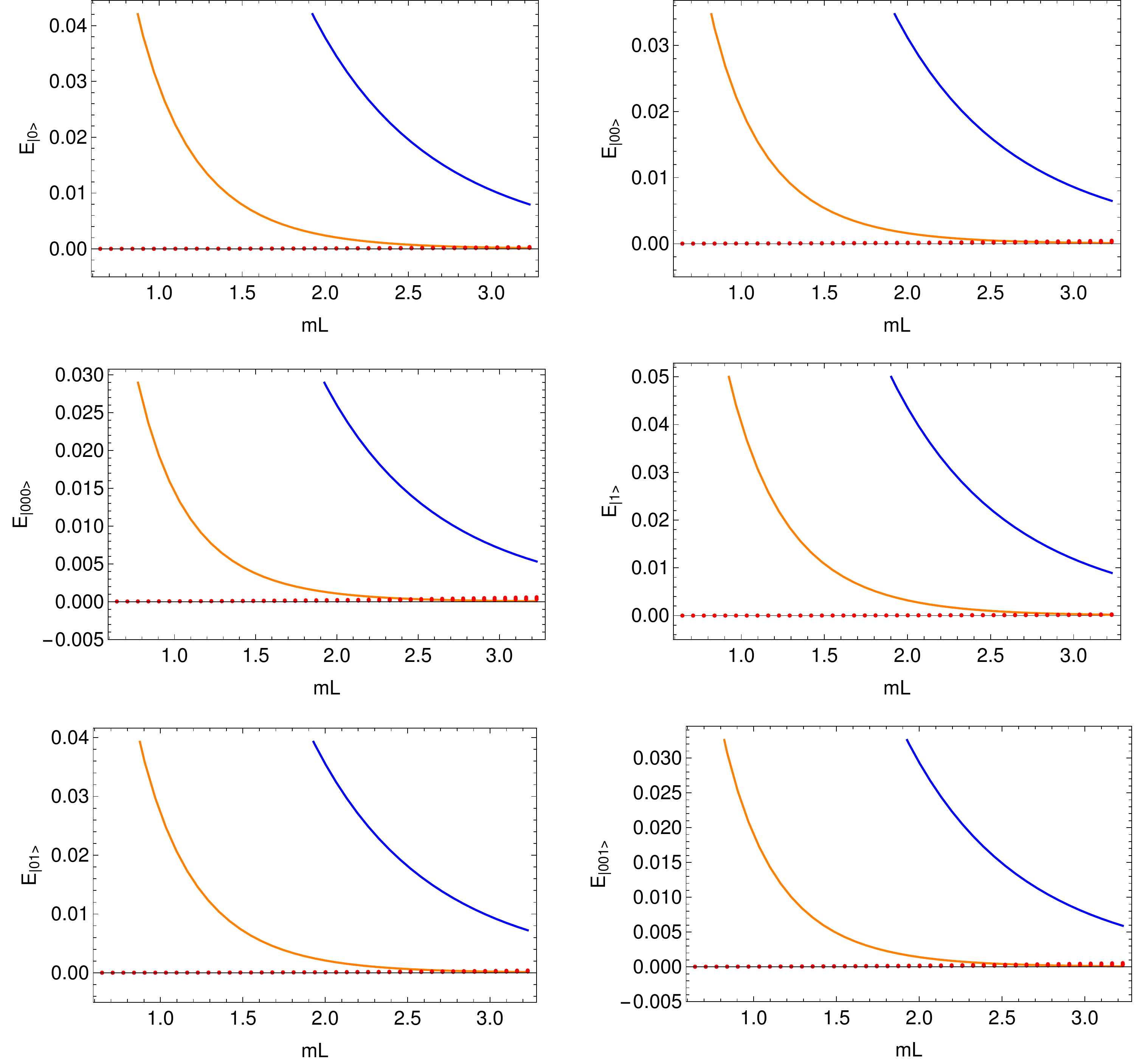}
\par\end{centering}
\caption{Volume dependence of the energy of some relevant low-lying states
at $b=1$. Again, TBA data are subtracted from each data set. TCSA
data is shown by red dots. The difference between the BY lines and
TBA (blue) and between BY+leading Lüscher and TBA (orange) are also
indicated. Once again, we see reassuring agreement between the TBA
and TCSA data (the difference being close to $0$, as shown.)\label{fig:fig4}}
\end{figure}

To make the results more transparent, we chose to depict the difference
of the quantities of interest from some reference data. In Figs. \ref{fig:fig3}-\ref{fig:fig4},
the results obtained by numerically solving the TBA system (\ref{eq:shGTBA})-(\ref{eq:shGE})
are subtracted from each other data sets (in the case of the TCSA
points, the energy density (\ref{eq:edens}) is also taken into account).
Note that we label the states by the corresponding Bethe Ansatz quantization
numbers. The Bethe-Yang lines are calculated via (\ref{eq:shGeps0})-(\ref{eq:shGE0}),
while the exponential corrections follow from (\ref{eq:shGeps1})-(\ref{eq:shGE1}).

From these plots it is clear that in the volume range $\text{1-3}$
the numerical results can be trusted and that adding the leading exponential
correction to the Bethe-Yang results considerably improved the precision.
We expect similar behaviors for finite volume form factors.

\newpage

\subsubsection*{Finite volume form factors}

\begin{figure}[h]
\begin{centering}
\begin{tabular}{cc}
\includegraphics[width=7cm]{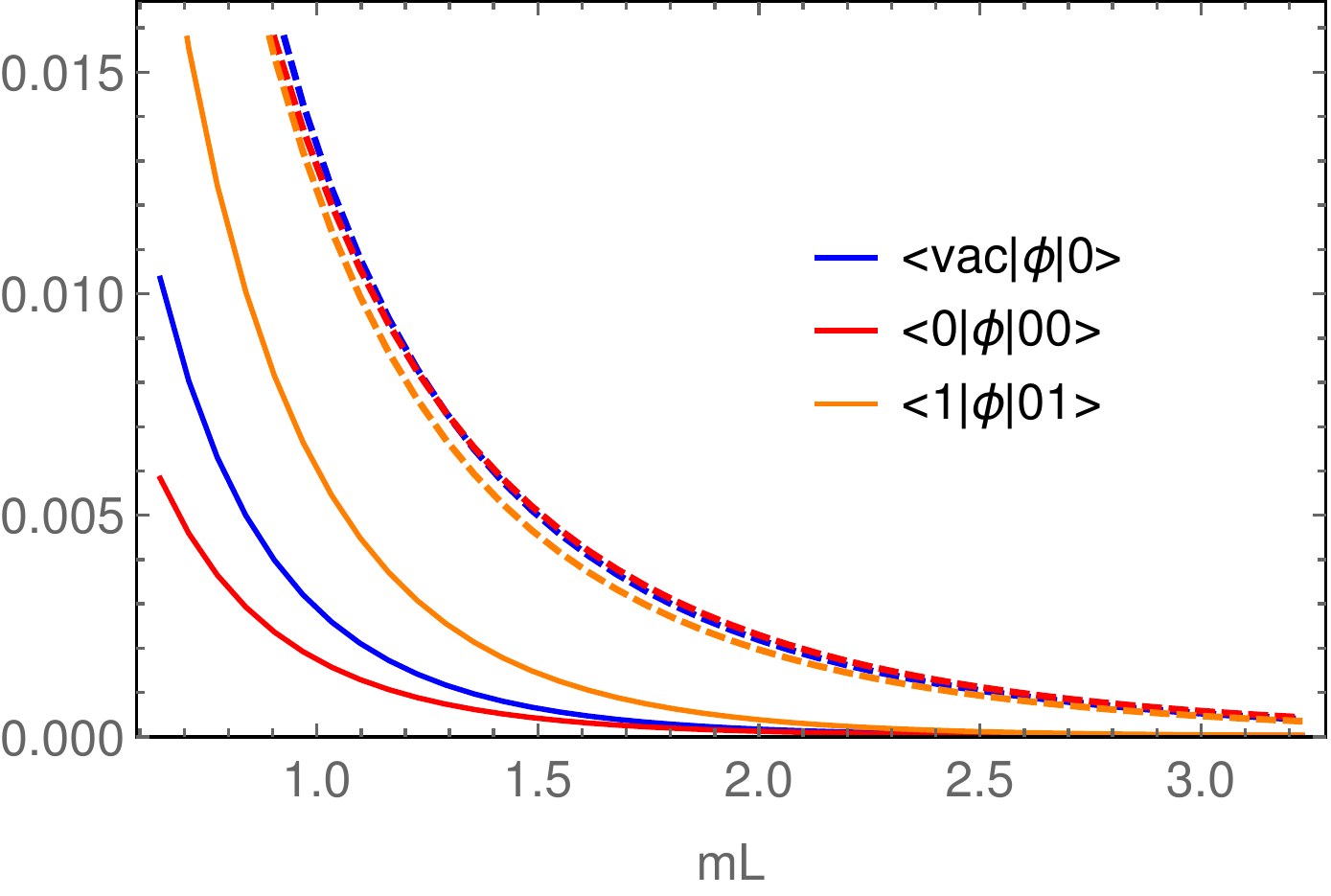} & \includegraphics[width=7cm]{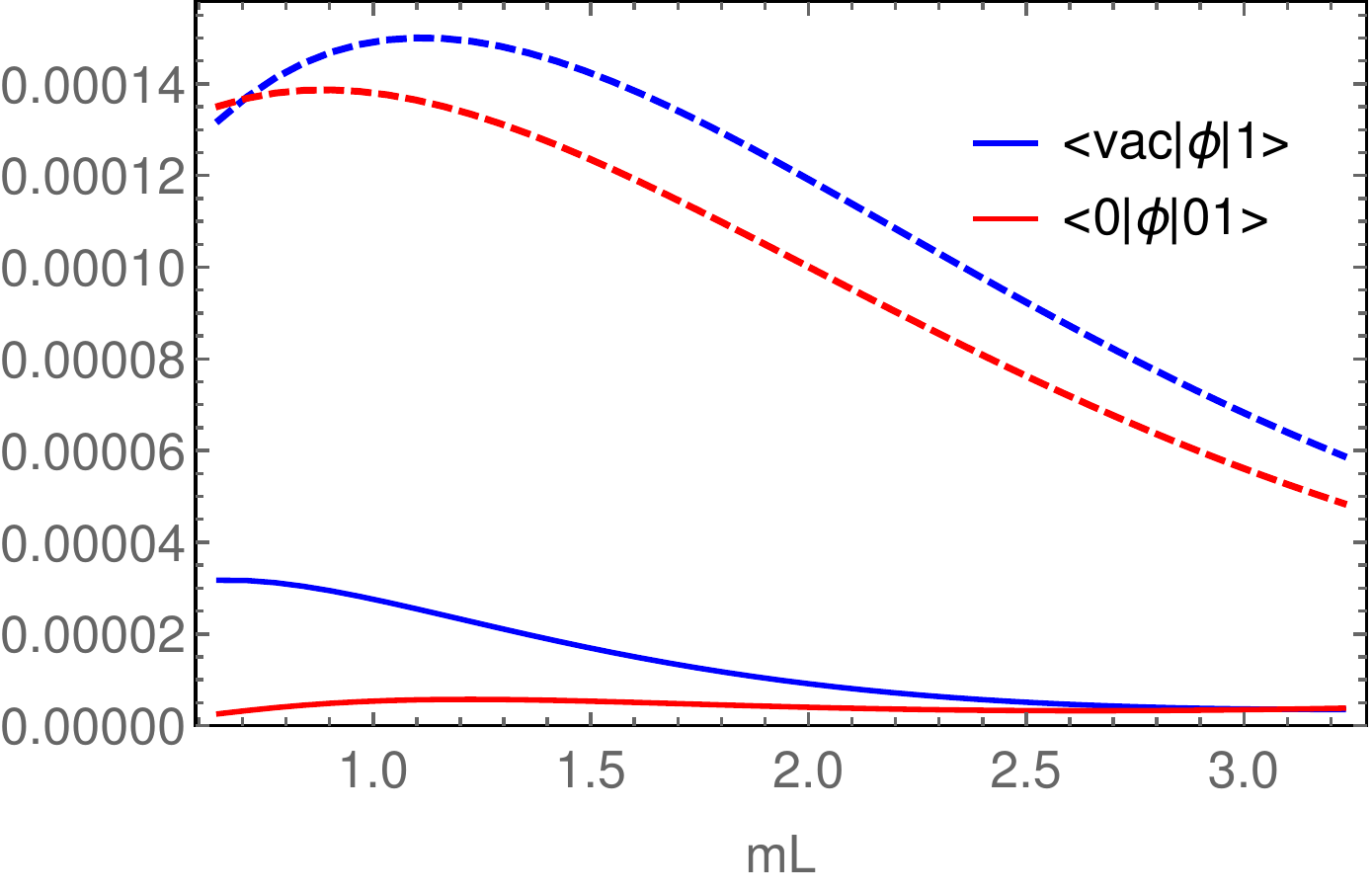}\tabularnewline
\includegraphics[width=7cm]{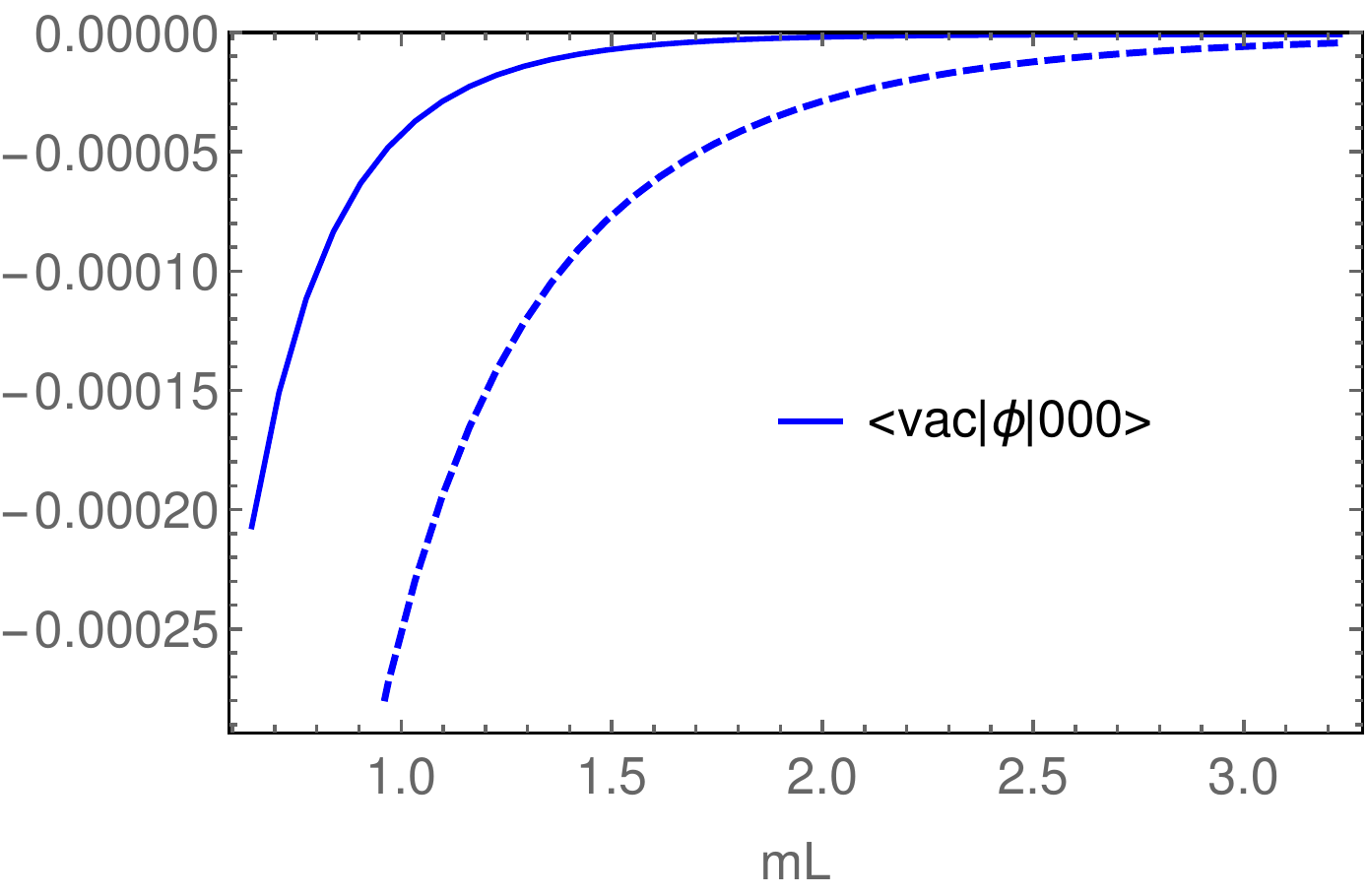} & \includegraphics[width=7cm]{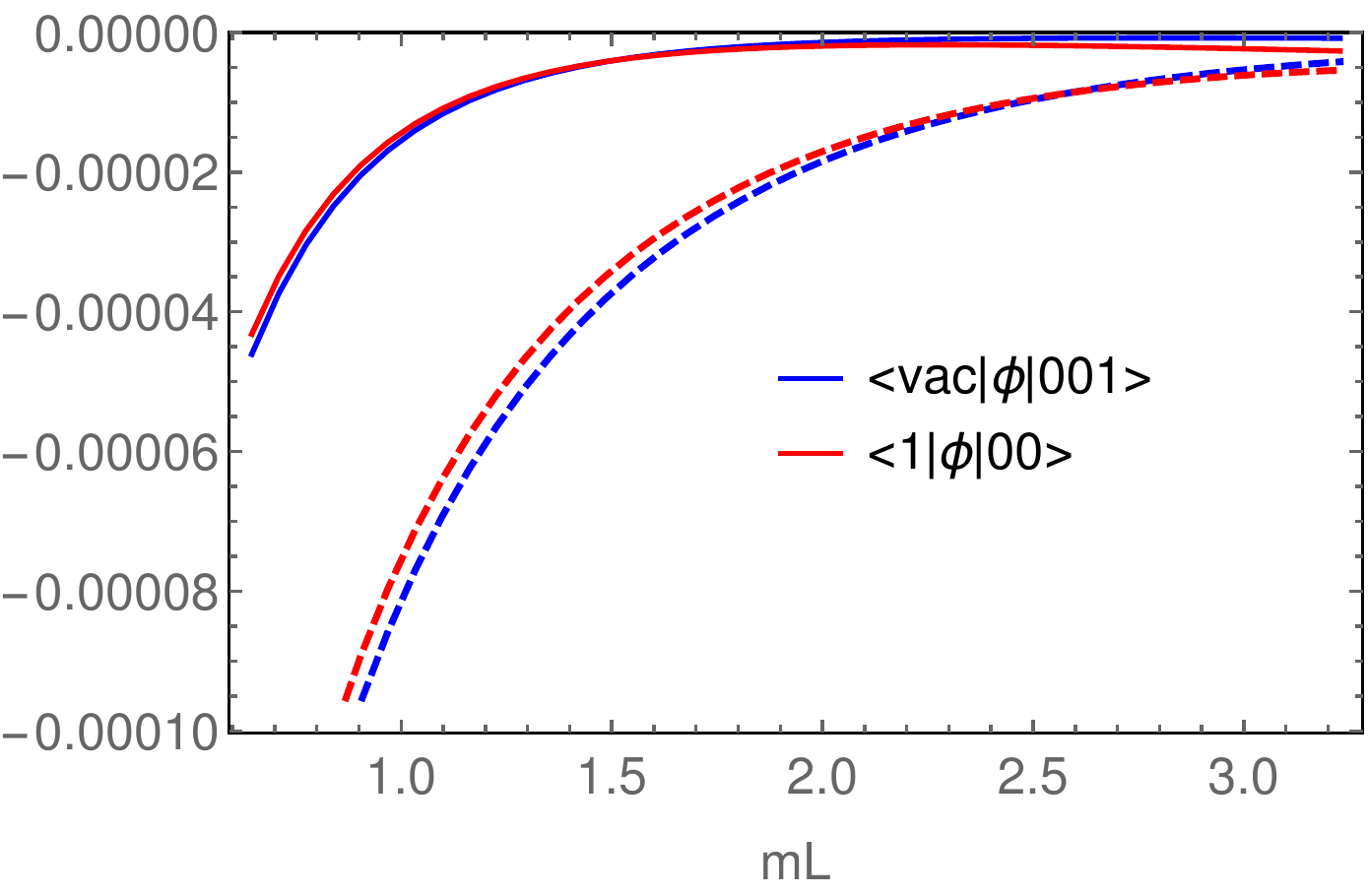}\tabularnewline
\end{tabular}
\par\end{centering}
\caption{Theoretical predictions for various finite volume form factors of
the field $\phi$, as compared to numerical TCSA data (the latter
is subtracted from each data set) at $b=1$. Polynomial (Pozsgay-Takács)
results are depicted by dashed curves, while the results containing
the first exponential corrections, conjectured by the present article,
are shown by continuous curves.\label{fig:fig5}}
\end{figure}

\begin{figure}[h]
\begin{centering}
\begin{tabular}{cc}
\includegraphics[width=7cm]{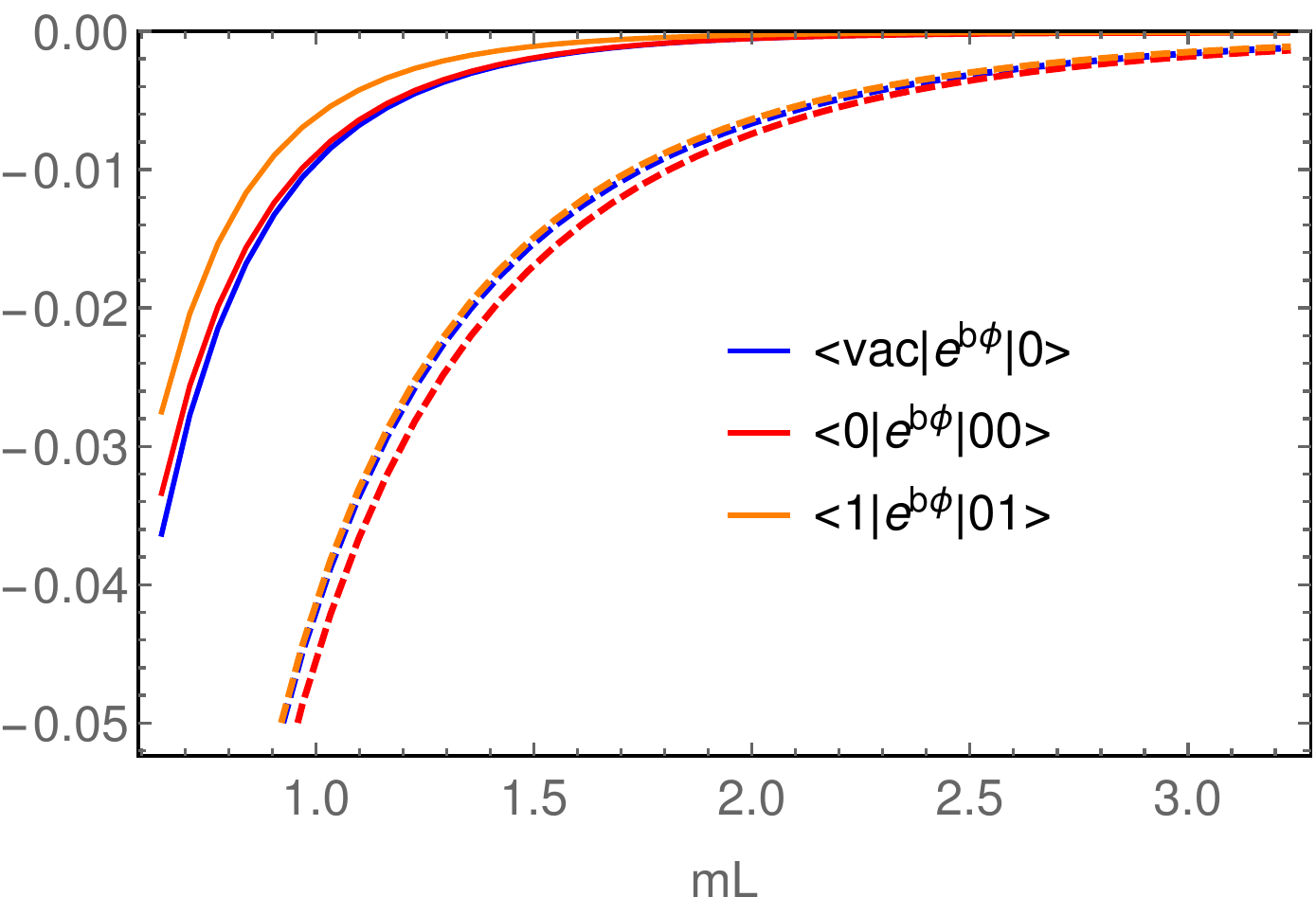} & \includegraphics[width=7cm]{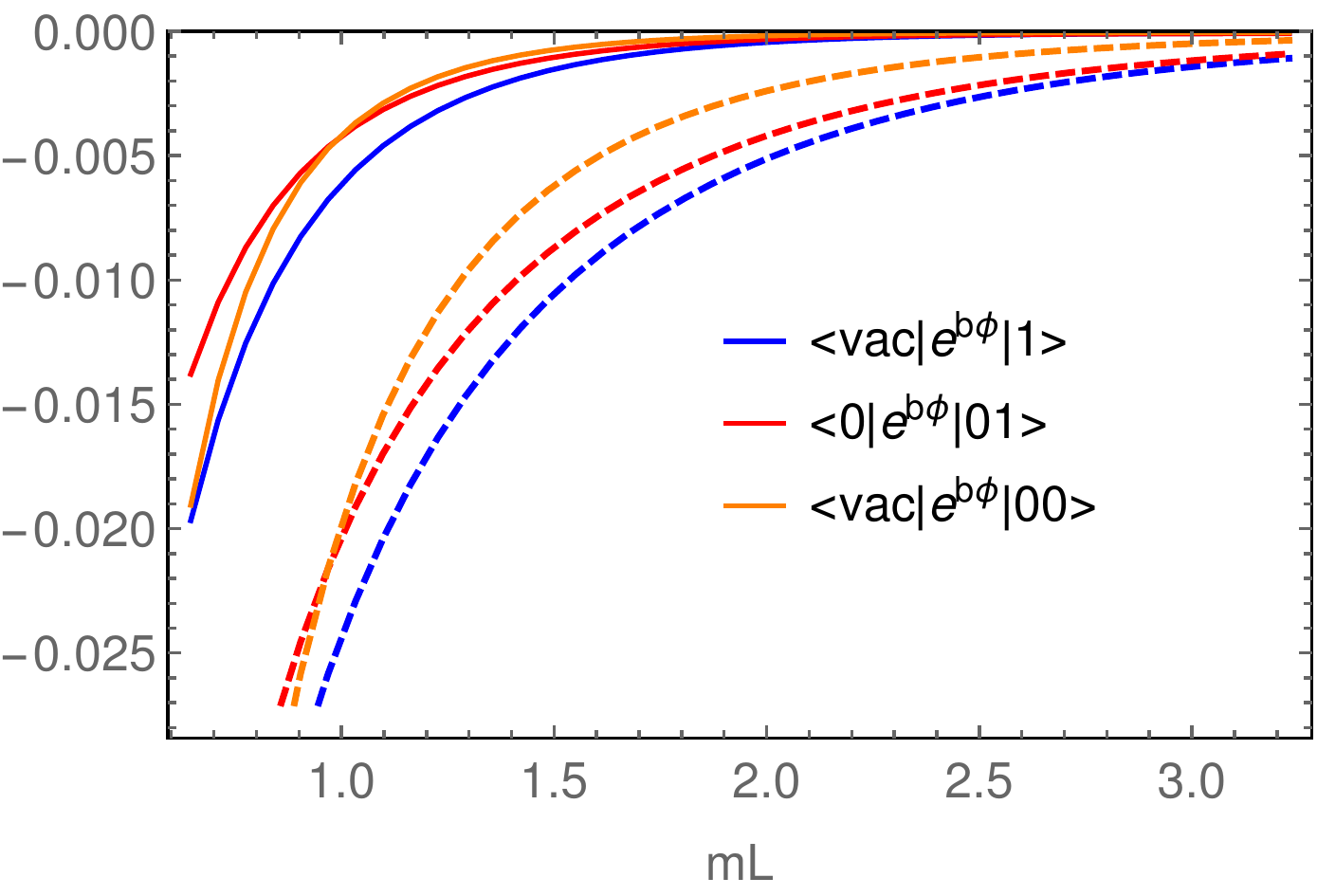}\tabularnewline
\includegraphics[width=7cm]{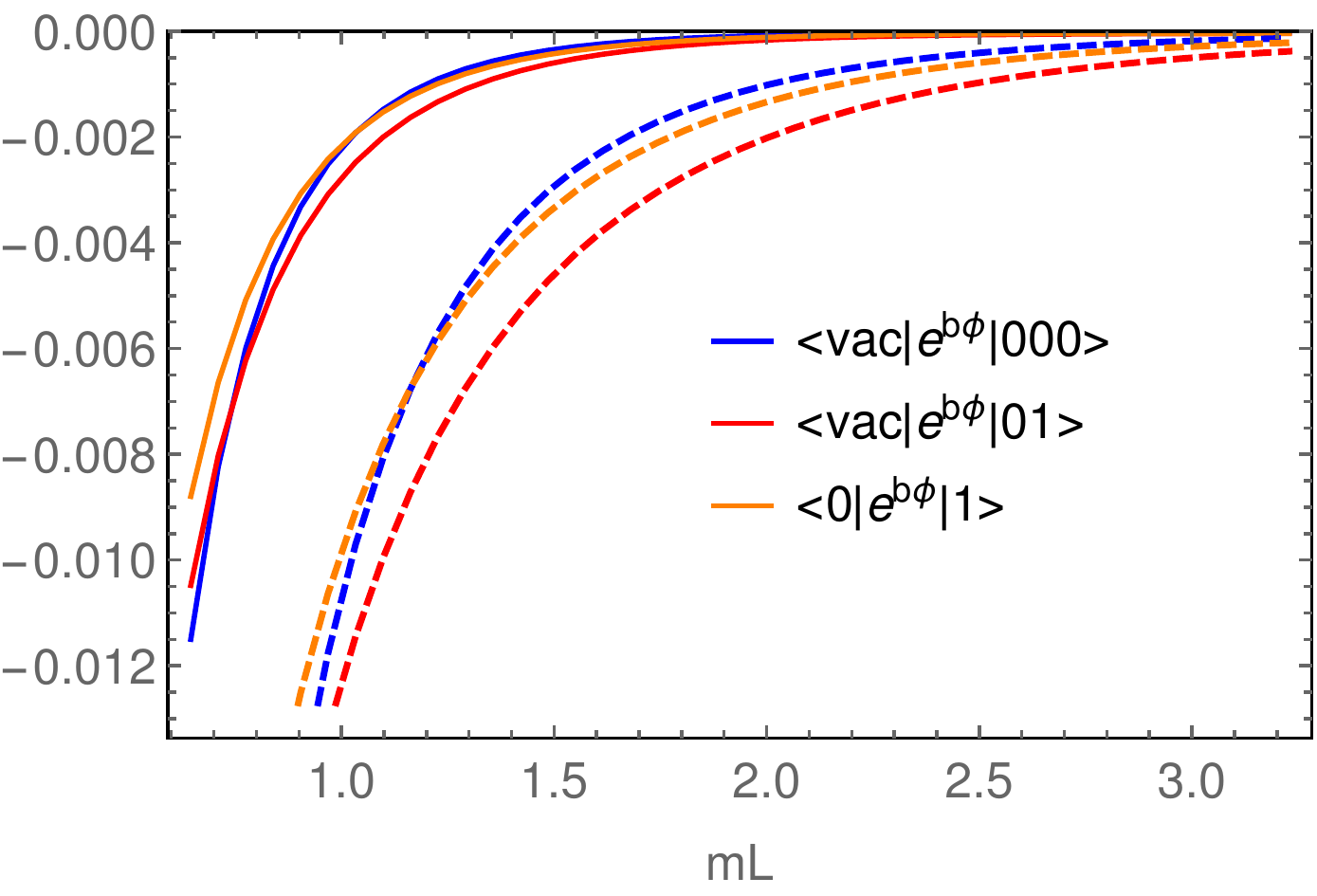} & \includegraphics[width=7cm]{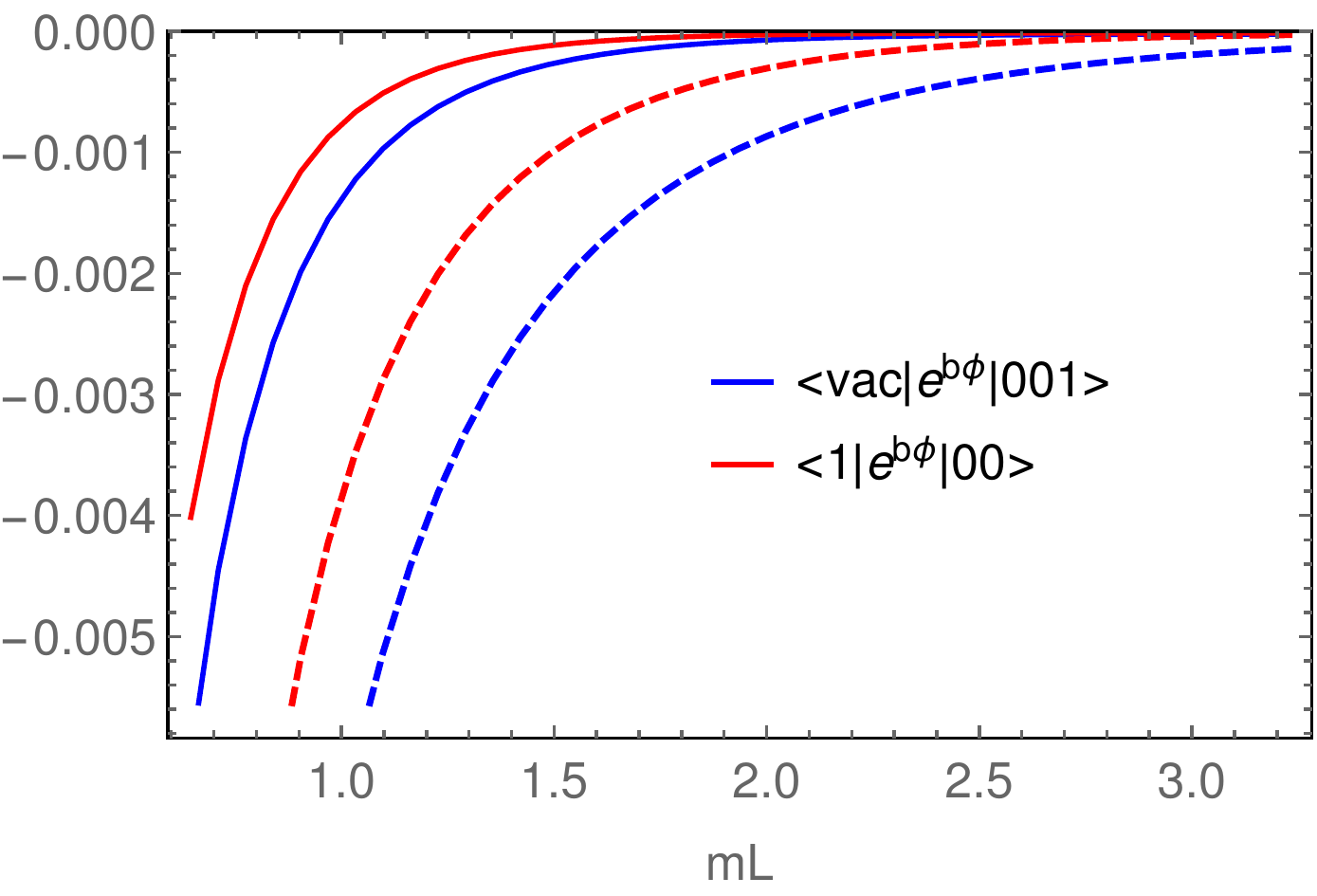}\tabularnewline
\end{tabular}
\par\end{centering}
\caption{Theoretical predictions for various finite volume form factors of
the primary operator $e^{b\varphi}$, as compared to numerical TCSA
data (the latter is subtracted from each data set) at $b=1$. Polynomial
(Pozsgay-Takács) results are depicted by dashed curves, while the
results containing the first exponential corrections, conjectured
by the present article, are shown by continuous curves.\label{fig:fig6}}
\end{figure}

In order to check our results in subsection (\ref{subsec:Fterm})
we analyze the finite size form factors of the elementary field and
the exponential operators $\mathcal{O}_{k}=:e^{kb\varphi}:$. For
both cases the infinite volume form factors have the form (\ref{eq:FFparam})
with
\begin{equation}
P_{n}^{\mathcal{O}_{k}}=\det_{i,j}\vert[i-j+k]\sigma_{2i-j}\vert\quad;\qquad[k]=\frac{\sin k\pi p}{\sin\pi p}
\end{equation}
where we used the basis of symmetric polynomials defined by 
\begin{equation}
\prod_{i=1}^{n}(z+x_{i})=\sum_{k=0}^{n}\sigma_{k}z^{n-k}
\end{equation}
The normalization for the elementary field is given by $H_{2n+1}^{\varphi}=\sqrt{\frac{Z\left(b\right)}{2}}\left(\frac{4\sin\pi p}{f(i\pi)}\right)^{n}$,
where $Z\left(b\right)$ is the wavefunction renormalization constant
\cite{Karowski:1978vz} 
\begin{equation}
Z\left(b\right)=\frac{8\pi^{2}p^{2}g}{b^{2}\sin\left(\pi p\right)f\left(i\pi\right)}
\end{equation}
In the plots regarding the finite volume form factors (Figures \ref{fig:fig5}
and \ref{fig:fig6}), the numerical TCSA data is subtracted. Then
the ``error'' of the polynomial (\ref{eq:FF0}) approximation (more
precisely, its difference from TCSA numerics), as calculated from
(\ref{eq:FFparam}) and (\ref{eq:FF0}), is shown by dashed lines.
Solid curves depict the results of the present paper ((\ref{eq:FF_F0}),
(\ref{eq:FF_Fterm})).

\begin{figure}[h]
\begin{centering}
\begin{tabular}{cc}
\includegraphics[width=7cm]{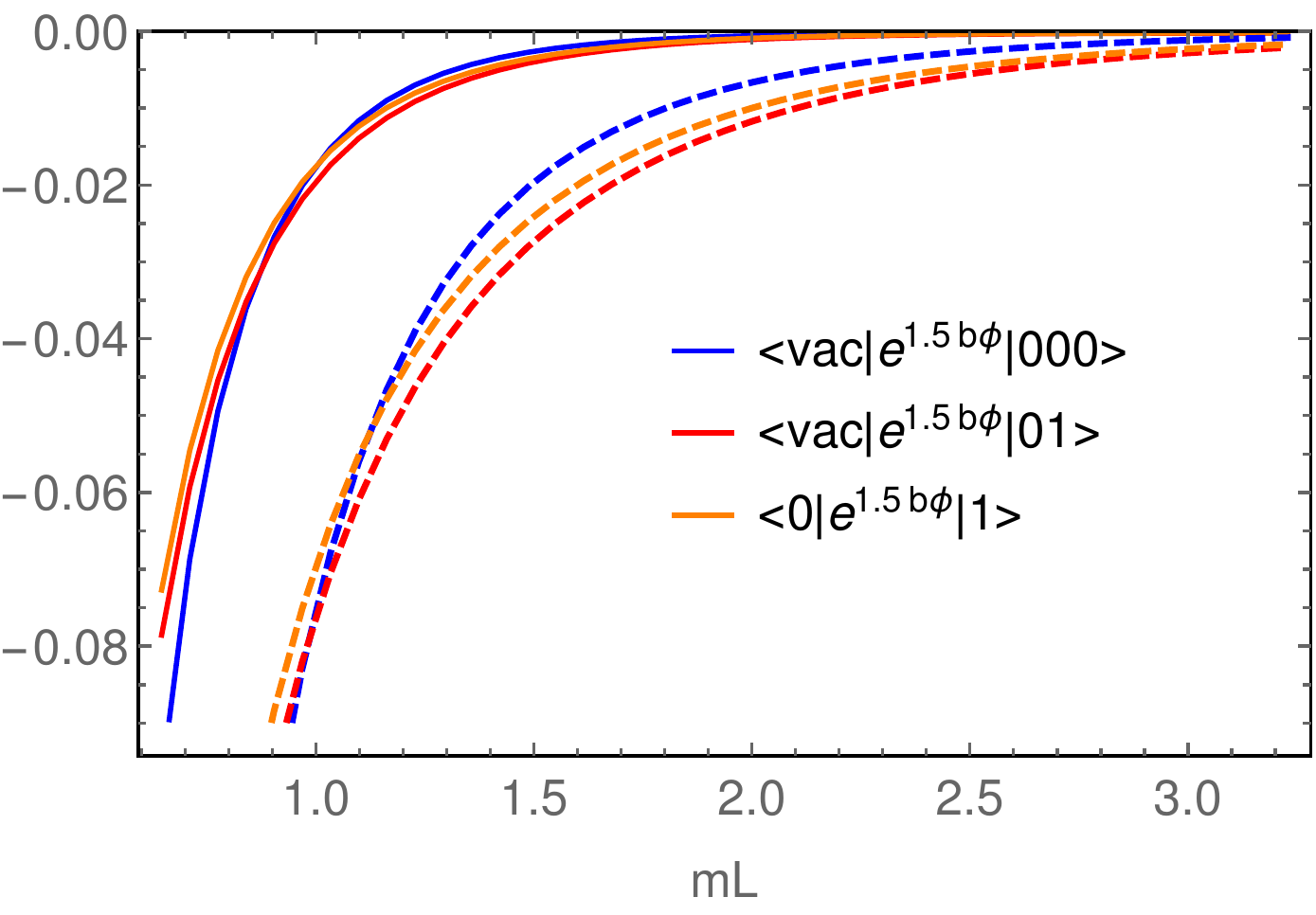} & \includegraphics[width=7cm]{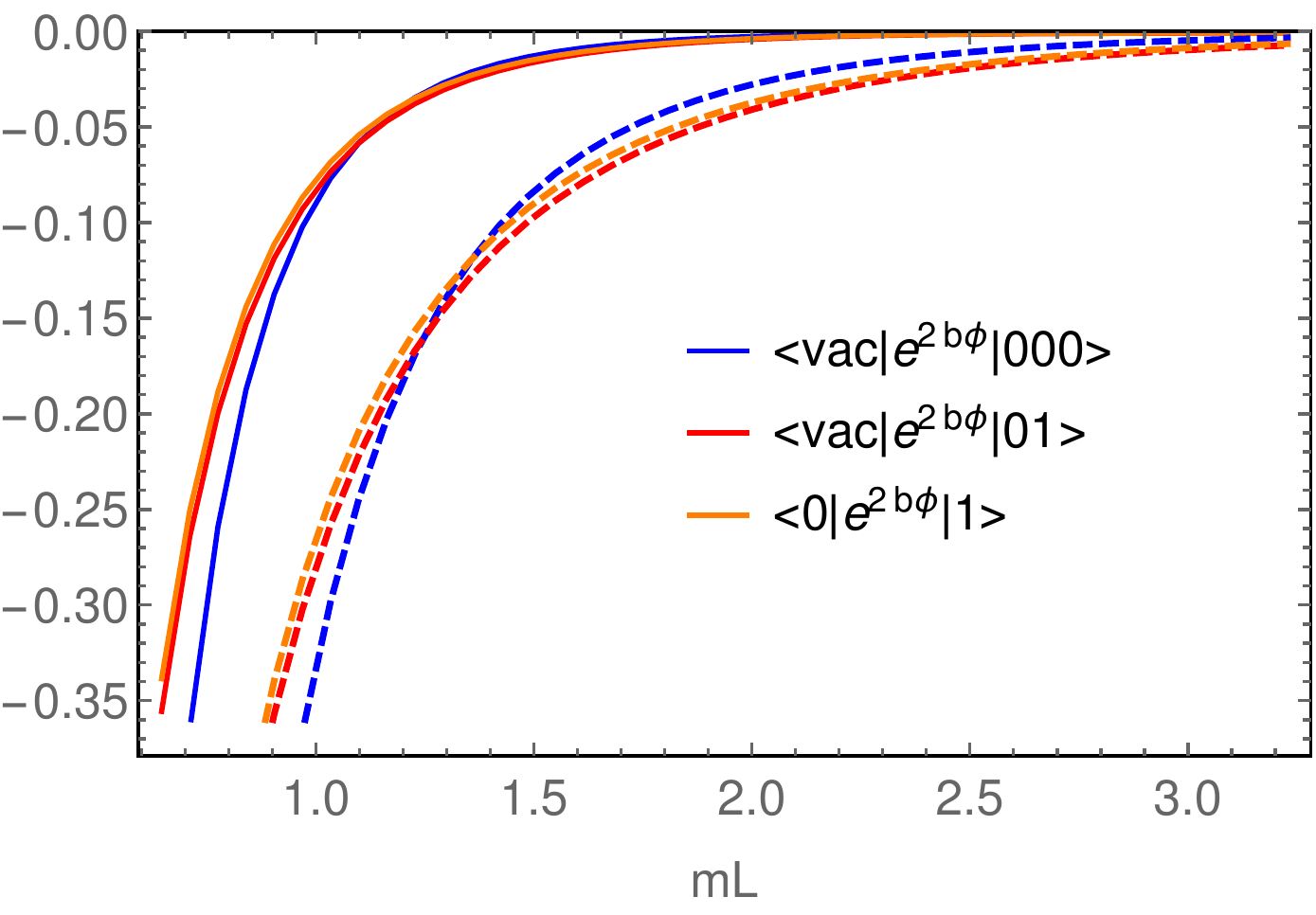}\tabularnewline
\end{tabular}
\par\end{centering}
\caption{Additional checks to some form factors of $e^{1.5b\varphi}$ and $e^{2b\varphi}$.
Polynomial (Pozsgay-Takács) results are depicted by dashed curves,
while the results containing the first exponential corrections, conjectured
by the present article, are shown by continuous curves. Again, the
TCSA results are subtracted from each data set.\label{fig:fig7}}
\end{figure}

The normalization for the exponentials is given by $H_{n}^{k}=\left\langle \mathcal{O}_{k}\right\rangle \left(\frac{4\sin\pi p}{f(i\pi)}\right)^{\frac{n}{2}}$,
where $\left\langle \mathcal{O}_{k}\right\rangle $ is given by the
Lukyanov-Zamolodchikov formula \cite{Lukyanov:1996jj}:
\begin{align}
\left\langle \mathcal{O}_{k}\right\rangle  & =m^{-\frac{k^{2}b^{2}}{4\pi g}}\left[\frac{\Gamma\left(\frac{1-p}{2}\right)\Gamma\left(1+\frac{p}{2}\right)}{4\sqrt{\pi}}\right]^{-\frac{k^{2}b^{2}}{4\pi g}}\\
 & \exp\intop_{0}^{\infty}\frac{dt}{t}\left[-\frac{\sinh^{2}\left(\frac{kb^{2}t}{4\pi g}\right)}{2\sinh\left(\frac{b^{2}t}{8\pi g}\right)\sinh t\cosh\left[\left(1+\frac{b^{2}}{8\pi g}\right)t\right]}+\frac{k^{2}b^{2}}{4\pi g}e^{-2t}\right]\nonumber 
\end{align}
On figure \ref{fig:fig7} we also present some additional checks for
the operators $e^{1.5b\varphi}$ and $e^{2b\varphi}$.

As shown on all of the plots including the theoretically predicted
leading exponential correction improved the data the same way as the
similar corrections improved the energy spectrum. This is a very strong
support for our conjectured $F$-term formulae. Let us see the analogous
results for the scaling Lee-Yang theory.

\subsection{Scaling Lee-Yang theory}

The scaling Lee-Yang theory is the only relevant perturbation of the
conformal Lee-Yang model:

\begin{equation}
S=S_{LY}+\lambda\int d^{2}z\,\Phi(z,\bar{z})
\end{equation}
The conformal Lee-Yang model is the simplest non-unitary minimal model
with central charge $c=-\frac{22}{5}$. There are two highest weight
representations: one corresponds to the identity operator and the
other one to the perturbing field $\Phi(z,\bar{z})$ with dimension
$(-\frac{1}{5},-\frac{1}{5})$. The Hamiltonian can be written on
the plane as

\begin{equation}
H=\frac{2\pi}{L}(L_{0}+\bar{L}_{0}-\frac{c}{12})+\lambda\biggl(\frac{L}{2\pi}\biggr)^{2+2/5}2\pi\delta_{P}\Phi(0,0)
\end{equation}
The parameter $\lambda$ is related to the mass of the scattering
particles as 
\begin{equation}
m=\frac{2^{\frac{19}{5}}\sqrt{\pi}(\Gamma(\frac{3}{5})\Gamma(\frac{4}{5})\lambda)^{\frac{5}{12}}}{5^{\frac{5}{16}}\Gamma(\frac{2}{3})\Gamma(\frac{5}{6})}
\end{equation}

\begin{figure}[h]
\centering{}\includegraphics[width=10cm]{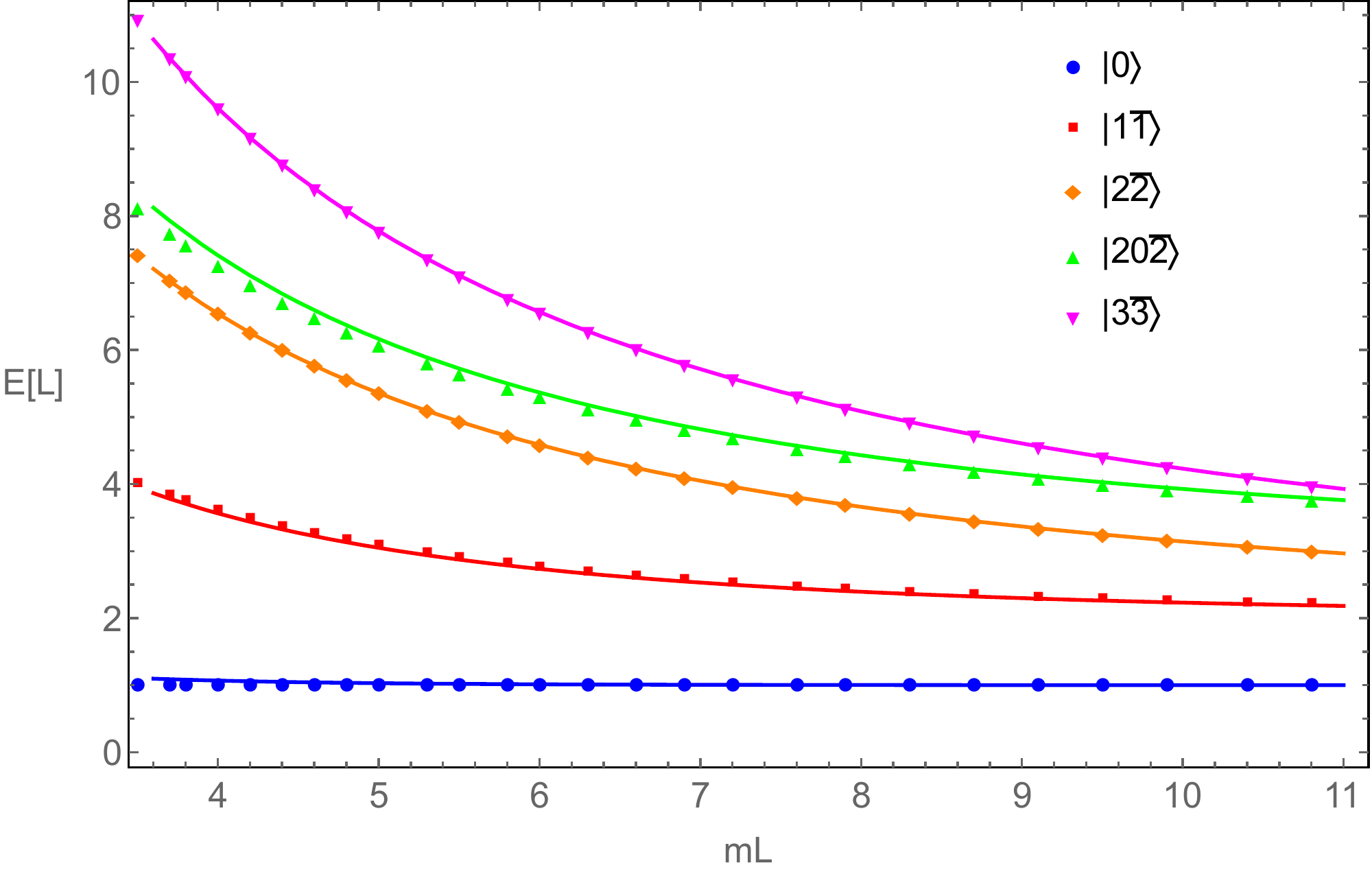}\caption{\label{fig:LYspectrum}Low lying finite size energy spectrum of the
scaling Lee-Yang model obtained from TCSA with the ground state energy
density subtracted. Continuous lines are the results of the Bethe-Yang
equations with the various quantization numbers, while discrete data
originates from TCSA.}
\end{figure}
The conformal Hilbert space is generated by acting with the negative
Virasoro modes on the highest weight states. On this space $L_{0}$
and $\bar{L}_{0}$ act diagonally and the matrix elements of $\Phi(0,0)$
can be calculated exactly. Similarly to the analysis of the sinh-Gordon
model, we truncate the Hilbert space and diagonalize $H$ on this
space to obtain the energy spectrum and the form factors of the perturbing
operator. The ground state energy density in the TCSA and the TBA
formulation are different. To relate the TCSA to the TBA results the
ground state energy density 
\begin{equation}
\epsilon=-\frac{m^{2}}{4\sqrt{3}}\text{ }
\end{equation}
has to be subtracted. The finite size spectrum obtained by the TCSA
method after the subtraction looks like in Figure \ref{fig:LYspectrum}.

\begin{figure}
\begin{centering}
\includegraphics[width=10cm]{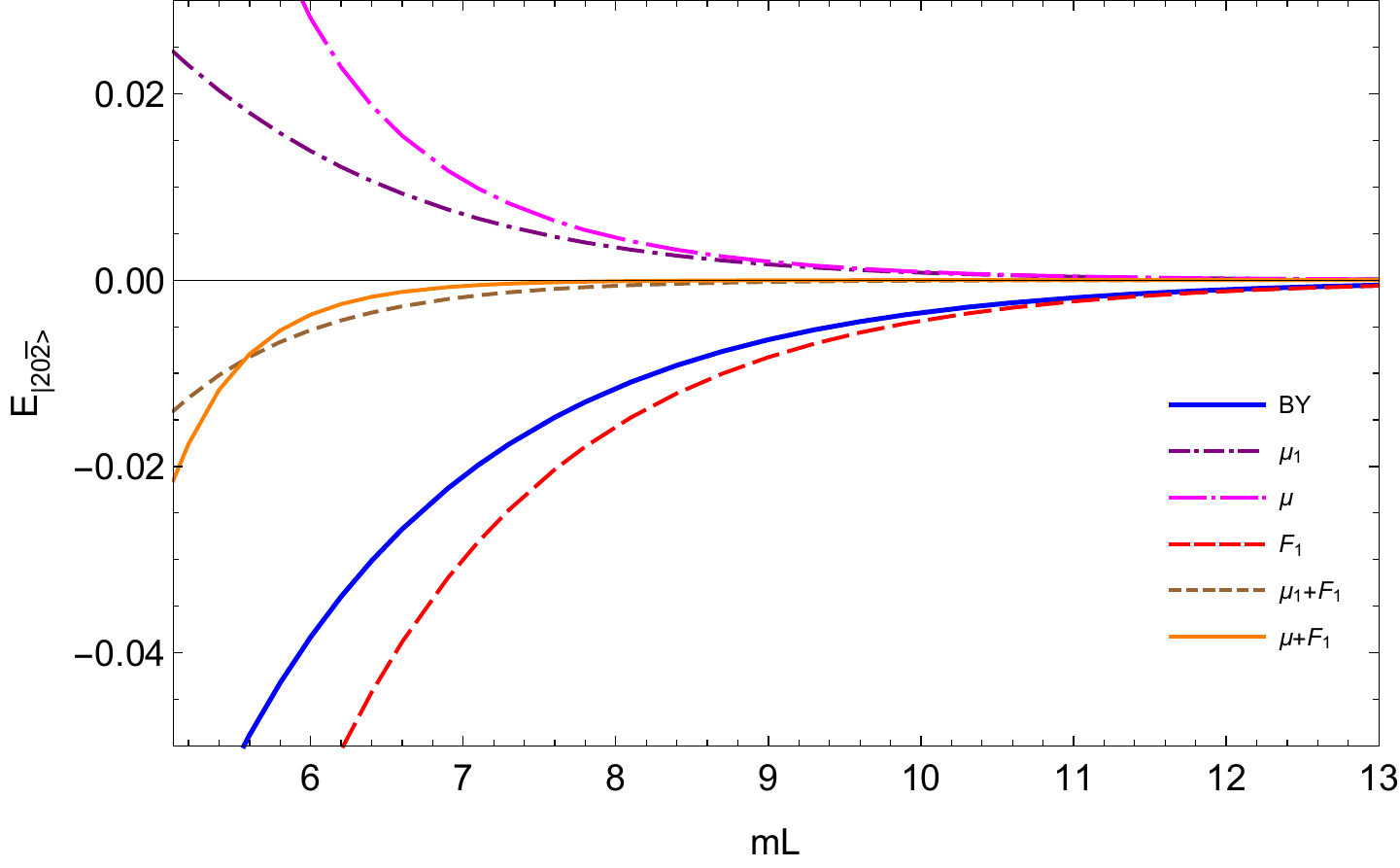}
\par\end{centering}
\caption{Exponential finite size energy correction for the state labeled by
the quantum numbers $\{2,0,-2\}$. ``$F_{1}$'' is the F-term correction,
while ``$\mu_{1}$'' is the leading $\mu$-term correction. ``$\mu$''
contains all the $\mu$-term corrections summed up, while ``$\mu+F_{1}$''
adds the $F$-term correction to this.}

\label{fig:LYenexp}
\end{figure}

In order to visualize the various finite size corrections we subtract
the numerical TCSA data from the theoretical curves as shown on Figure
\ref{fig:LYenexp}. In the domain investigated we compared the TCSA
data to the exact TBA result and found that it's precision was $10^{-5}$.
Thus there is no visual difference in subtracting TCSA compared to
the exact results. For the form factors we do not know the exact results,
therefore we can only subtract the TCSA data and this is the reason
why we followed this approach here. The Bethe-Yang correction (\ref{eq:shGQ0})
contains the polynomial volume corrections. The $F_{1}$ term is the
leading $F$-term correction (\ref{eq:shGE1}), while $\mu_{1}$ is
the leading $\mu$-term correction (\ref{eq:mu1}). The $\mu$ correction
sums up all the $\mu$-terms by solving (\ref{eq:LYQQ0}) for the
constituents. Then we combine these $\mu$-terms corrections with
the leading $F$-term corrections.

Neither the $F$-term nor the $\mu$-term correction gives a good
approximation for volumes $7-13$, however their sum is very close
to result of the numerics. The best approximation arises from combining
the summed up $\mu$-term correction with the leading $F$-term correction.

\begin{figure}
\includegraphics[width=7cm]{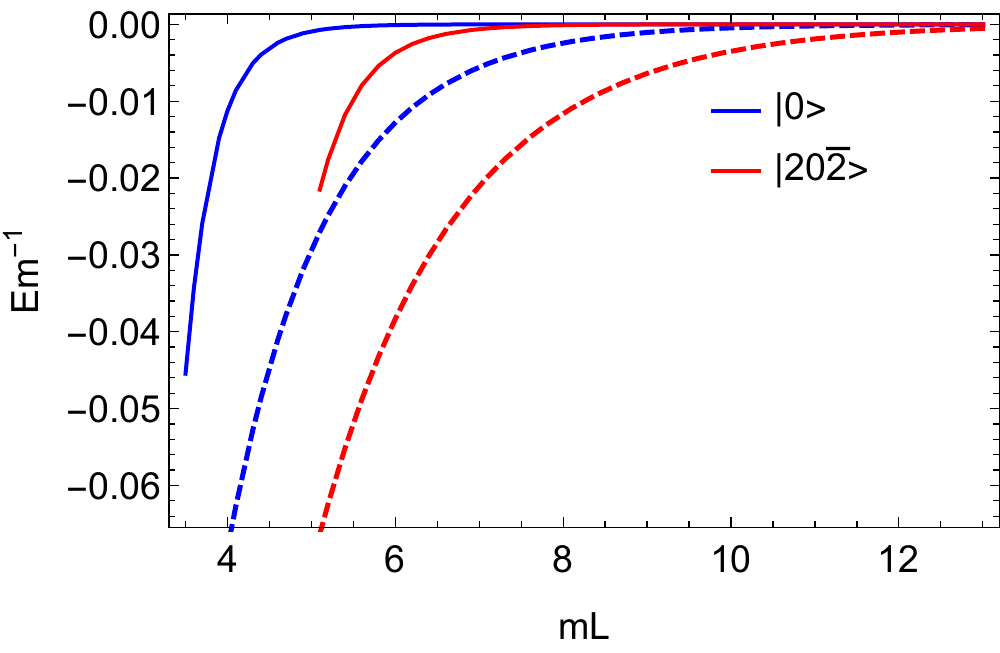}\hspace{1cm}\includegraphics[width=7.3cm]{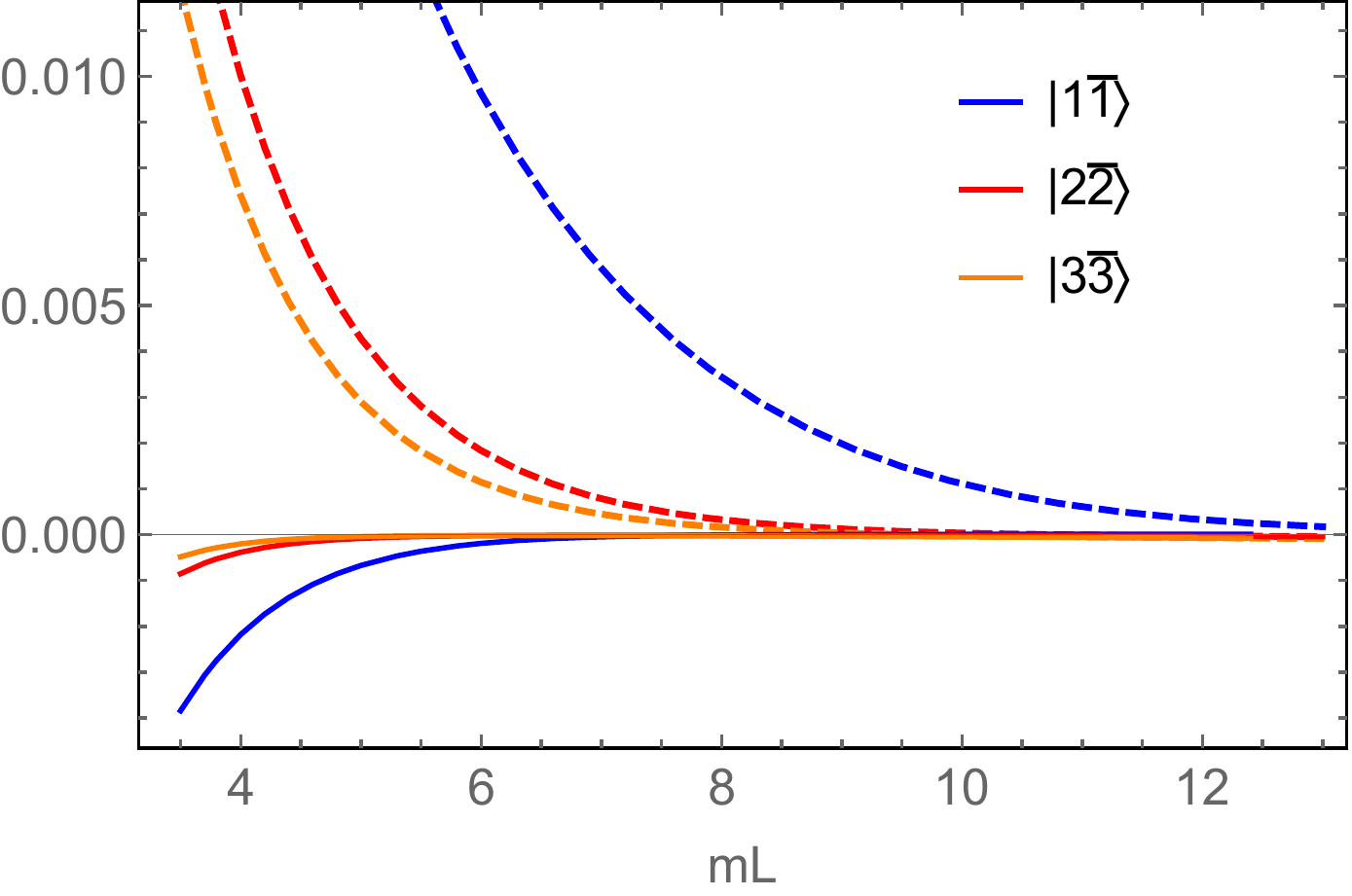}

\caption{\label{fig:LYEnergybest}Demonstration of the polynomial and exponential
volume corrections for the energy. Dashed lines are the difference
between the Bethe-Yang energies and TCSA data and solid lines are
the difference of the best $\mu+F$ approximation and the TCSA data.
The energy cut is $16$.}
\end{figure}
The best results are demonstrated on figure (\ref{fig:LYEnergybest}).
These suggest that we understand the finite size correction of the
energy levels very well, thus we now turn to the investigation of
finite volume form factors.

\begin{figure}[h]
\begin{centering}
\includegraphics[width=7cm]{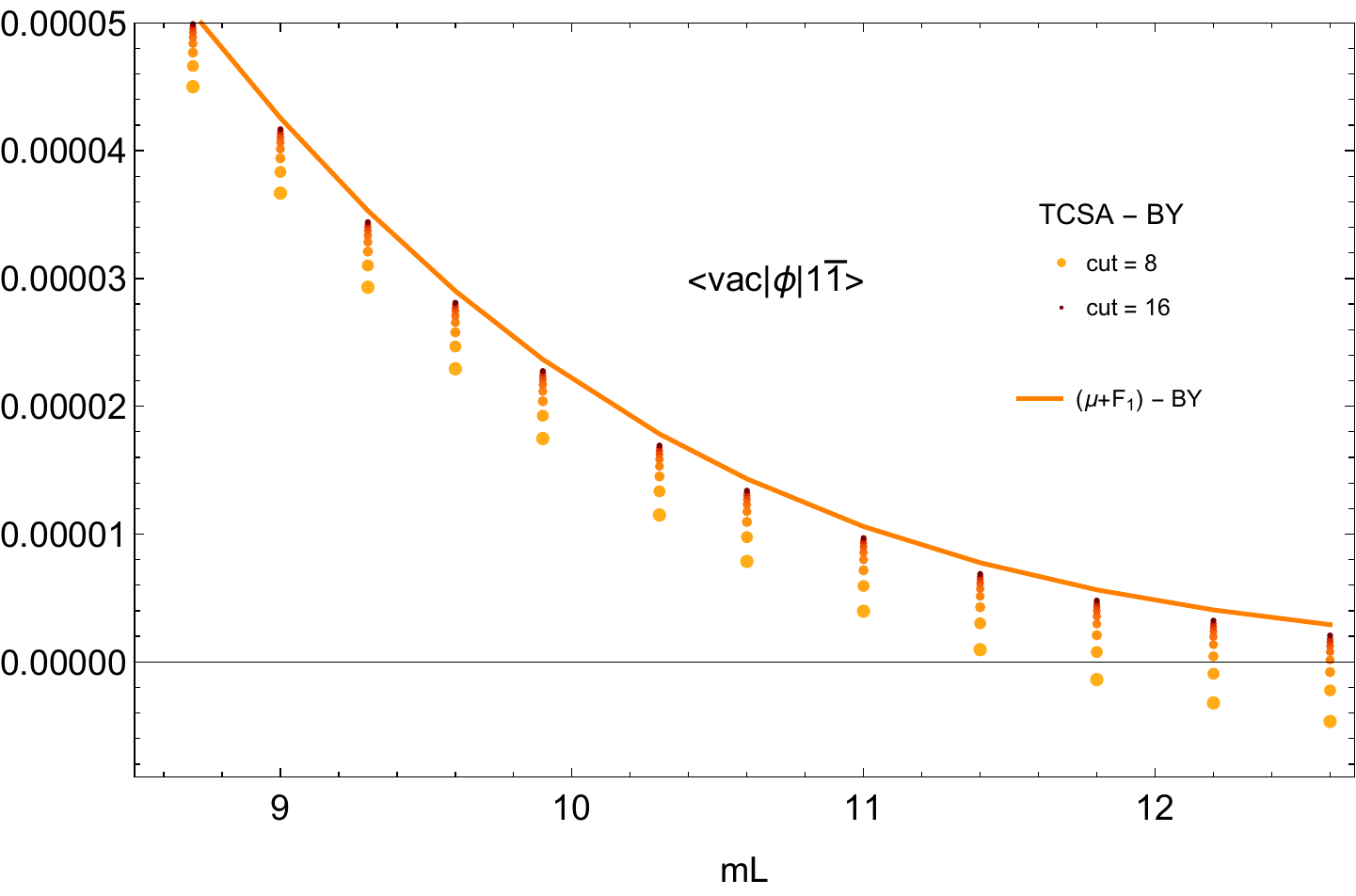}\hspace{1cm}\includegraphics[width=7cm]{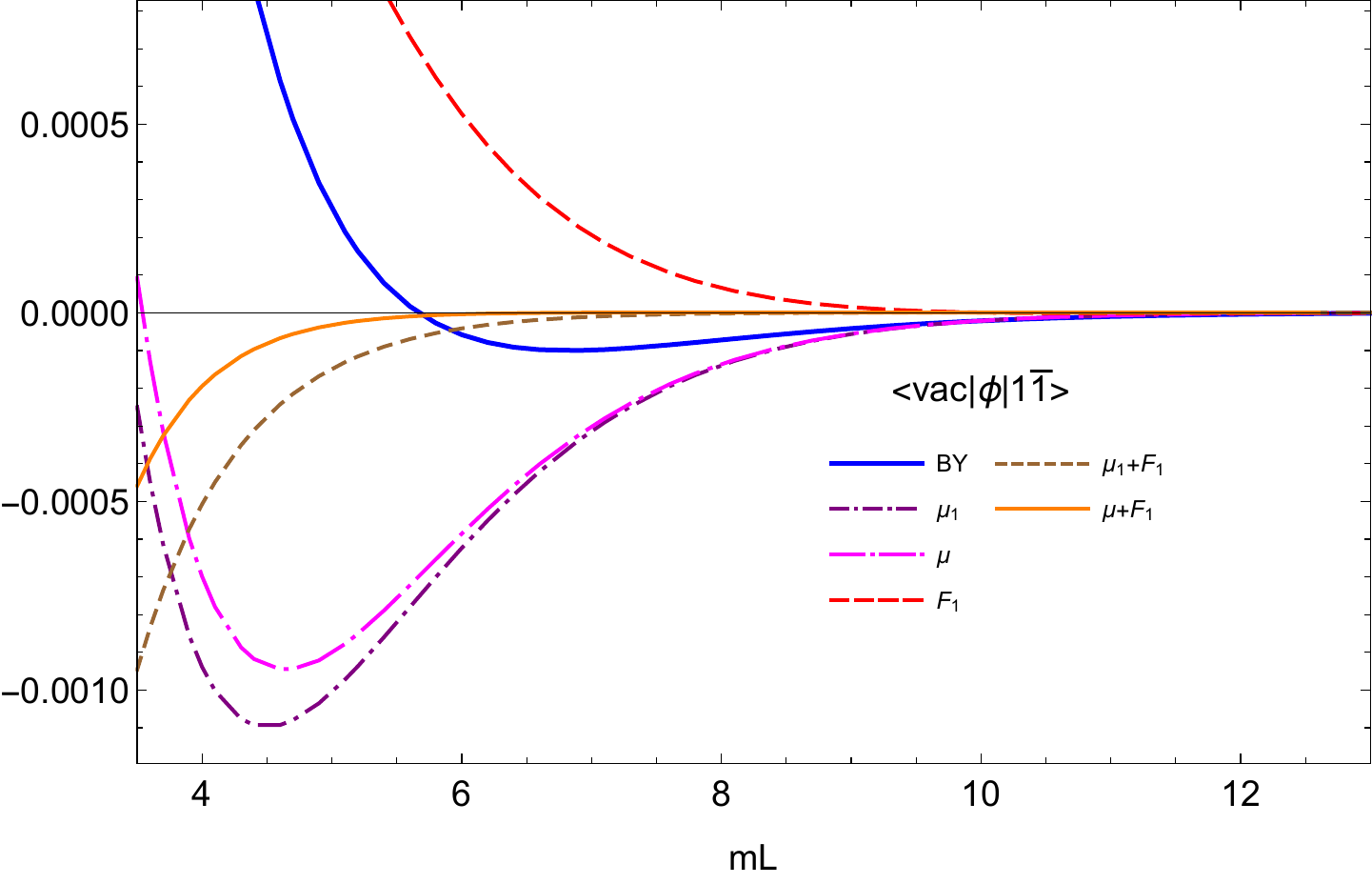}
\par\end{centering}
\caption{Left: Cut dependence of the TCSA data. The difference of the TCSA
and the BY data at different cuts is marked by dots with various colours
and the difference between the $\mu+F_{1}$ and BY data is represented
with a solid line. Right: Volume corrections of the form factor $\langle\mathrm{vac}\vert\Phi\vert\{1,-1\}\rangle$.
Every line shows the absolute value of the difference between the
given theoretical curve and the TCSA data.}

\label{LYFVFF01}
\end{figure}
The infinite volume form factors of the perturbing operators are given
by (\ref{eq:FFparam}) with
\begin{equation}
\langle\Phi\rangle=\frac{3^{\frac{9}{10}}\Gamma(\frac{1}{3})^{\frac{36}{5}}}{(2\pi)^{\frac{14}{5}}5^{\frac{1}{4}}\Gamma(\frac{1}{5})\Gamma(\frac{2}{5})}\quad;\qquad H_{n}=\left(\frac{\Gamma}{\sqrt{f(i\pi)}}\right)^{n}
\end{equation}
and with $P_{1}=1$, $P_{2}=\sigma_{1}$ and for $n>2$ 
\begin{equation}
P_{n}(x_{1},\dots,x_{n})=\sigma_{1}\sigma_{n-1}\det_{ij}\left|\sigma_{3i-2j+1}\right|
\end{equation}
where we used the basis of symmetric polynomials. 

Similarly to the energy spectrum we subtract the TCSA data from the
various theoretical curves. Such a result is displayed on Figure \ref{LYFVFF01}.
The BY curves are the result of (\ref{eq:FF0}), the leading $\mu$-term
correction denoted as $\mu_{1}$ is given by (\ref{eq:inoutmu}) and
the leading $F$-term correction, $F_{1},$ by (\ref{eq:FF_Fterm}).
Plugging the solution of (\ref{eq:LYQQ0}) into the asymptotic formula
(\ref{eq:Fconstituent}) the form factor $\mu$-terms can be summed
up. This correction is denoted by $\mu$. Finally, adding the leading
$F$-term to the $\mu$-terms leads to the best approximations.

\begin{figure}
\includegraphics[width=7cm]{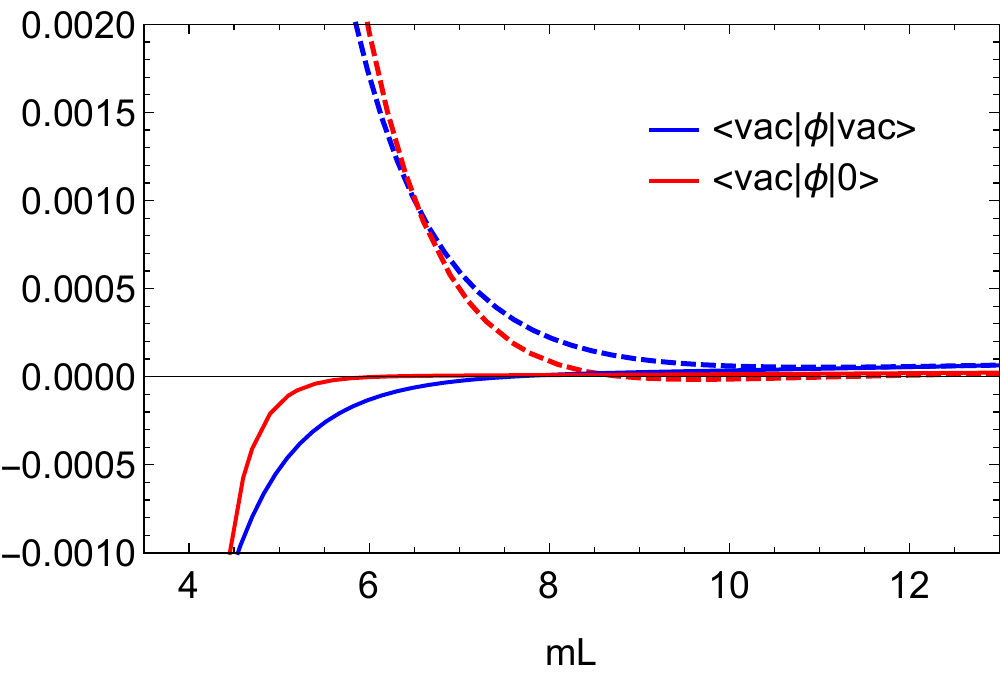}\hspace{1cm}\includegraphics[width=7cm]{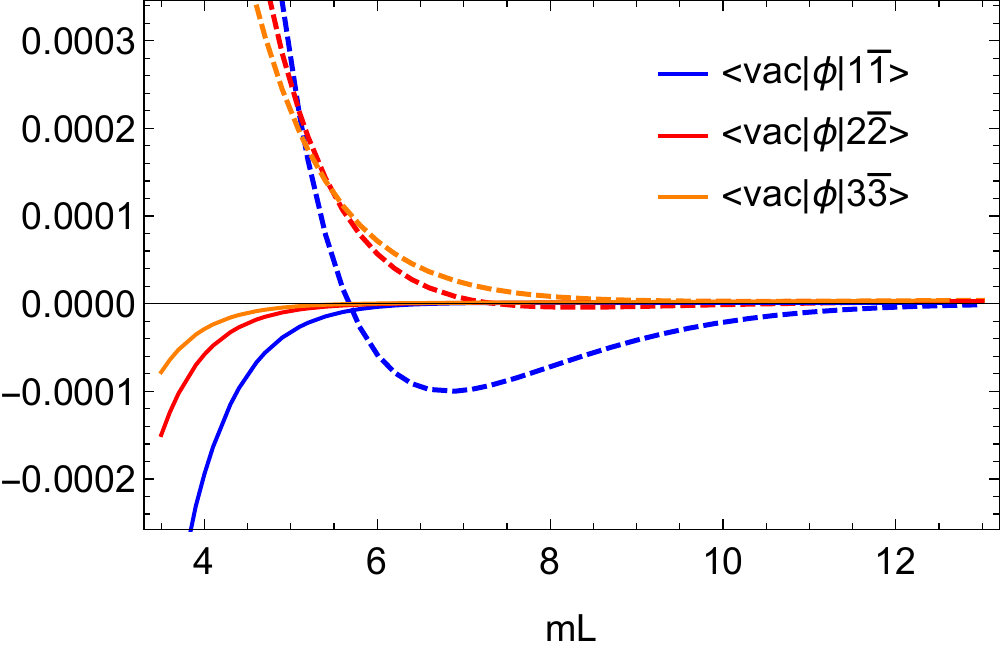}

\includegraphics[width=7cm]{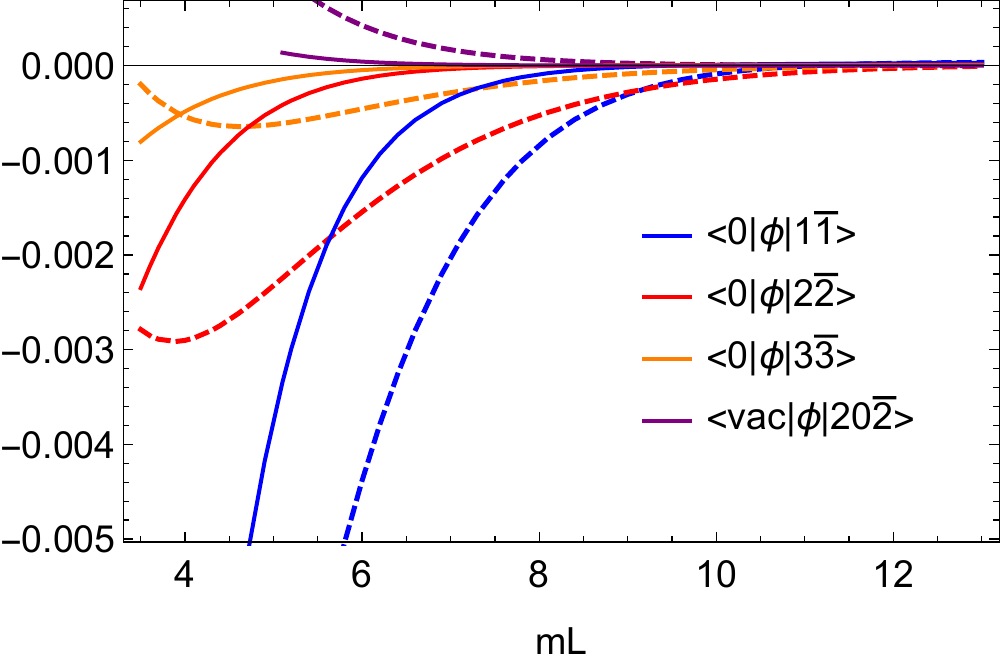}\hspace{1cm}\includegraphics[width=7cm]{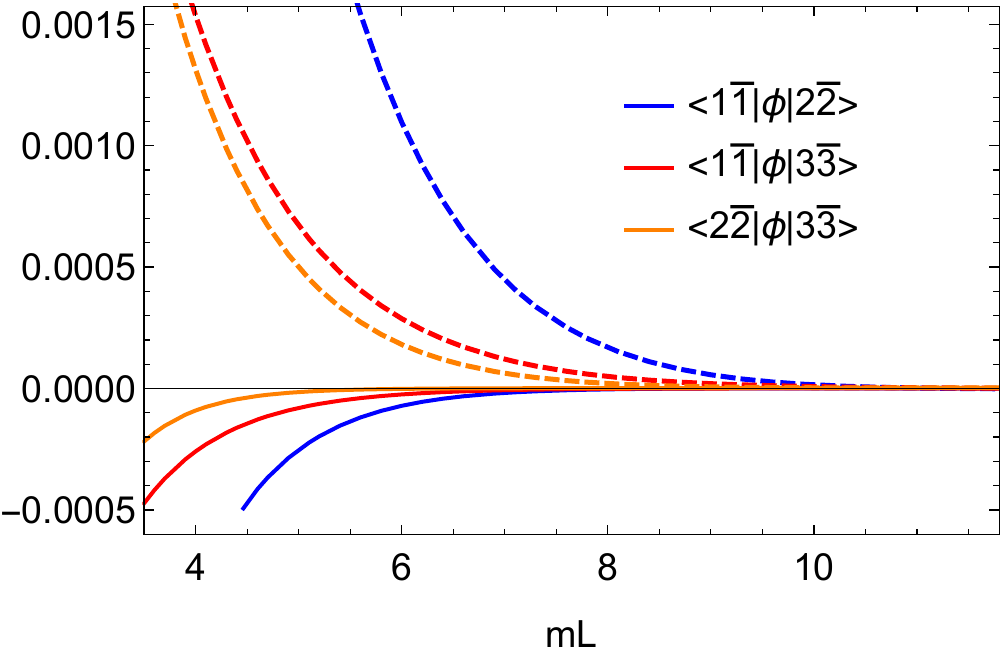}

\caption{\label{fig:LYFFbest}Representative figures for the polynomial and
exponential corrections. Dashed lines are the absolute values of the
difference between the BY and TCSA data, while solid lines are the
same for the best available exponential corrections. For the $\langle\mathrm{vac}\vert\Phi\vert\mathrm{vac}\rangle$
diagonal matrix element we show the F-term only. For $\langle\mathrm{vac}\vert\Phi\vert\{0\}\rangle,\langle\mathrm{vac}\vert\Phi\vert\{n,-n\}\rangle$
the best available correction is ``$\mu+F_{1}$'', while for $\langle\{0\}\vert\Phi\vert\{n,-n\}\rangle,\langle\mathrm{vac}\vert\Phi\vert\{2,0,-2\}\rangle,\langle\{m,-m\}\vert\Phi\vert\{n,-n\}\rangle;\;n,m=1,2,3$:
best available correction is ``$\mu_{1}+F_{1}$''}
\end{figure}

On figure (\ref{fig:LYFFbest}) we demonstrate the BY and the best
$\mu+F_{1}$ corrections for various states.

In summary, the data presented gives a strong evidence for the correctness
of our exponential finite size corrections.

\section{Conclusion}

In this paper we presented the leading exponentially small volume
corrections for non-diagonal form factors in diagonally scattering
theories. In theories with bound states the leading correction is
the $\mu$-term, which we derived using the asymptotic finite volume
form factor and the assumption that particles are composed of their
constituents. The $F$-term is universal in the sense that it is present
in theories both with and without bound states, providing the next
and leading exponential correction, respectively. We derived the $F$-term
formally and tested the results in various ways.

We showed that by taking appropriate residues of the integral for
the rapidities of the virtual particles we can completely recover
the $\mu$-term correction. We checked that taking the diagonal limit
of the form factors, by sending one rapidity to infinity based on
\cite{Bajnok:2017bfg} reproduced the diagonal result of \cite{Pozsgay:2014gza}.
By developing numerical methods to ``measure'' the finite volume
form factors we tested the finite size corrections in the sinh-Gordon
and Lee-Yang models, where we found convincing confirmation of our
formulae.

\begin{figure}
\begin{centering}
\includegraphics[height=6cm]{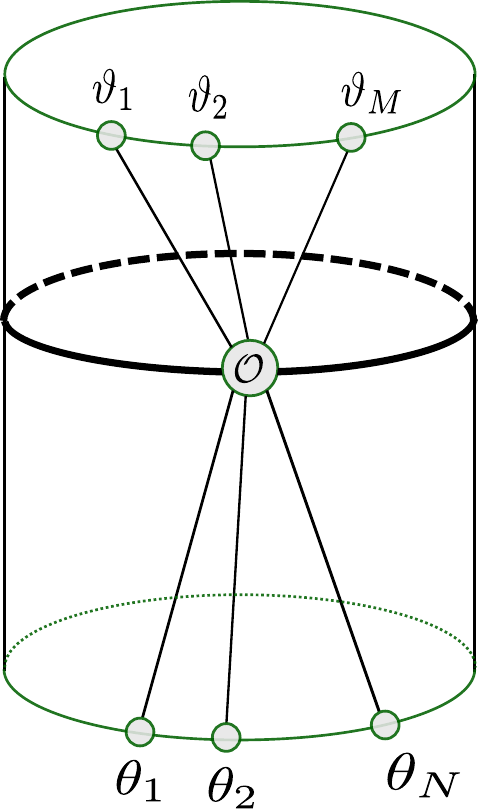}
\par\end{centering}
\caption{Graphical representation of the form factor $F$-term L\"uscher correction.
A virtual particle pair appears from the vacuum and after travelling
around the world is absorbed by the operator.}

\label{fig:LuscherF}
\end{figure}
Figure \ref{fig:LuscherF} visualizes the physical picture behind
the $F$-term: First a virtual particle-anti-particle pair appears
from the finite volume vacuum, then one of them travels around the
world and finally both are absorbed by the operator. Since the infinite
volume form factor with a particle-anti-particle pair is singular
we had to regulate the appearing amplitude. Our complicated derivation
and checks resulted in the proper definition of this \emph{regulated
}form factor. We found that we had to subtract the kinematical singularity
of the form factor in a symmetric way (\ref{eq:regFFgen}). This \emph{regulated}
form factor has very nice properties. Its phase is the same as the
original form factor's for which we are calculating the correction.
Its singularities on the upper and lower half planes are related to
each other, in such a way that the $\mu$-term corrections are correctly
reproduced when the residues are taken . In the simplest non-trivial
(vacuum-one-particle) form factor it is real and reproduces our previous
results \cite{Bajnok:2018fzx}, which we derived using a finite volume
analogue of the LSZ reduction formula of the two-point function.

We approached the problem of calculating the partition function and
evaluating the asymptotic form factor for bound states through systematic
large volume expansions. Clearly this method can be used at higher
orders and the resulting finite size form factors give the building
blocks of the finite size or finite temperature correlation functions.
These can be used in statistical physical or solid state systems as
well as in the AdS/CFT duality. Although the expansion can be useful
for practical applications, for obtaining exact results the series
has to be summed up. In this respect the integral equation derived
recently for diagonal form factors in the sinh-Gordon theory can be
useful \cite{Bajnok:2019yik}.

\subsection*{Acknowledgments}

We would like to thank Aron Bodor for the collaboration at an early
stage of this work. Furthermore, Janos Balog, Arpad Hegedus, Fedor
Smirnov and Gabor Takacs for the useful discussions. M.L. thanks to
Robert Konik and Giuseppe Mussardo for the useful discussions regarding
the TCSA method employed (see also footnote \ref{fn:footnote_mussardo}).
This research was supported by the NKFIH research Grant K116505, the
NKFIH Summer Student Internship Program and the Lendulet Program of
the Hungarian Academy of Sciences.

\appendix

\section{Derivation of the form factors' $\mu$-term correction\label{sec:muterm}}

In this Appendix we calculate the $\mu$-term correction for form
factors in the scaling Lee-Yang model. We start with the order $\left(0\right)$
form factor 
\begin{equation}
\langle0\vert\mathcal{O}\vert\{\theta_{\pm}\}\rangle_{L}=\frac{F_{2N}(\theta_{1+},\theta_{1-},\ldots\theta_{N+},\theta_{N-})}{\sqrt{\prod\limits _{k}S(\theta_{k+,k-})\rho_{2N}(\{\theta_{\pm}\})\prod\limits _{i<j,s,k}S(\theta_{is,jk})}}
\end{equation}
evaluated at $\theta_{j\pm}=\bar{\theta}_{j\pm}^{(\mu)}=\bar{\theta}_{j}^{(\mu)}\pm i(u+\delta\bar{u}_{j})$.
Our aim is to expand this expression at leading order in $\delta\bar{u}_{j}$.
We first multiply both the numerator and the denominator by $\left(\prod_{k}\frac{2\delta\bar{u}_{k}}{\Gamma}\right)$
in order to ensure that the expansion of the numerator starts with
the form factor $F_{N}(\{\bar{\theta}^{(\mu)}\})$:

\begin{equation}
\prod_{k}\left(\frac{2\delta\bar{u}_{k}}{\Gamma}\right)F_{2N}(\{\bar{\theta}_{\pm}^{(\mu)}\})=F_{N}(\{\bar{\theta}^{(\mu)}\})+\sum_{k}\left(\frac{2\delta\bar{u}_{k}}{\Gamma}\right)F_{N,k}^{b}(\{\bar{\theta}^{(0)}\})+\dots
\end{equation}
In evaluating the denominator we first combine the rows and columns
of $\rho_{2N}$ to obtain the derivatives of $Q$ and $\bar{Q}$ with
respect to $\partial$ and $\bar{\partial}$ as:

\begin{align}
\rho_{2N}(\{\theta_{\pm}\}) & =\det\left\{ \frac{\partial\left(Q_{1+},Q_{1-},\ldots,Q_{N+},Q_{N-}\right)}{\partial\left(\theta_{1+},\theta_{1-},\ldots,\theta_{N+},\theta_{N-}\right)}\right\} =\frac{1}{4^{N}}\begin{vmatrix}\left[\bar{\partial}\bar{Q}\right] & \left[\bar{\partial}Q\right]\\
\left[\partial\bar{Q}\right] & \left[\partial Q\right]
\end{vmatrix}\nonumber \\
 & =\frac{1}{4^{N}}\det\left[\bar{\partial}\bar{Q}\right]\det\left\{ \left[\partial Q\right]-\left[\partial\bar{Q}\right]\left[\bar{\partial}\bar{Q}\right]^{-1}\left[\bar{\partial}Q\right]\right\} 
\end{align}
where $\left[\bar{\partial}\bar{Q}\right]_{ij}=\left[\bar{\partial}\bar{Q}\right]_{ji}=\bar{\partial}_{i}\bar{Q}_{j}^{(0)}(\{\theta_{\pm}\})$
with $\bar{\partial}_{i}=\partial_{i+}-\partial_{i-}$. Similarly
$\left[\bar{\partial}Q\right]_{ij}=\left[\partial\bar{Q}\right]_{ji}=\bar{\partial}_{i}Q_{j}^{(0)}(\{\theta_{\pm}\})$
and $\left[\partial Q\right]_{ij}=\left[\partial Q\right]_{ji}=\partial_{i}Q_{j}^{(0)}(\{\theta_{\pm}\})$.
Up to the next-to-leading order we can use that 
\begin{equation}
\left(\frac{2\delta\bar{u}_{k}}{\Gamma}\right)^{2}S(\theta_{k+,k-})=\phi(2i(u+\delta\bar{u}_{k}))^{-1}+O((\delta\bar{u})^{3})=2\delta\bar{u}_{k}+\left(\frac{2\delta\bar{u}_{k}}{\Gamma}\right)^{2}S_{0}+O((\delta\bar{u})^{3})
\end{equation}
thus there are poles of type $\delta\bar{u}_{k}^{-1}$ in the diagonal
elements of $[\bar{\partial}\bar{Q}]$ originating from 
\begin{equation}
\bar{\partial}_{j}\bar{Q}_{j}^{(0)}(\{\theta_{\pm}\})=4\phi(2i(u+\delta\bar{u}_{j}))+\partial_{j}Q_{j}^{(0)}(\{\theta_{\pm}\})
\end{equation}
Expanding up to next-to-leading order gives 
\begin{align}
\left(\prod_{k}\frac{1}{4\phi(2i(u+\delta\bar{u}_{k}))}\right)\det\left\{ \text{diag}[\{4\phi(2i(u+\delta\bar{u}_{k}))\}]+[\partial Q]\right\} \det\left\{ \left[\partial Q\right]-\left[\partial\bar{Q}\right]\left[\bar{\partial}\bar{Q}\right]^{-1}\left[\bar{\partial}Q\right]\right\}  & =\nonumber \\
\left(1+\sum_{k}\frac{1}{2}\partial_{k}Q_{k}^{(0)}(\{\theta\})\delta\bar{u}_{k}\right)\det\left\{ \partial_{i}Q_{j}^{\left(0\right)}(\{\theta_{\pm}\})-\sum_{k}\frac{1}{2}\left[\partial\bar{Q}\right]_{ik}\left[\bar{\partial}Q\right]_{kj}\delta\bar{u}_{k}\right\} \label{eq:inter}
\end{align}
where we used that at leading order $\left[\bar{\partial}\bar{Q}\right]^{-1}=\mathrm{diag}\left(\frac{\delta\bar{u}_{1}}{2},\ldots,\frac{\delta\bar{u}_{n}}{2}\right)$.
It is natural to introduce the density of states corresponding to
the quantization of order $(\mu$): $\rho_{N}^{\left(\mu\right)}(\{\theta\})$
as in (\ref{eq:rhomu}). Evaluating now expression (\ref{eq:inter})
at the solutions one can show that

\begin{equation}
\prod_{k}\left(\frac{2\delta\bar{u}_{k}}{\Gamma}\right)^{2}S(2i(u+\delta\bar{u}_{k}))\rho_{2N}(\{\bar{\theta}_{\pm}^{(0)}\})=\rho_{N}^{\left(\mu\right)}(\{\bar{\theta}^{(\mu)}\})\left(1+\sum_{k}\partial_{k}Q_{k}^{(0)}(\{\bar{\theta}^{(0)}\})\delta\bar{u}_{k}\right)
\end{equation}
where we also used that 
\begin{equation}
\det\left[\partial Q\right]\left\{ 1-\frac{1}{2}\sum_{k}\partial_{k}Q_{k}^{\left(0\right)}(\{\theta\})\delta\bar{u}_{k}\right\} =\det\left\{ \left[\partial Q\right]_{ij}-\frac{1}{2}\sum_{k}\left[\partial Q\right]_{ik}\left[\partial Q\right]_{kj}\delta\bar{u}_{k}\right\} 
\end{equation}
Finally, we expand the product of scattering matrices as 
\begin{equation}
S(\theta_{i+,j+})S(\theta_{i+,j-})S(\theta_{i-,j+})S(\theta_{i-,j-})=S(\theta_{i,j})\left(1+\bar{\partial}_{i}Q_{j}^{(0)}(\{\theta_{\pm}\})\delta u_{i}(\{\theta\})-\partial_{i}\bar{Q}_{j}^{\left(0\right)}(\{\theta_{\pm}\})\delta u_{j}(\{\theta\})\right)
\end{equation}
Collecting all factors the $\mu$-term of the finite volume form factor
can be parametrized as 
\begin{equation}
\langle0\vert\mathcal{O}\vert\{n\}\rangle_{L}=\frac{F_{N}(\{\bar{\theta}^{(\mu)}\})+\delta^{(\mu)}F_{N}(\{\bar{\theta}^{(\mu)}\})}{\sqrt{\prod\limits _{k<j}S(\bar{\theta}_{k,j}^{(\mu)})\rho_{N}^{(\mu)}(\{\bar{\theta}^{(\mu)}\})}}
\end{equation}
where the $\mu$-term correction takes the form 
\begin{align}
\delta^{(\mu)}F_{N}(\{\bar{\theta}^{(\mu)}\}) & =\sum_{k}\left\{ \frac{2}{\Gamma}F_{N,k}^{b}(\{\bar{\theta}^{(0)}\})-\frac{1}{2}\partial_{k}Q_{k}^{(0)}(\{\bar{\theta}^{(0)}\})F_{N}(\{\bar{\theta}^{(0)}\})\right\} \delta\bar{u}_{k}\nonumber \\
 & +\frac{1}{2}\sum_{k}\sum_{j<k}\left[\phi_{-}(\bar{\theta}_{j,k}^{(0)})\left(\delta\bar{u}_{j}+\delta\bar{u}_{k}\right)\right]F_{N}(\{\bar{\theta}^{(0)}\})
\end{align}
where we introduced $\phi_{-}(\theta)=\phi(\theta+2iu)-\phi(\theta-2iu)$.

\section{Formal derivation of the form factors' $F$-term correction\label{sec:Fterm}}

\begin{figure}
\begin{centering}
\includegraphics[width=5cm]{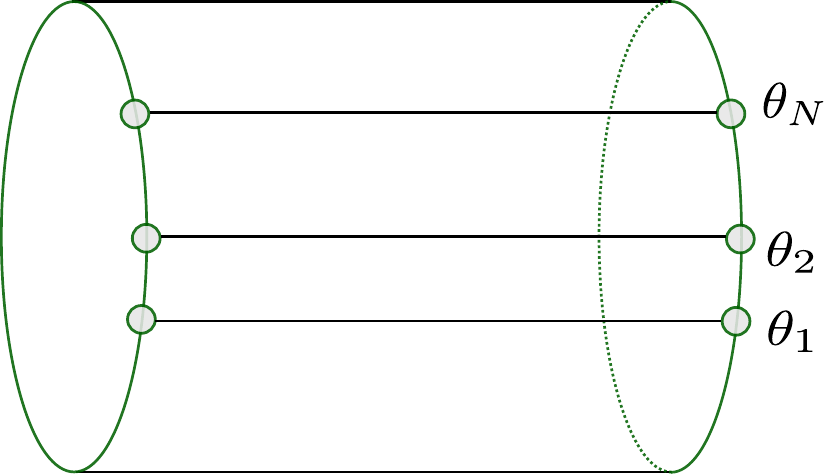}
\par\end{centering}
\caption{Graphical representation of the excited state partition function.
A physical particle with rapidity $\theta_{j}$ serves as a defect
operator with transmission factor $T(u)=S(\frac{i\pi}{2}+\theta_{j}-u)$.}

\label{expart}
\end{figure}

In this Appendix we give a formal derivation of the leading exponential
correction of the non-diagonal form factor. We work in a theory without
bound states and focus on the F-term correction only. As we explained
in the main text we have to evaluate the following expression

\begin{equation}
\frac{\mbox{Tr}(e^{-LH}\mathcal{O}_{N,M})}{\sqrt{\mbox{Tr}_{N}(e^{-LH})\mbox{Tr}_{M}(e^{-LH})}}\label{eq:ZOnm}
\end{equation}
where the normalization is related to the excited state partition
function 
\begin{equation}
Z_{N}=\mbox{Tr}_{N}(e^{-LH})
\end{equation}
which can be represented graphically as shown on Figure \ref{expart}.

In the general non-diagonal case, i.e. in formula (\ref{eq:ZOnm})
on the left half space we have the defect operator of the outgoing
state and on the right half that of the incoming excited state. These
half spaces are taken into account by the square roots in the normalization.
This normalization factor is also understood as the removal of the
operator in the trace, with keeping the incoming and outgoing particle
lines.

In the analysis of thermal two-point function we showed for small
particle numbers that infinite and finite volume normalizations can
be equivalent if the $\delta$-function is regularized properly \cite{Bajnok:2018fzx}.
By following the same steps we introduce two complete systems of states
to evaluate the trace:

\begin{equation}
\mbox{Tr}(e^{-LH}\mathcal{O}_{N,M})=\sum_{\nu,\mu}\langle\nu\vert O_{N,M}\vert\mu\rangle\langle\mu\vert\nu\rangle e^{-E_{\nu}L}
\end{equation}
In the following we evaluate the leading 1-particle contribution.
For the numerator we have
\begin{equation}
\mbox{Tr}(e^{-LH}O_{N,M})=F_{N+M}+\int\frac{du}{2\pi}\int\frac{dv}{2\pi}\langle v\vert O_{N,M}\vert u\rangle\langle u\vert v\rangle e^{-mL\cosh v}\label{eq:1pOnm}
\end{equation}
while for the normalization factor we obtain 
\begin{equation}
\mbox{Tr}_{N}(e^{-LH})=1+\int\frac{du}{2\pi}\int\frac{dv}{2\pi}\langle u\vert v\rangle\prod_{j}S(\frac{i\pi}{2}+\theta_{j}-v)\langle v\vert u\rangle e^{-mL\cosh v}
\end{equation}
Since infinite volume states are normalized to Dirac delta functions
$\langle u\vert v\rangle=2\pi\delta(u-v)$ we have to calculate the
square of the $\delta$-function. This is an ambiguous quantity, but
based on experience from the evaluation of the 2-point function we
regulate the $\delta$-function as 
\begin{equation}
2\pi\delta(u-v)=\frac{i}{u-v+i\epsilon}-\frac{i}{u-v-i\epsilon}
\end{equation}
Then we shift the $v$ contour from the real line above $i\epsilon$.
On the shifted integral the $\epsilon\to0$ limit can be taken, such
that due to the previous $\delta$-function no contribution will survive.
Thus we only need the residue term at $v=u+i\epsilon$. Calculating
the residue for the partition function we obtain: 
\begin{equation}
\mbox{Tr}_{N}(e^{-LH})=1+\int\frac{du}{2\pi}\frac{1}{\epsilon}\prod_{j}S(\frac{i\pi}{2}+\theta_{j}-u)e^{-mL\cosh u}+O(\epsilon)
\end{equation}
in which there is no $O(1)$ term. This is consistent with the usual
evaluation of the partition function based on finite volume regularization
where $mR\cosh u$ appears, instead of $\frac{1}{\epsilon}$ .

\begin{figure}
\begin{centering}
\includegraphics[width=5cm]{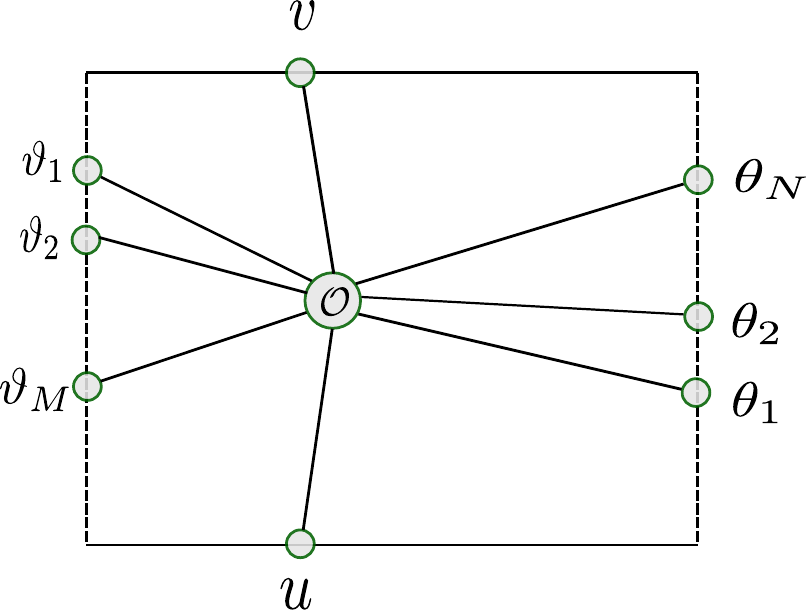}
\par\end{centering}
\caption{Graphical representation of the matrix element $\langle v\vert O_{n,m}\vert u\rangle$,
which is the matrix element of the non-local operator ${\cal O}_{n,m}$.}

\label{FOnm}
\end{figure}

Let us focus on the form factor contribution. The matrix element $\langle v\vert O_{n,m}\vert u\rangle$
can be represented graphically as on Figure \ref{FOnm}.

Using the crossing properties of the form factors $\langle v\vert O_{N,M}\vert u\rangle$
can be written as 
\begin{eqnarray}
\langle v\vert O_{N,M}\vert u\rangle & = & F_{N+M+2}(v+i\pi-i\epsilon,\{\vartheta+\frac{i\pi}{2}\},u,\{\theta-\frac{i\pi}{2}\})+\\
 &  & 2\pi\delta(v-u)\prod_{j}S(\frac{i\pi}{2}+\vartheta_{j}-u)F_{N+M}(\{\vartheta+\frac{i\pi}{2}\},\{\theta-\frac{i\pi}{2}\})\nonumber 
\end{eqnarray}
Alternatively, using the permutation property of the infinite volume
form factors we can write
\begin{eqnarray}
\langle v\vert O_{N,M}\vert u\rangle & = & \prod_{j}S(\frac{i\pi}{2}+\vartheta_{j}-u)F_{N+M+2}(v+i\pi-i\epsilon,u,\{\vartheta+\frac{i\pi}{2}\},\{\theta-\frac{i\pi}{2}\})+\nonumber \\
 &  & 2\pi\delta(v-u)\prod_{j}S(\frac{i\pi}{2}+\vartheta_{j}-u)F_{N+M}(\{\vartheta+\frac{i\pi}{2}\},\{\theta-\frac{i\pi}{2}\})
\end{eqnarray}
Separating the singular part of the form factor as
\begin{align}
F_{N+M+2}(v+i\pi,u,\{\vartheta+\frac{i\pi}{2}\},\{\theta-\frac{i\pi}{2}\}) & =\frac{i}{v-u}\biggl(1-\frac{\prod_{k}S(\frac{i\pi}{2}+\theta_{k}-u)}{\prod_{j}S(\frac{i\pi}{2}+\vartheta_{j}-u)}\biggr)F_{N+M}(\{\vartheta+\frac{i\pi}{2}\},\{\theta-\frac{i\pi}{2}\})\nonumber \\
 & +F_{N+M+2}^{c}(v+i\pi,u,\{\vartheta+\frac{i\pi}{2}\},\{\theta-\frac{i\pi}{2}\})
\end{align}
and introducing the connected part of the form factor, such that combining
this with the $\delta$-term we obtain: 
\begin{eqnarray}
\langle v\vert O_{N,M}\vert u\rangle & = & \prod_{j}S(\frac{i\pi}{2}+\vartheta_{j}-u)F_{N+M+2}^{c}(v+i\pi,u,\{\vartheta+\frac{i\pi}{2}\},\{\theta-\frac{i\pi}{2}\})+\\
 &  & \left(\frac{i\prod_{j}S(\frac{i\pi}{2}+\vartheta_{j}-u)}{v-u+i\epsilon}-\frac{i\prod_{k}S(\frac{i\pi}{2}+\theta_{k}-u)}{v-u-i\epsilon}\right)F_{N+M}(\{\vartheta+\frac{i\pi}{2}\},\{\theta-\frac{i\pi}{2}\})\nonumber 
\end{eqnarray}
Plugging this expression back to eq. (\ref{eq:1pOnm}) and evaluating
the $v$ integral the same way we did for the partition functions
we obtain the singular term 
\begin{equation}
\frac{F_{N+M}(\{\vartheta+\frac{i\pi}{2}\},\{\theta-\frac{i\pi}{2}\})}{2\epsilon}\int\frac{du}{2\pi}\left\{ \prod_{j}S(\frac{i\pi}{2}+\vartheta_{j}-u)+\prod_{k}S(\frac{i\pi}{2}+\theta_{k}-u)\right\} e^{-mL\cosh u}
\end{equation}
This only cancels the singular term coming from the $\sqrt{Z_{N}Z_{M}}$
normalizations. The O(1) term gives the finite size correction
\begin{eqnarray}
\int\frac{du}{2\pi}\left\{ \prod_{j}S(\frac{i\pi}{2}+\vartheta_{j}-u)F_{N+M+2}^{c}(u+i\pi,u,\{\vartheta+\frac{i\pi}{2}\},\{\theta-\frac{i\pi}{2}\})-\hspace{4cm}\right.\\
\left.\frac{F_{N+M}(\{\vartheta+\frac{i\pi}{2}\},\{\theta-\frac{i\pi}{2}\})}{2}imL\sinh u\left(\prod_{j}S(\frac{i\pi}{2}+\vartheta_{j}-u)-\prod_{k}S(\frac{i\pi}{2}+\theta_{k}-u)\right)\right\} e^{-mL\cosh u}\nonumber 
\end{eqnarray}
which after integration by parts gives the regulated expression 
\begin{align}
F_{N+M+2}^{r}(u+i\pi,\{\vartheta+\frac{i\pi}{2}\},u,\{\theta-\frac{i\pi}{2}\})=\lim_{\epsilon\to0}\biggl\{ F_{N+M+2}(u+i\pi+\frac{\epsilon}{2},\{\vartheta+\frac{i\pi}{2}\},u-\frac{\epsilon}{2},\{\theta-\frac{i\pi}{2}\})\hspace{2cm}\nonumber \\
-\frac{i}{\epsilon}\bigl(\prod_{j}S(\vartheta_{j}+\frac{i\pi}{2}-u)-\prod_{j}S(\theta_{k}+\frac{i\pi}{2}-u)\bigr)F_{N+M}(\{\vartheta+\frac{i\pi}{2}\},\{\theta-\frac{i\pi}{2}\})\biggr\}\hspace{2cm}\label{eq:regFFgen}
\end{align}
 equivalent to(\ref{eq:FF_Fterm}).

\section{Relation between $\mu$- and $F$-terms\label{sec:muFrel}}

In this Appendix we show that, similarly to the Bethe-Yang equations
and energy formulas, the $\mu$-term form factor corrections can be
obtained by taking appropriate residues of the $F$-terms corrections.
Due to the additive structure of the correction (\ref{eq:inoutmu})
we analyze the form factor only containing incoming particles. We
start with the $F$-term formula
\begin{align}
\langle0\vert\mathcal{O}\vert\{n_{j}\}\rangle_{L} & =\frac{F_{N}(\{\bar{\theta}^{(1)}\})+\delta^{(F)}F_{N}(\{\bar{\theta}^{\left(1\right)}\})}{\sqrt{\prod_{i<j}S(\bar{\theta}_{i,j}^{(1)})\rho_{N}^{(1)}(\{\bar{\theta}^{(1)}\})}}+\dots
\end{align}
where the $F$-term correction contains the regulated form factor

\begin{equation}
\delta^{(F)}F_{N}(\{\theta\})=\int duF_{N+2}^{r}(v+i\pi,v,\{\theta-\frac{i\pi}{2}\})e^{-mL\cosh v}
\end{equation}
As we already observed in the case of the energy correction, by summing
half the residues at $v=\theta_{k}+iu-\frac{i\pi}{2}$ and subtracting
half the residues at the complex conjugate points $v=\theta_{k}-iu+\frac{i\pi}{2}$
in each $F$-term integral, the $\mu$-term expressions can be obtained.
In the particular case of the Bethe-Yang equation, $Q_{k}^{(\mu)}(\{\theta\})$
can be obtained from $Q_{k}^{(1)}(\{\theta\})$ using this method.
This implies that at order $(\mu)$ the solutions $\bar{\theta}_{j}^{(1)}$
will be replaced by the solutions $\bar{\theta}_{j}^{(\mu)}$ and
$\rho_{N}^{(1)}(\{\bar{\theta}^{(1)}\})$ by $\rho_{N}^{(\mu)}(\{\bar{\theta}^{(\mu)}\})$,
respectively. Thus, we only need to show that the residue of $\delta^{(F)}F_{N}(\{\bar{\theta}^{(1)}\})$
will reproduce $\delta^{(\mu)}F_{N}(\{\bar{\theta}^{(\mu)}\})$. 

Instead of the symmetric definition (\ref{eq:regFFgen}) of the regulated
form factor we can use alternative formulations depending on how we
take the limit. We introduce two connected form factors $(F^{c},F^{\bar{c}}$)
as:
\begin{equation}
F_{N+2}^{r}(v+i\pi,v,\{\theta\})=F_{N+2}^{c/\bar{c}}(v+i\pi,v,\{\theta\})\pm\frac{i}{2}\partial_{v}\left(\prod_{j}S(v-\theta_{j})\right)F(\{\theta\})\label{eq:FregFcbar}
\end{equation}
The definition
\begin{align}
F_{N+2}^{\alpha}(v+i\pi,v,\{\theta\}) & \equiv\lim_{\epsilon\to0}\{F_{N+2}(v+i\pi+\epsilon\left(\frac{1+\alpha}{2}\right),v-\epsilon\left(\frac{1-\alpha}{2}\right),\{\theta\})-\nonumber \\
 & -\frac{i}{\epsilon}\left(1-\prod_{j}S(v-\theta_{j})\right)F_{N}(\{\theta\})\}
\end{align}
summarizes the various subtracted form factors, which can be obtained
as: $c\leftrightarrow\alpha=1,\ r\leftrightarrow\alpha=0,\ \bar{c}\leftrightarrow\alpha=-1$.
These alternative choices are simpler to deal with, since $F_{N+2}^{\bar{c}}$
contains only a simple pole at $v=\theta_{k}+iu-\frac{i\pi}{2}$ (the
same is true for $F_{N+2}^{c}$ at $v=\theta_{k}-iu+\frac{i\pi}{2}$),
while the derivative term in (\ref{eq:FregFcbar}) gives always a
second order pole. The distribution of the poles in the direct evaluation
of $F_{N+2}^{r}$ from eq. (\ref{eq:regFFgen}) is less clear. We
focus on the residue at $v=\theta_{k}+iu-\frac{i\pi}{2}$ as the other
one is related to this by complex conjugation and investigate the
singularity structure of $F_{N+2}(v+i\pi,v-\epsilon,\{\theta-\frac{i\pi}{2}\})$
near the pole $v=\theta_{k}+iu-\frac{i\pi}{2}+i\delta$, i.e. in $\delta$.
Using the monodromy axioms we can move the virtual particles to sandwich
$\theta_{k}-\frac{i\pi}{2}$ : 
\begin{align}
F_{N+2}(v+i\pi,v-\epsilon,\{\theta-\frac{i\pi}{2}\})=\hspace{8.5cm}\nonumber \\
\prod_{j:j<k}S(\frac{i\pi}{2}+v-\theta_{j}-\epsilon)\prod_{j:j>k}S(\frac{i\pi}{2}+v-\theta_{j})F_{N+2}(\ldots,v-\epsilon,\theta_{k}-\frac{i\pi}{2},v-i\pi,\ldots)\label{eq:prefactor}
\end{align}
Let us define $\hat{\theta}_{k\pm}\equiv\theta_{k}\pm iu+i\delta$.
These arguments, having shifted by $i\frac{\pi}{2}$, take the form
$(\dots,\hat{\theta}_{k+}-\epsilon,\theta_{k},\hat{\theta}_{k+}-i\pi,\dots)$.
For scalar operators an overall $-i\frac{\pi}{2}$ shift in the arguments
of the form factors has no effect. Then we use the dynamical pole
axiom in the second and third of these arguments; after that we repeat
the same in $\hat{\theta}_{k+}-\epsilon,\hat{\theta}_{k-}$, too:
\begin{align}
F_{N+2}(...,\hat{\theta}_{k+}-\epsilon,\theta_{k},\hat{\theta}_{k+}-i\pi,...) & =\frac{i\Gamma}{-i\delta}F_{N+1}(...,\hat{\theta}_{k+}-\epsilon,\hat{\theta}_{k-},...)+\mathcal{O}(\delta^{0})=\nonumber \\
 & =\frac{i\Gamma}{-i\delta}\left[\frac{i\Gamma}{-\epsilon}F_{N}(\ldots,\theta_{k},\ldots)+\hat{F}_{N,k+}^{b}(\ldots,\theta_{k},\ldots)+\mathcal{O}(\epsilon)\right]+\mathcal{O}(\delta^{0}).
\end{align}
Since we explicitly subtracted the $\epsilon^{-1}$ singularity in
the first and third arguments (due to the definition of $F_{N+2}^{\bar{c}}$)
the term proportional to $(\epsilon\delta)^{-1}$ disappears in the
$\epsilon\to0$ limit. So the remaining singularity is a simple $\delta^{-1}$
pole proportional to $\hat{F}_{N,k+}^{b}(\{\theta\})$, where $\hat{F}_{N,k\pm}^{b}$
and $F_{N,k}^{b}$ (see (\ref{eq:bkff})) are analogous to $F^{c/\bar{c}}$
and $F^{r}$ regarding the type of subtraction of the dynamical singularity.
Notice that the prefactor in (\ref{eq:prefactor}) contains $\mathcal{O}(\epsilon)$
terms, and their product with $(\epsilon\delta)^{-1}$ contributes
to the residue of $\delta^{-1}$ in $F_{N+2}^{\bar{c}}$ as well.
The second, derivative term in (\ref{eq:FregFcbar}) gives a $\delta^{-2}$
pole. The sum of these (after reordering, up to $\mathcal{O}(\delta^{0})$)
can be expressed in terms of S-matrices as ($\hat{\theta}_{k+,j}\equiv\hat{\theta}_{k+}-\theta_{j}$)
\begin{align}
F_{N+2}^{r}(v+i\pi,v,\{\theta-\frac{i\pi}{2}\})\vert_{v=\hat{\theta}_{k+}-\frac{i\pi}{2}}=-\frac{i}{2}S'(\hat{\theta}_{k+,k})\prod_{j\neq k}S(\hat{\theta}_{k+,j})F_{N}(\{\theta\})+\prod_{j}S(\hat{\theta}_{k+,j})\Gamma^{-1}\hat{F}_{N,k+}^{b}(\{\theta\})\nonumber \\
+\frac{i}{2}\left(\left[\prod_{j:j<k}S(\hat{\theta}_{k+,j})\right]^{'}\prod_{j:j\geq k}S(\hat{\theta}_{k+,j})-\prod_{j:j\leq k}S(\hat{\theta}_{k+,j})\left[\prod_{j:j>k}S(\hat{\theta}_{k+,j})\right]^{'}\right)F_{N}(\{\theta\})
\end{align}
where, by taking the residues, the first term gives $\frac{i}{2}\partial_{k}\delta u_{k+}$,
the second $\delta u_{k+}$, and in the second line we get $-\frac{1}{2}\sum_{j<k}i\partial_{j}\delta u_{k+}+\frac{1}{2}\sum_{j>k}i\partial_{j}\delta u_{k+}$.
We can repeat the same steps for the complex conjugate pole at $v=\theta_{k}-iu+\frac{i\pi}{2}$,
starting from $F_{N+2}^{c}.$ In the end we can take half the difference
of the two contributions (see Figure \ref{fig:contour}): 
\begin{align}
\delta^{(\mu)}F_{N}(\{\theta\}) & =\frac{1}{2}\sum_{k,\pm}\pm i\text{Res}_{v\to\theta_{k}\mp(\frac{i\pi}{2}-iu)}\left\{ F_{N+2}^{r}(v+i\pi,v,\{\theta_{j}-\frac{i\pi}{2}\})e^{-mL\cosh v}\right\} =\nonumber \\
 & =\sum_{k,\pm}\bigl\{\Gamma^{-1}\hat{F}_{N,k\pm}^{b}(\{\theta\})\delta u_{k\pm}\mp\frac{i}{2}F_{N}(\{\theta\})(-\partial_{k}+\sum_{j:j<k}\partial_{j}-\sum_{j:j>k}\partial_{j})\delta u_{k\pm}\bigr\}
\end{align}
Using that $\partial_{j}u_{k\pm}=\pm i\partial_{j}Q_{k\pm}^{(0)}(\{\theta_{\pm}\})\delta u_{j\pm}$
we evaluate the expression at the leading order solution $\theta_{j}=\bar{\theta}_{j}^{(0)}$
to obtain
\begin{align}
\delta^{(\mu)}F_{N}(\{\bar{\theta}^{(0)}\}) & =\sum_{k}\left\{ \frac{2}{\Gamma}F_{N,k}^{b}(\{\bar{\theta}^{(0)}\})-\frac{1}{2}\partial_{k}Q_{k}^{(0)}(\{\bar{\theta}^{(0)}\})F_{N}(\{\bar{\theta}^{(0)}\})\right\} \delta\bar{u}_{k}\nonumber \\
 & +\frac{1}{2}\sum_{j<k}\left[\phi_{-}(\bar{\theta}_{j,k}^{(0)})\left(\delta\bar{u}_{j}+\delta\bar{u}_{k}\right)\right]F_{N}(\{\bar{\theta}^{(0)}\})
\end{align}
where we exploited that $\hat{F}_{N,k+}^{b}(\{\theta\})+\hat{F}_{N,k-}^{b}(\{\theta\})=2F_{N,k}^{b}(\{\theta\})$.
This is exactly the same form as what we obtained earlier (\ref{eq:FFmu}).

\section{Relating the massive boson scheme to the massless one\label{sec:relate_m_mless}}

In this Appendix we relate the two alternative descriptions of the
sinh-Gordon theory based on the perturbation of the massless and the
massive free boson theories. First we consider the free massive boson
on the cylinder as a perturbation of the massless one as in eq. (\ref{eq:Hm0}).
Let us introduce a new set of creation operators $\alpha_{n}$ as
\begin{equation}
a_{n}=\begin{cases}
-i\sqrt{n}\alpha_{n} & \quad n>0\\
i\sqrt{\left|n\right|}\alpha_{\left|n\right|}^{\dagger} & \quad n<0
\end{cases};\quad\bar{a}_{n}=\begin{cases}
-i\sqrt{n}\alpha_{-n} & \quad n>0\\
i\sqrt{\left|n\right|}\alpha_{-\left|n\right|}^{\dagger} & \quad n<0
\end{cases}
\end{equation}
We perform a Bogoliubov transformation, which acts on the massless
Fock states with a unitary operator
\begin{equation}
U=\exp\left\{ -\sum_{m>0}\chi_{m}\left(\alpha_{m}\alpha_{-m}-\alpha_{m}^{\dagger}\alpha_{-m}^{\dagger}\right)\right\} \quad;\qquad e^{2\chi_{n}}=\frac{\left|k_{n}\right|}{\omega_{n}}
\end{equation}
The creation operators transform according to
\begin{equation}
U\alpha_{n}^{\dagger}U^{\dagger}=d_{n}^{\dagger};\quad\alpha_{n}=\cosh\chi_{n}d_{n}+\sinh\chi_{n}d_{-n}^{\dagger}
\end{equation}
We would like to emphasize that obtaining the massive vacuum by acting
$U$ on the massless ground state indicates that the massive basis
is significantly different from the massless one (from the truncated
space point of view). The field operator (\ref{eq:field}) is expressed
in terms of the new massive creation operators $d_{n}$ as in eq.
(\ref{eq:phi_d}). Finally, introducing $\varphi_{0}=\left(2MLg\right)^{-1/2}\left(d_{0}+d_{0}^{\dagger}\right)$,
the Hamiltonian (\ref{eq:H0m}) becomes the free massive boson Hamiltonian
(\ref{eq:free_massive_H}). The vacuum energy contribution $\tilde{E}_{0}^{\prime}$
appears due to the difference between normal ordering with respect
to the mode operators $\alpha_{n}$ and $d_{n}$.

Considering the sinh-Gordon model as a perturbation of the massive
boson the normal ordering is chosen at infinite volume (\ref{eq:massive_scheme}),
i.e. $::_{M,\infty}$ means normal ordering with respect to the modes
$d_{k}$ in infinite volume. Our goal is to connect the bare parameter
$M$ to the bare coupling $\mu$ in the conformal plus zero mode scheme.
As a first step, $H^{\left(L\rightarrow\infty\right)}$ needs to be
connected to the Hamiltonian on the cylinder. This is achieved by
requiring that the perturbation has the same behavior in the UV for
both, i.e. the Hamiltonian density expressed in terms of bare fields
takes the same form for all volumes (\ref{eq:Hmass}). Let us assume
that we have temporarily introduced an UV momentum cutoff $\Lambda$.
We use the BCH formula
\begin{equation}
e^{X+Y}=e^{-\frac{1}{2}\left[X,Y\right]}e^{X}e^{Y},\quad\text{if}\quad\left[X,\left[X,Y\right]\right]=\left[Y,\left[X,Y\right]\right]=0
\end{equation}
to relate
\begin{equation}
e^{b\varphi\left(0,0\right)}=e^{\frac{b^{2}}{2}\left[\varphi_{+},\varphi_{-}\right]_{M,L}^{\left(\Lambda\right)}}:e^{b\varphi\left(0,0\right)}:_{M,L}=e^{\frac{b^{2}}{2}\left[\varphi_{+},\varphi_{-}\right]_{M,\infty}^{\left(\Lambda\right)}}:e^{b\varphi\left(0,0\right)}:_{M,\infty}
\end{equation}
In the relation between the two normal ordered quantities the limit
$\Lambda\rightarrow\infty$ can be taken leading to (\ref{eq:rho}).
Note that the coefficient diverges in the limit $L\rightarrow0$.
Then, we bring the zero mode exponentials out of normal ordering
\[
:e^{b\varphi_{0}}:_{M,L}=e^{-\frac{b^{2}}{4gLM}}e^{b\varphi_{0}}
\]
 and (keeping in mind an UV cutoff again) we can obtain ($\tilde{\varphi}=\varphi-\varphi_{0}$)
\begin{equation}
:e^{b\tilde{\varphi}\left(x\right)}:_{M,L}=e^{\frac{b^{2}}{4gL}\sum_{q\neq0}\left(\frac{1}{\left|k_{q}\right|}-\frac{1}{\omega_{q}}\right)}:e^{b\tilde{\varphi}\left(x\right)}:_{0,L}
\end{equation}
where the sum in the exponent has an integral representation
\begin{equation}
\frac{1}{L}\sum_{q\neq0}\left(\frac{1}{\left|k_{q}\right|}-\frac{1}{\omega_{q}}\right)=\frac{1}{ML}+\frac{1}{\pi}\ln\frac{ML}{4\pi}-2\rho\left(ML\right)+\frac{\gamma_{E}}{\pi}
\end{equation}
Comparing (\ref{eq:Hmass}) with (\ref{eq:Hm0}) we arrive at the
relations (\ref{eq:muM}) and (\ref{eq:E0M}).

\end{document}